%% file: medima-template.tex
\documentclass[times,twocolumn,final]{elsarticle}

\usepackage{medima}
\usepackage{framed,multirow}
\usepackage{comment}
\usepackage{amssymb}
\usepackage{latexsym}
\usepackage{amsmath}
\usepackage{enumitem}
\usepackage[toc,page]{appendix}
\usepackage[switch]{lineno} 
\usepackage{url}
\usepackage{xcolor}
\usepackage{subcaption}

\DeclareMathOperator*{\argmin}{arg\,min}
\usepackage[colorlinks = true,
            linkcolor = tealblue,
            urlcolor  = tealblue,
            citecolor = tealblue,
            anchorcolor = tealblue]{hyperref}
\setcitestyle{authoryear,open={},close={}}
\usepackage{mathtools}

\DeclarePairedDelimiter\floor{\lfloor}{\rfloor}

\definecolor{newcolor}{rgb}{.0, .502, .675}

\makeatletter
\AtBeginDocument{\def\@citecolor{newcolor}}
\AtBeginDocument{\def\@linkcolor{newcolor}}
\AtBeginDocument{\def\@anchorcolor{newcolor}}
\AtBeginDocument{\def\@filecolor{newcolor}}
\AtBeginDocument{\def\@urlcolor{newcolor}}
\AtBeginDocument{\def\@menucolor{newcolor}}
\AtBeginDocument{\def\@pagecolor{newcolor}}
\makeatother

\begin{document}

\verso{Junyu Chen \textit{et~al.}}

\begin{frontmatter}

\title{TransMorph: Transformer for unsupervised medical image registration}

\author[1,2]{Junyu \snm{Chen}\corref{cor1}}
\ead{jchen245@jhmi.edu}
\cortext[cor1]{Corresponding author} 
\author[1,2]{Eric C. \snm{Frey}}
\ead{efrey@jhmi.edu}
\author[5]{Yufan \snm{He}}
\ead{yufanh@nvidia.com}
\author[3]{William P. \snm{Segars}}
\ead{paul.segars@duke.edu}
\author[4]{Ye \snm{Li}}
\ead{gary.li@mgh.harvard.edu}
\author[1]{Yong \snm{Du}}
\ead{duyong@jhmi.edu}
\address[1]{Russell H. Morgan Department of Radiology and Radiological Science, Johns Hopkins Medical Institutes, Baltimore, MD, USA}
\address[2]{Department of Electrical and Computer Engineering, Johns Hopkins University, Baltimore, MD, USA}
\address[3]{Carl E. Ravin Advanced Imaging Laboratories, Department of Radiology, Duke University Medical Center, Durham, NC, USA}
\address[4]{Center for Advanced Medical Computing and Analysis, Massachusetts General Hospital and Harvard Medical School, Boston, MA, USA}
\address[5]{NVIDIA Corporation, Bethesda, MD, USA}

\received{xx xx 20xx}
\finalform{xx xx 20xx}
\accepted{xx xx 20xx}
\availableonline{xx xx 20xx}

\begin{abstract}
In the last decade, convolutional neural networks (ConvNets) have been a major focus of research in medical image analysis. However, the performances of ConvNets may be limited by a lack of explicit consideration of the long-range spatial relationships in an image. Recently Vision Transformer architectures have been proposed to address the shortcomings of ConvNets and have produced state-of-the-art performances in many medical imaging applications. Transformers may be a strong candidate for image registration because their substantially larger receptive field enables a more precise comprehension of the spatial correspondence between moving and fixed images. Here, we present TransMorph, a hybrid Transformer-ConvNet model for volumetric medical image registration. This paper also presents diffeomorphic and Bayesian variants of TransMorph: the diffeomorphic variants ensure the topology-preserving deformations, and the Bayesian variant produces a well-calibrated registration uncertainty estimate. We extensively validated the proposed models using 3D medical images from three applications: inter-patient and atlas-to-patient brain MRI registration and phantom-to-CT registration. The proposed models are evaluated in comparison to a variety of existing registration methods and Transformer architectures. Qualitative and quantitative results demonstrate that the proposed Transformer-based model leads to a substantial performance improvement over the baseline methods, confirming the effectiveness of Transformers for medical image registration.
\end{abstract}


\end{frontmatter}
\thispagestyle{empty}

\section{Introduction}
\label{sec1}
\input{introduction}
\begin{figure*}[!t]
\centering
\includegraphics[width=.98\textwidth]{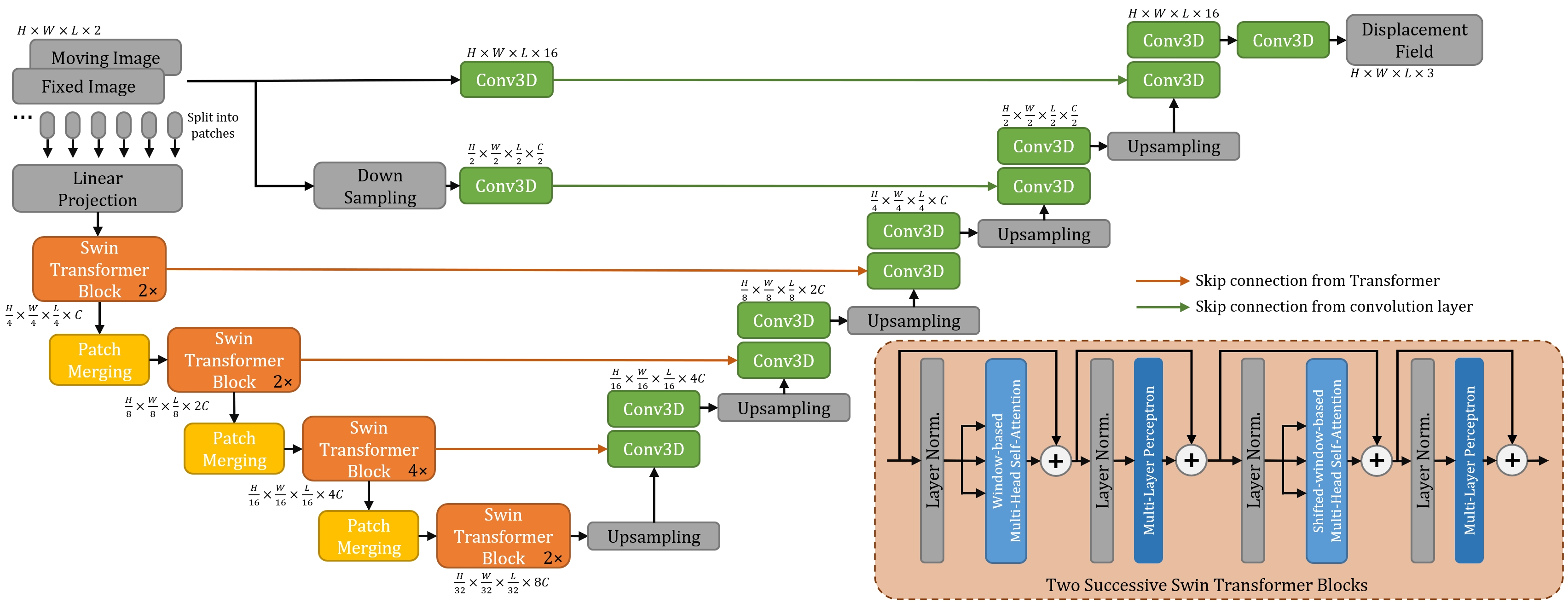}
\caption{The architecture of the proposed \texttt{TransMorph} registration network.\label{fig:net_arch}}
\end{figure*}

\begin{figure}[!h]
\centering
\includegraphics[width=.47\textwidth]{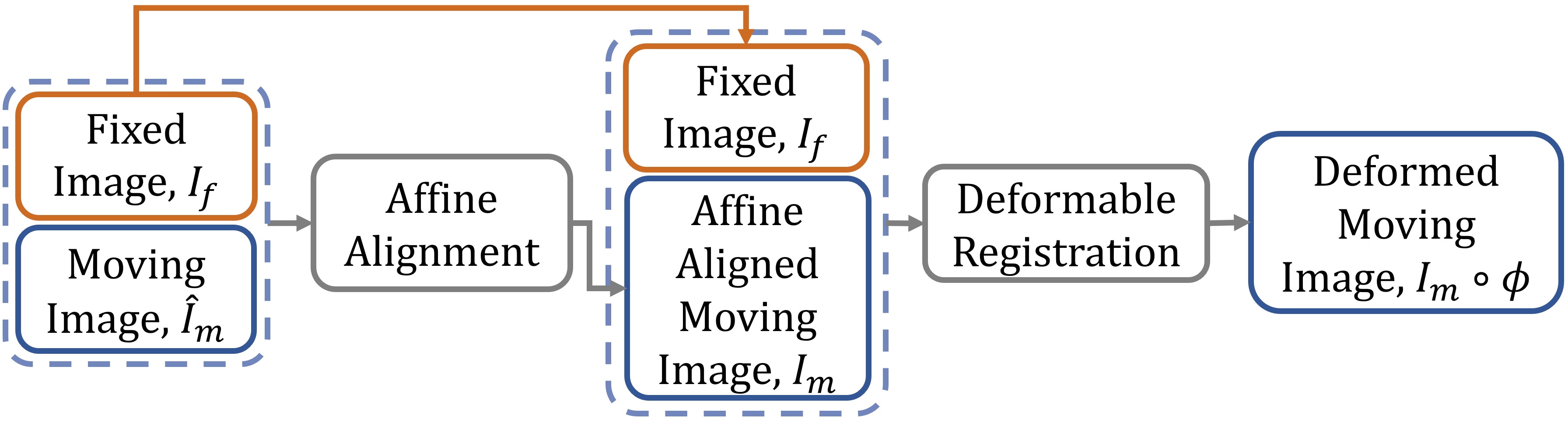}
\caption{The conventional paradigm of image registration.\label{fig:reg_pipe}}
\end{figure}

\section{Related Work}
\label{sec:rel_work}
\input{rel_works}

\section{Methods}
\label{sec:methods}
The conventional paradigm of image registration is shown in Fig. \ref{fig:reg_pipe}. The moving and fixed images, denoted respectively as $\hat{I}_m$ and $I_f$, are first affinely transformed into a single coordinate system. The resulting affine-aligned moving image is denoted as $I_m$. Subsequently, $I_m$ is warped to $I_f$ using a deformation field, $\phi$, generated by a DIR algorithm (i.e., $I_m\circ\phi$). Fig. \ref{fig:overall} presents an overview of the proposed method. Here, both the affine transformation and the deformable registration are performed using Transformer-based neural networks. The affine Transformer takes $\hat{I}_m$ and $I_f$ as inputs and computes a set of affine transformation parameters (e.g., rotation angle, translation, etc.). These parameters are used to affinely align $\hat{I}_m$ with $I_f$ via an affine transformation function, yielding an aligned image $I_m$. Then, a DIR network computes a deformation field $\phi$ given $I_m$ and $I_f$, which warps $I_m$ using a spatial transformation function (i.e., $\hat{I}_f=I_m\circ\phi$). During training, the DIR network may optionally include supplementary information (e.g., anatomical segmentation). The network architectures, the loss and regularization functions, and the variants of the method are described in detail in the following sections.

\begin{figure}[!h]
\centering
\includegraphics[width=.47\textwidth]{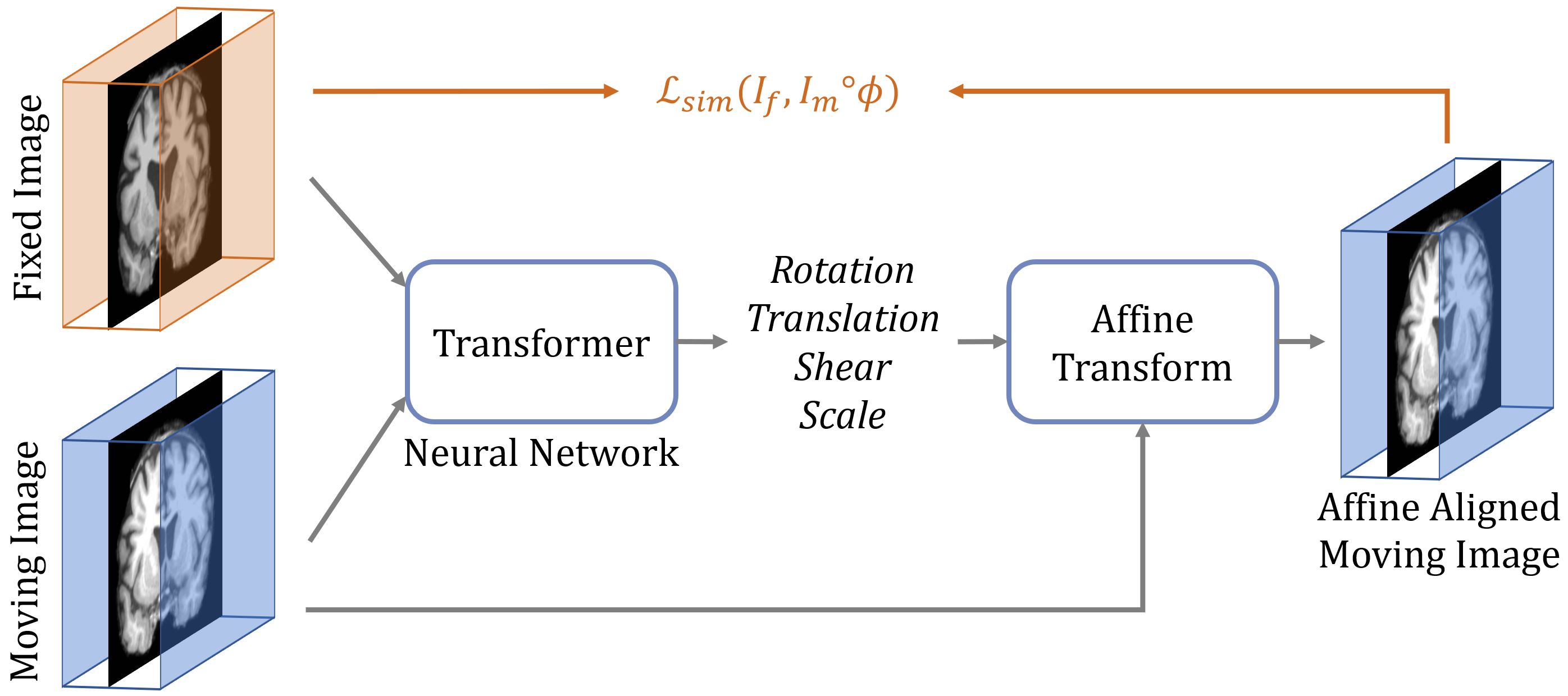}
\caption{The framework of the proposed Transformer-based affine model.}
\end{figure}
\subsection{Affine Transformation Network}

Affine transformation is often used as the initial stage in image registration because it facilitates the optimization of the following more complicated DIR processes~(\cite{de2019deep}). An affine network examines a pair of moving and fixed images globally and produces a set of transformation parameters that aligns the moving image with the fixed image. Here, the architecture of the proposed Transformer-based affine network is a modified Swin Transformer~(\cite{liu2021swin}) that takes two 3D volumes as the inputs (i.e., $I_f$ and $\hat{I}_m$) and generates 12 affine parameters: three rotation angles, three translation parameters, three scaling parameters, and three shearing parameters. The details and a visualization of the architecture are shown in Fig. \ref{fig:affine_arch} in the Appendix. We reduced the number of parameters in the original Swin Transformer due to the relative simplicity of affine registration. The specifics of the Transformer's architecture and parameter settings are covered in a subsequent section.
\begin{figure*}[!h]
\centering
\includegraphics[width=.92\textwidth]{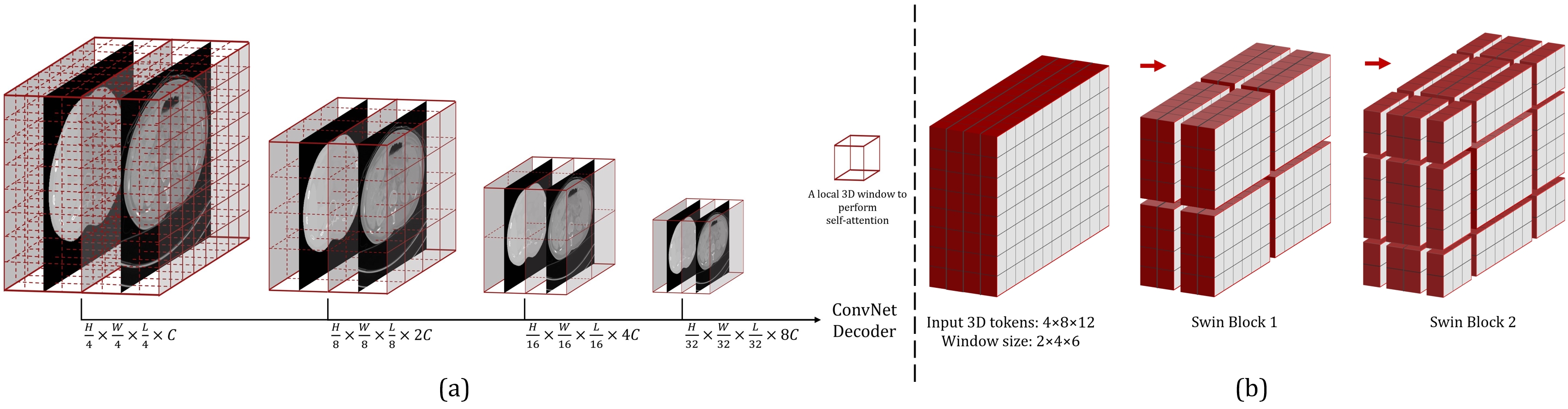}
\caption{(a): Swin Transformer creates hierarchical feature maps by merging image patches. The self-attention is computed within each local 3D window (the red box). The feature maps generated at each resolution are sent into a ConvNet decoder to produce an output. (b): The 3D cyclic shift of local windows for shifted-window-based self-attention computation. \label{fig:swin_feat}}
\end{figure*}
\subsection{Deformable Registration Network}
\label{sec:reg_net}
Fig. \ref{fig:net_arch} shows the network architecture of the proposed \texttt{TransMorph}. The encoder of the network first splits the input moving and fixed volumes into non-overlapping 3D patches, each of size $2\times P\times P\times P$, where $P$ is typically set to 4~(\cite{dosovitskiy2020image,liu2021swin,dong2021cswin}). We denote the $i^{th}$ patch as $x_p^i$, where $i\in\{1,...,N\}$ and $N=\frac{H}{P}\times\frac{W}{P}\times\frac{L}{P}$ is the total number of patches. Each patch is flattened and regarded as a ``token", and then a linear projection layer is used to project each token to a feature representation of an arbitrary dimension (denoted as $C$):
\begin{linenomath}
\begin{equation}
\label{eqn:embding}
    \mathbf{z}_0 = [x_p^1\mathbf{E};x_p^2\mathbf{E};...;x_p^N\mathbf{E}],
\end{equation}
\end{linenomath}
where $\mathbf{E}\in\mathbb{R}^{2P^3\times C}$ denotes the linear projection, and the output $\mathbf{z}_0$ has a dimension of $N\times C$. 

Because the linear projection operates on image patches and does not keep the token's location relative to the image as a whole, previous Transformer-based models often added a positional embedding to the linear projections in order to integrate the positional information into tokens, i.e. $\mathbf{z}_0 + \mathbf{E}_{pos}$~(\cite{vaswani2017attention, dosovitskiy2020image, liu2021swin, dong2021cswin}). Such Transformers were primarily designed for image classification, where the output is often a vector describing the likelihood of an input image being classified as a certain class. Thus, if the positional embedding is not employed, the Transformer may lose the positional information. However, for pixel-level tasks such as image registration, the network often includes a decoder that generates a dense prediction with the same resolution as the input or target image. The spatial correspondence between voxels in the output image is enforced by comparing the output with the target image using a loss function. Any spatial mismatches between output and target would contribute to the loss and be backpropagated into the Transformer encoder. The Transformer should thereby inherently capture the tokens' positional information. In this work, we observed, as will be shown in section \ref{sec:pos_embd}, that positional embedding is not necessary for image registration, and it only adds extra parameters to the network without improving performance.

Following the linear projection layer, several consecutive stages of patch merging and Swin Transformer blocks~(\cite{liu2021swin}) are applied on the tokens $\mathbf{z}_0$. The Swin Transformer blocks outputs the same number of tokens as the input, while the patch merging layers concatenate the features of each group of $2\times2\times2$ neighboring tokens, thus they reduce the number of tokens by a factor of $2\times2\times2=8$ (e.g., $H\times W\times L\times C\xrightarrow{}\frac{H}{2}\times\frac{W}{2}\times\frac{L}{2}\times 8C$). Then, a linear layer is applied on the $8C$-dimensional concatenated features to produce features each of $2C$-dimension. After four stages of Swin Transformer blocks and three stages of patch merging in between the Transformer stages (i.e., orange boxes in Fig. \ref{fig:net_arch}), the output dimension at the last stage of the encoder is $\frac{H}{32}\times\frac{W}{32}\times\frac{L}{32}\times 8C$. The decoder consists of successive upsampling and convolutional layers with the kernel size of $3\times3$. Each of the upsampled feature maps in the decoding stage was concatenated with the corresponding feature map from the encoding path via skip connections, then followed by two consecutive convolutional layers. As shown in Fig. \ref{fig:net_arch}, the Transformer encoder can only provide feature maps up to a resolution of $\frac{H}{P}\times\frac{W}{P}\times\frac{L}{P}$ owing to the nature of patch operation (denoted by the orange arrows). Hence, Transformer may fall short of delivering high-resolution feature maps and aggregating local information at lower layers~(\cite{raghu2021vision}). To address this shortcoming, we employed two convolutional layers using the original and downsampled image pair as inputs to capture local information and generate high-resolution feature maps. The outputs of these layers were concatenated with the feature maps in the decoder to produce a deformation field. The output deformation field, $\phi$, was generated the application of sixteen $3\times3$ convolutions. Except for the last convolutional layer, each convolutional layer is followed by a Leaky Rectified Linear Unit~(\cite{maas2013rectifier}) activation. Finally, the spatial transformation function~(\cite{jaderberg2015spatial}) is used to apply a nonlinear warp to the moving image $I_m$ with the deformation field $\phi$ (or the displacement field $\mathbf{u}$) provided by the network.

In the next subsections, we discuss the Swin Transformer block, the spatial transformation function, and the loss functions in detail.

\subsubsection{3D Swin Transformer Block}
Swin Transformer~(\cite{liu2021swin}) can generate hierarchical feature maps at various resolutions by using patch merging layers, making it ideal for usage as a general-purpose backbone for pixel-level tasks like image registration and segmentation. Swin Transformer's most significant component, apart from patch merging layers, is the shifted window-based self-attention mechanism. Unlike ViT~(\cite{dosovitskiy2020image}), which computes the relationships between a token and all other tokens at each step of the self-attention modules. Swin Transformer computes self-attention within the evenly partitioned non-overlapping local windows of the original and the lower resolution feature maps (as shown in Fig. \ref{fig:swin_feat} (a)). In contrast to the original Swin Transformer, this work uses rectangular-parallelepiped windows to accommodate non-square images, and each has a shape of $M_x\times M_y \times M_z$. At each resolution, the first Swin Transformer block employs a regular window partitioning method, beginning with the top-left voxel, and the feature maps are evenly partitioned into non-overlapping windows of size $M_x\times M_y \times M_z$. The self-attention is then calculated locally within each window. To introduce connections between neighboring windows, the Swin Transformer uses a shifted window design: in the successive Swin Transformer blocks, the windowing configuration shifts from that of the preceding block, by displacing the windows in the preceding block by ($\floor*{\frac{M_x}{2}}\times \floor*{\frac{M_y}{2}} \times \floor*{\frac{M_z}{2}}$) voxels. As illustrated by an example in Fig. \ref{fig:swin_feat} (b), the input feature map has $4\times8\times12$ voxels. With a window size of $2\times4\times6$, the feature map is evenly partitioned into $2\times2\times2=8$ windows in the first Swin Transformer block (``Swin Block 1" in Fig. \ref{fig:swin_feat} (b)). Then, in the next block, the windows are shifted by $(\floor*{\frac{2}{2}}\times \floor*{\frac{4}{2}} \times \floor*{\frac{6}{2}})=(1\times2\times3)$, and the number of windows becomes $3\times3\times3=27$. We extended the original 2D efficient batch computation (i.e., cyclic shift)~(\cite{liu2021swin, liu2021video}) to 3D and applied it to the 27 shifted windows, keeping the final number of windows for attention computation at 8. With the windowing-based attention, two consecutive Swin Transformer blocks can be computed as:
\begin{linenomath}
\begin{equation}
\label{eqn:sw_block}
    \begin{split}
        \hat{\mathbf{z}}_\ell &= \text{W-MSA}(\text{LN}(\mathbf{z}_{\ell-1}))+\mathbf{z}_{\ell-1},\\
        \mathbf{z}_\ell &= \text{MLP}(\text{LN}(\hat{\mathbf{z}}_\ell))+\hat{\mathbf{z}}_{\ell},\\
        \hat{\mathbf{z}}_{\ell+1} &= \text{SW-MSA}(\text{LN}(\mathbf{z}_{\ell}))+\mathbf{z}_{\ell},\\
        \mathbf{z}_{\ell+1} &= \text{MLP}(\text{LN}(\hat{\mathbf{z}}_{\ell+1}))+\hat{\mathbf{z}}_{\ell+1},\\
    \end{split}
\end{equation}
\end{linenomath}
where W-MSA and SW-MSA denote, respectively, window-based multi-head self-attention and shifted-window-based multi-head self-attention modules; MLP denotes the multi-layer perceptron module~(\cite{vaswani2017attention}); $\hat{\mathbf{z}}_\ell$ and $\mathbf{z}_\ell$ denote the output features of the (S)W-MSA and the MLP module for block $\ell$, respectively. The self-attention is computed as:
\begin{linenomath}
\begin{equation}
\label{eqn:atten}
    \mathbf{A}(Q,K,V)=\text{softmax}(\frac{QK^\top}{\sqrt{d}}+B)V,
\end{equation}
\end{linenomath}
where $Q, K, V\in\mathbb{R}^{M_xM_yM_z\times d}$ are \textit{query}, \textit{key}, \textit{value} matrices, $d$ denotes the dimension of \textit{query} and \textit{key} features, $M_xM_yM_z$ is the number of tokens in a 3D window, and $B$ represents the relative position of tokens in each window. Since the relative position between tokens along each axis (i.e., $x, y, z$) can only take values from $[-M_{x,y,z}+1, M_{x,y,z}-1]$, the values in $B$ are taken from a smaller bias matrix $\hat{B}\in\mathbb{R}^{(2M_x-1)\times(2M_y-1)\times(2M_z-1)}$. For the reasons given previously, we will show in section \ref{sec:pos_embd} that positional bias $B$ is not needed for the proposed network and that it just adds extra parameters without improving registration performance.

\subsubsection{Loss Functions}
The overall loss function for network training derives from the energy function of traditional image registration algorithms (i.e., Eqn. (\ref{eqn:E})). The loss function consists of two parts: one computes the similarity between the deformed moving and the fixed images, and another one regularizes the deformation field so that it is smooth:
\begin{linenomath}
\begin{equation}
\label{eqn:loss}
    \mathcal{L}(I_f, I_m, \phi)=\mathcal{L}_{sim}(I_f, I_m, \phi)+\lambda\mathcal{R}(\phi),
\end{equation}
\end{linenomath}
where $\mathcal{L}_{sim}$ denotes the image fidelity measure, and $\mathcal{R}$ denotes the deformation field regularization.
\paragraph{Image Similarity Measure}
In this work, we experimented with two widely-used similarity metric for $\mathcal{L}_{sim}$. The first was the mean squared error, which was the mean of the squared difference in voxel values between $I_f$ and $I_m$:
\begin{linenomath}
\begin{equation}
\label{eqn:MSE}
    MSE(I_f, I_m, \phi)=\frac{1}{\Omega}\sum_{\mathbf{p}\in\Omega}\vert I_f(\mathbf{p})-[I_m\circ\phi](\mathbf{p})\vert^2,
\end{equation}
\end{linenomath}
where $\mathbf{p}$ denotes the voxel location, and $\Omega$ represents the image domain.

Another similarity metric used was the local normalized cross-correlation between $I_f$ and $I_m$:
\begin{linenomath}
\begin{equation}
\label{eqn:lncc}
\begin{split}
    &LNCC(I_f, I_m, \phi)=\\
    &\sum_{\mathbf{p}\in\Omega}\frac{\left(\sum_{\mathbf{p}_i}(f(\mathbf{p}_i)-\bar{f}(\mathbf{p}))([I_m\circ\phi](\mathbf{p}_i)-[\bar{I}_m\circ\phi](\mathbf{p}))\right)^2}{\left(\sum_{\mathbf{p}_i}(f(\mathbf{p}_i)-\bar{f}(\mathbf{p}))^2\right)\left(\sum_{\mathbf{p}_i}([I_m\circ\phi](\mathbf{p}_i)-[\bar{I}_m\circ\phi](\mathbf{p}))^2\right)},\\
\end{split}
\end{equation}
\end{linenomath}
where $\bar{I}_f(\mathbf{p})$ and $\bar{I}_m(\mathbf{p})$ denotes the mean voxel value within the local window of size $n^3$ centered at voxel $\mathbf{p}$. We used $n=9$ in the experiments.  
\paragraph{Deformation Field Regularization} Optimizing the similarity metric alone would encourage $I_m\circ\phi$ to be visually as close as possible to $I_f$. The resulting deformation field $\phi$, however, might not be smooth or realistic. To impose smoothness in the deformation field, a regularizer $\mathcal{R}(\phi)$ was added to the loss function. $\mathcal{R}(\phi)$ encourages the displacement value in a location to be similar to the values in its neighboring locations. Here, we experimented with two regularizers. The first was the diffusion regularizer~\cite{balakrishnan2019voxelmorph}:
\begin{linenomath}
\begin{equation}
\label{Eqn:diffusion}
\mathcal{R}_{diffusion}(\phi)=\sum_{\mathbf{p}\in\Omega}\Vert\nabla\mathbf{u}(\mathbf{p})\Vert^2,
\end{equation}
\end{linenomath}
where $\mathbf{u}(\mathbf{p})$ is the spatial gradients of the displacement field $\mathbf{u}$. The spatial gradients were approximated using forward differences, that is, $\frac{\partial\mathbf{u}(\mathbf{p})}{\partial\{x,y,z\}}\approx\mathbf{u}(p_{\{x,y,z\}}+1)-\mathbf{u}(p_{\{x,y,z\}})$.

The second regularizer was bending energy~(\cite{rueckert1999nonrigid}), which penalizes sharply curved deformations, thus, it may be helpful for abdominal organ registration. Bending energy operates on the second derivative of the displacement field $\mathbf{u}$, and it is defined as:
\begin{linenomath}
\begin{equation}
\label{eqn:bending}
\begin{split}
    &\mathcal{R}_{bending}(\phi)=\sum_{\mathbf{p}\in\Omega}\Vert\nabla^2\mathbf{u}(\mathbf{p})\Vert^2=\sum_{\mathbf{p}\in\Omega}\Bigg[\left(\frac{\partial^2\mathbf{u}(\mathbf{p})}{\partial x^2}\right)^2+\left(\frac{\partial^2\mathbf{u}(\mathbf{p})}{\partial y^2}\right)^2+\\
    &\left(\frac{\partial^2\mathbf{u}(\mathbf{p})}{\partial z^2}\right)^2+2\left(\frac{\partial^2\mathbf{u}(\mathbf{p})}{\partial xz}\right)^2+2\left(\frac{\partial^2\mathbf{u}(\mathbf{p})}{\partial xy}\right)^2+2\left(\frac{\partial^2\mathbf{u}(\mathbf{p})}{\partial yz}\right)^2\Bigg],
\end{split}
\end{equation}
\end{linenomath}
where the derivatives were estimated using the same forward differences that were used previously.

\paragraph{Auxiliary Segmentation Information}
\label{sec:aux_seg}
When the organ segmentations of $I_f$ and $I_m$ are available, \texttt{TransMorph} may leverage this auxiliary information during training to improve the anatomical mapping between $I_m\circ\phi$ and $I_f$. A loss function $\mathcal{L}_{seg}$ that quantifies the segmentation overlap is added to the overall loss function (Eqn. \ref{eqn:loss}): 
\begin{linenomath}
\begin{equation}
\label{eqn:loss_w_seg}
    \mathcal{L}(I_f, I_m, \phi)=\mathcal{L}_{sim}(I_f, I_m, \phi)+\lambda\mathcal{R}(\phi) +\gamma\mathcal{L}_{seg}(s_f, s_m, \phi),
\end{equation}
\end{linenomath}
where $s_f$ and $s_m$ represent, respectively, the organ segmentation of $I_f$ and $I_m$, and $\gamma$ is a weighting parameter that controls the strength of $\mathcal{L}_{seg}$. In the field of image registration, it is common to use Dice score~(\cite{dice1945measures}) as a figure of merit to quantify registration performance. Therefore, we directly minimized the Dice loss~(\cite{milletari2016v}) between $s^k_f$ and $s^k_m$, where $k$ represents the $k^{th}$ structure/organ:
\begin{linenomath}
\begin{equation}
\label{eqn:dice}
\begin{split}
    Dice&(s_f, s_m, \phi)=1-\\
    &\frac{1}{K}\sum_k\frac{2\sum_{\mathbf{p}\in\Omega}s^k_f(\mathbf{p})[s^k_m\circ\phi](\mathbf{p})}{\sum_{\mathbf{p}\in\Omega}\left(s_f^k(\mathbf{p})\right)^2+\sum_{\mathbf{p}\in\Omega}\left([s^k_m\circ\phi](\mathbf{p})\right)^2}.
\end{split}
\end{equation}
\end{linenomath}
To allow backpropagation of the Dice loss, we used a method similar to that described in~(\cite{balakrishnan2019voxelmorph}), in which we designed $s_f$ and $s_m$ as image volumes with $K$ channels, each channel containing a binary mask defining the segmentation of a specific structure/organ. Then, $s_m\circ\phi$ is computed by warping the $K$-channel $s_m$ with $\phi$ using linear interpolation so that the gradients of $\mathcal{L}_{seg}$ can be backpropagated into the network. 

\subsection{Probabilistic and B-spline Variants}
\label{sec:prob_bspl}
In this section, we demonstrate that by simply altering the decoder, \texttt{TransMorph} can be used in conjunction with the concepts from prior research to ensure a diffeomorphic deformation such that the resulting deformable mapping is continuous, differentiable, and topology-preserving. The diffeomorphic registration was achieved using the \textit{scaling-and-squaring} approach (described in section \ref{sec:ss_diff}) with a stationary velocity field representation~(\cite{arsigny2006log}). Two existing diffeomorphic models, \texttt{VoxelMorph-diff}~(\cite{dalca2019unsupervised}) and \texttt{MIDIR}~(\cite{qiu2021learning}), have been adopted as bases for the proposed \texttt{TransMorph} diffeomorphic variants, designated by \texttt{TransMorph-diff} (section \ref{sec:prob_diff}) and \texttt{TransMorph-bspl} (section \ref{sec:bspline_diff}), respectively. The architectures of the two variants are shown in Fig. \ref{fig:transmorph_var}. The detailed derivation of these two variants are listed in Appendix.

\texttt{TransMorph-diff} was trained using the same loss functions as \texttt{VoxelMorph-diff}~(\cite{dalca2019unsupervised}):
\begin{linenomath}
\begin{equation}
\label{eqn:L_prob}
    \begin{split}
        &\mathcal{L}_{prob.}(I_f, I_m, \phi_\mathbf{u}; \psi)\\&=-\mathbb{E}_{\mathbf{u}\sim q_{\psi}}\left[\log p(I_f|\mathbf{u}, I_m)\right]+\text{KL}\left[q_\psi(\mathbf{u}|I_f, I_m)\Vert p(\mathbf{u})\right]\\
        &=\frac{1}{2\sigma^2}\Vert I_f-I_m\circ\phi_\mathbf{u}\Vert^2+\frac{1}{2}\Bigg[\text{tr}(\lambda\mathbf{D}\mathbf{\Sigma}_\psi-\log \mathbf{\Sigma}_\psi)+\mathbf{\mu}^\top_\psi\Lambda_\mathbf{u}\mathbf{\mu}_\psi\Bigg],
    \end{split}
\end{equation}
\end{linenomath}
and when anatomical label maps are available:
\begin{linenomath}
\begin{equation}
\label{eqn:l_prob_w_seg}
    \begin{split}
        \mathcal{L}_{prob.\ w/\ aux.}&(I_f, s_f, I_m, s_m, \phi_\mathbf{u}; \psi)\\
        &=\frac{1}{2\sigma^2}\Vert I_f-I_m\circ\phi_\mathbf{u}\Vert^2+\frac{1}{2\sigma_s^2}\Vert s_f-s_m\circ\phi_\mathbf{u}\Vert^2\\
        &\ \ \ \ +\frac{1}{2}\Bigg[\text{tr}(\lambda\mathbf{D}\mathbf{\Sigma}_\psi-\log \mathbf{\Sigma}_\psi)+\mathbf{\mu}^\top_\psi\Lambda_\mathbf{u}\mathbf{\mu}_\psi\Bigg].
    \end{split}
\end{equation}
\end{linenomath}
However, it is important to note that in~(\cite{dalca2019unsupervised}), $s_f$ and $s_m$ represent \textit{anatomical surfaces} obtained from label maps. In contrast, we directly used the \textit{label maps} as $s_f$ and $s_m$ in this work. They were image volumes with multiple channels, each channel contained a binary mask defining the segmentation of a certain structure/organ.
\begin{figure}[!h]
\centering
\includegraphics[width=.48\textwidth]{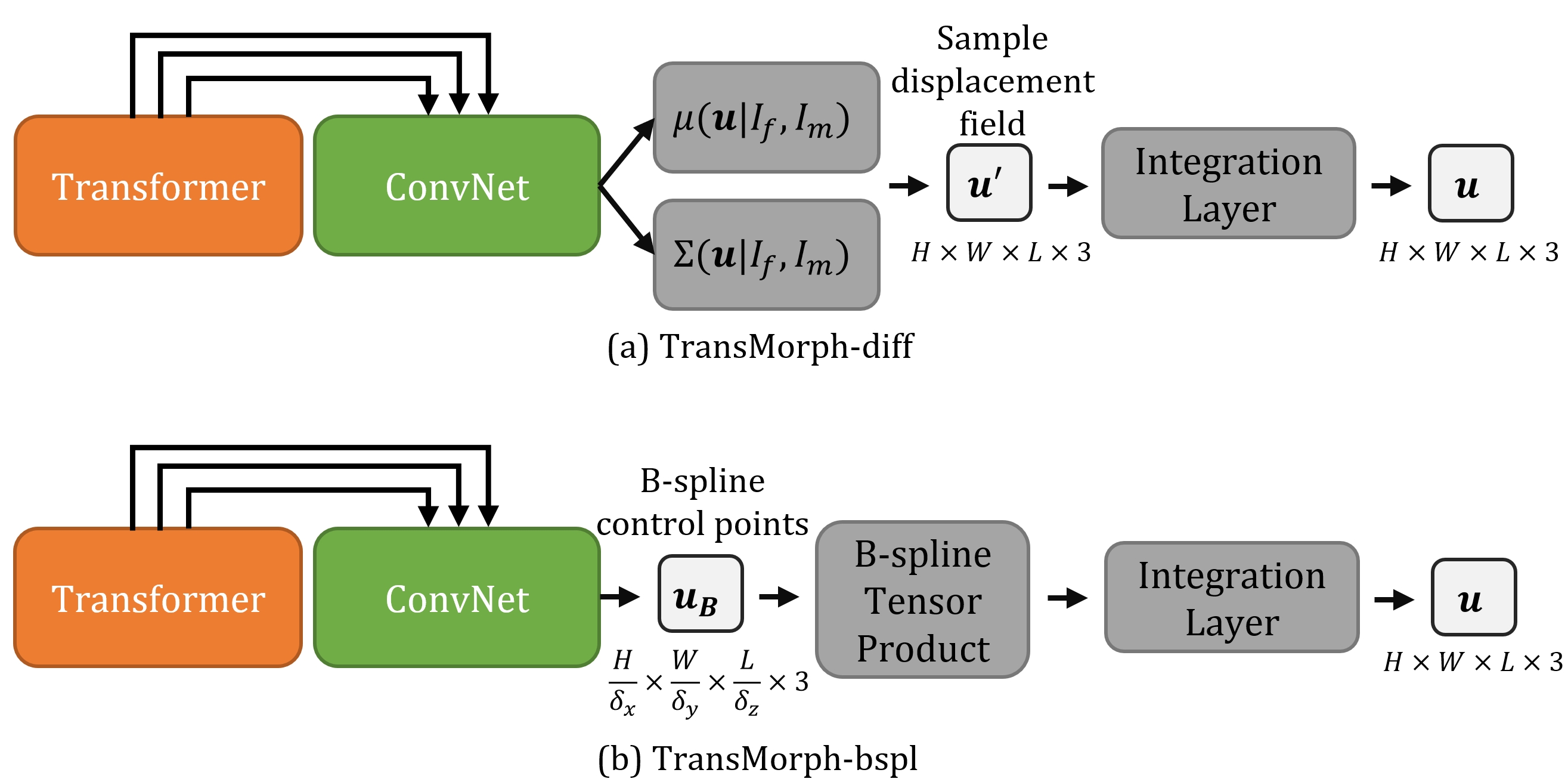}
\caption{The probabilistic and B-spline variants of TransMorph. (a): The architecture of the probabilistic diffeomorphic \texttt{TransMorph}. (b): The architecture of the B-spline diffeomorphic \texttt{TransMorph}.\label{fig:transmorph_var}}
\end{figure}
\begin{table*}[!ht]
\centering
\caption{The ablation study of \texttt{TransMorph} models with skip connections and positional embedding. ``Conv. skip." denotes the skip-connections from convolutional layers (indicated by green arrows in Fig. \ref{fig:net_arch}); ``Trans. skip," denotes the skip-connections from the Transformer blocks (indicated by orange arrows in Fig. \ref{fig:net_arch}); ``lrn. positional embedding" denotes the learnable positional embedding; ``sin. positional embedding" denotes the sinusoidal positional embedding.}
\begin{tabular}{ c | c  c  c c}
 \hline
 Model & Conv. skip.  & Trans. skip. & Parameters (M)\\
 \hline 
 w/o conv. skip. & \checkmark & -& 46.70\\
 \hline
 w/o Trans. skip. & - & \checkmark  & 41.55\\
  \hline
 w/o positional embedding & \checkmark & \checkmark & 46.77\\
 \hline
 w/ shuffling & \checkmark & \checkmark & 46.77\\
 \hline
 w/ rel. positional bias &  \checkmark &  \checkmark &  46.77\\
 \hline
 w/ lrn. positional embedding & \checkmark & \checkmark & 63.63\\
 \hline
 w/ sin. positional embedding & \checkmark & \checkmark & 46.77\\
 \hline

\end{tabular}
\label{table: TransMorph_skip_ablation}
\end{table*}
\begin{table*}[!ht]
\centering
\caption{The architecture hyperparameters of the \texttt{TransMorph} models used in the ablation study. ``Embed. Dimension" denotes the embedding dimension, $C$, in the very first stage (described in section \ref{sec:reg_net}); ``Swin-T." denotes Swin Transformer.}
\begin{tabular}{ c | c  c  c  c}
 \hline
 Model & Embed. Dimension  & Swin-T. block numbers  & Head numbers & Parameters (M)\\
 \hline
 \texttt{TransMorph} & 96 & \{2, 2, 4, 2\} & \{4, 4, 8, 8\} & 46.77\\
 \hline
 \texttt{TransMorph}-tiny & 6 & \{2, 2, 4, 2\} & \{4, 4, 8, 8\} & 0.24\\
 \hline
 \texttt{TransMorph}-small & 48 & \{2, 2, 4, 2\} & \{4, 4, 4, 4\} & 11.76\\
 \hline
 \texttt{TransMorph}-large & 128 & \{2, 2, 12, 2\} & \{4, 4, 8, 16\} & 108.34\\
 \hline
 \texttt{VoxelMorph}-huge & - & - & - & 63.25\\
 \hline
\end{tabular}
\label{table: TransMorph_sz_ablation}
\end{table*}
\subsection{Bayesian Uncertainty Variant}
\label{sec:bayes_variant}
In this section, we extend the proposed \texttt{TransMorph} to a Bayesian neural network (BNN) using the variational inference framework with Monte Carlo dropout~(\cite{gal2016dropout}), for which we refer readers to~(\cite{gal2016dropout, yang2017fast, yang2016fast}) for both theoretical and technical details. We denoted the resulting model as \texttt{TransMorph-Bayes}. In this model, Dropout layers were inserted into the Transformer encoder of the \texttt{TransMorph} architecture but not into the ConvNet decoder, in order to avoid imposing excessive regularity for the network parameters and thus decreasing performance. We added a dropout layer after each fully connected layer in the MLPs (Eqn. \ref{eqn:sw_block}) and after each self-attention computation (Eqn. \ref{eqn:atten}). Note that these are the locations where dropout layers are commonly used for Transformer training. We set the dropout probability $p$ to 0.15 to further avoid the network imposing an excessive degree of regularity on the network weights.

Both the transformation and appearance uncertainty can be estimated as the variability from the predictive mean (i.e., the variance), where the predictive mean of the deformation fields and the deformed images can be estimated by Monte Carlo integration~(\cite{gal2016dropout}):
\begin{linenomath}
\begin{equation}
\label{eqn:pred_mean_phi}
    \hat{\phi} = \frac{1}{T}\sum^T_{t=1}\phi_t,
\end{equation}
\end{linenomath}
and
\begin{linenomath}
\begin{equation}
\label{eqn:pred_mean}
    \hat{I}_f = \frac{1}{T}\sum^T_{t=1}I_m\circ\phi_t.
\end{equation}
\end{linenomath}
This is equivalent to averaging the output of $T$ forward passes through the network during inference, where $\phi_t$ represents the deformation field produced by $t^{th}$ forward pass. The transformation and appearance uncertainty can be estimated using the predictive variances of the deformation fields and the deformed images, respectively, as:
\begin{linenomath}
\begin{equation}
\label{eqn:pred_var_phi}
    \hat{\Sigma}^2_\phi = \frac{1}{T}\sum^T_{t=1}\left(\phi_t - \hat{\phi}_f\right)^2,
\end{equation}
\end{linenomath}
and
\begin{linenomath}
\begin{equation}
\label{eqn:pred_var_0}
    \hat{\Sigma}^2_f = \frac{1}{T}\sum^T_{t=1}\left(I_m\circ\phi_t - \hat{I}_f\right)^2.
\end{equation}
\end{linenomath}
\subsubsection{Appearance uncertainty calibration}
\label{sec:uncert_cal}
An ideal uncertainty estimate should be properly correlated to the inaccuracy of the registration results; that is, a high uncertainty value should indicate a large registration error, and vice versa. Otherwise, doctors/surgeons may be misled by the erroneous estimate of registration uncertainty and place unwarranted confidence in the registration results, resulting in severe consequences~(\cite{luo2019applicability, risholm2013bayesian, risholm2011estimation}). The appearance uncertainty given by Eqn. \ref{eqn:pred_var_0} is expressed as the variability from the mean model prediction. Such an appearance uncertainty estimation does not account for the systematic errors (i.e., bias) between the mean registration prediction and the target image; therefore, a low uncertainty value given by Eqn. \ref{eqn:pred_var_0} does not always guarantee an accurate registration result.

When the predicted uncertainty values closely corresponded to the expected model error, the uncertainty estimates are considered to be well-calibrated~(\cite{laves2019well, levi2019evaluating}). In an ideal scenario, the estimated registration uncertainty should completely reflect the actual registration error. For instance, if the predictive variance of a batch of registered images generated by the network is found to be 0.5, the expectation of the squared error should likewise be 0.5. Accordingly, if the expected model error is quantified by MSE, then the perfect calibration of appearance uncertainty may be defined as the following~(\cite{guo2017calibration,levi2019evaluating, laves2020uncertainty}):
\begin{linenomath}
\begin{equation}
    \mathbb{E}_{\hat{\Sigma}^2}\left[\Vert I_m\circ\phi-I_f\Vert^2|\hat{\Sigma}^2=\Sigma^2\right]=\Sigma^2\ \ \ \ \ \ \forall\left\{\Sigma^2\in\mathbb{R} \vert \Sigma^2\geq0\right\}.
\end{equation}
\end{linenomath}

In the conventional paradigm of Bayesian neural networks, the uncertainty estimate is derived from the predictive variance $\hat{\Sigma}^2$ relative to the predictive mean $\hat{I}_f$ as in Eqn. \ref{eqn:pred_var_0}. However, it can be shown that this predictive variance can be miscalibrated as a result of overfitting the training dataset (as shown in \ref{app:mis_cal}). Therefore, the uncertainty values estimated based on $\hat{\Sigma}^2_f$ in Eqn. \ref{eqn:pred_var_0} may be biased. This bias must be corrected in applications such as image denoising or classification~(\cite{laves2019well, guo2017calibration, kuleshov2018accurate, phan2018calibrating, laves2020uncertainty, pmlr-v121-laves20a}), such that the uncertainty values closely reflect the expected error. In image registration, however, the expected appearance error may be computed even during the test time since the target image is always known. Therefore, a perfectly calibrated appearance uncertainty quantification may be achieved without additional effort. Here, we propose to replace the predicted mean $\hat{I}_f$ with the target image $I_f$ in Eqn. \ref{eqn:pred_var_0}. Then, the appearance uncertainty is the equivalent to the expected error:
\begin{linenomath}
\begin{equation}
\label{eqn:pred_var_1}
    \Sigma^2_f=err(I_m\circ\phi) = \frac{1}{T}\sum^T_{t=1}\left(I_m\circ\phi_t - I_f\right)^2.
\end{equation}
\end{linenomath}
A comparison between the two appearance uncertainty estimate methods (i.e., $\hat{\Sigma}^2_f$ and $\Sigma^2_f$) is shown later in this paper.

\section{Experiments}
\label{sec:experiments}
\subsection{Datasets and Preprocessing}
Three datasets including over 1000 image pairs were used to thoroughly validate the proposed method. The details of each dataset are described in the following sections.
\subsubsection{Inter-patient Brain MRI Registration}
For the inter-patient brain MR image registration dataset, we used a dataset of 260 T1–weighted brain MRI images acquired at Johns Hopkins University. The images were anonymized and acquired under IRB approval. The dataset was split into 182, 26, and 52 (7:1:2) volumes for training, validation, and test sets. Each image volume was used as a moving image to form two image pairs by randomly matching it to two other volumes in the set (i.e., the fixed images). Then, the moving and fixed images were inverted to form another two image pairs, resulting in four registration pairings of $I_f$ and $I_m$. The final data comprises 768, 104, and 208 image pairs for training, validation, and testing, respectively. FreeSurfer~(\cite{fischl2012freesurfer}) was used to perform standard pre-processing procedures for structural brain MRI, including skull stripping, resampling, and affine transformation. The pre-processed image volumes were all cropped to size of $160\times192\times224$. Label maps including 30 anatomical structures were obtained using FreeSurfer for evaluating registration performances.

\subsubsection{Atlas-to-patient Brain MRI Registration}
We used a publicly available dataset to evaluate the proposed model with atlas-to-patient brain MRI registration task. A total number of 576 T1–weighted brain MRI images from the Information eXtraction from Images (IXI) database\footnote{\url{https://brain-development.org/ixi-dataset/}} was used as the fixed images. The moving image for this task was an atlas brain MRI obtained from~(\cite{kim2021cyclemorph}). The dataset was split into 403, 58, and 115 (7:1:2) volumes for training, validation, and test sets. FreeSurfer was used to pre-process the MRI volumes. We carried out the same pre-processing procedures we used for the previous dataset applied to the IXI dataset. All image volumes were cropped to size of $160\times192\times224$. Label maps of 30 anatomical structures were used to evaluate registration performances.

\subsubsection{Learn2Reg OASIS Brain MRI Registration}
We additionally evaluated \texttt{TransMorph} on a public registration challenge, OASIS~(\cite{marcus2007open, hoopes2021hypermorph}), obtained from the 2021 Learn2Reg challenge~(\cite{hering2021learn2reg}) for inter-patient registration. This dataset contains a total of 451 brain T1 MRI images, with 394, 19, and 38 images being used for training, validation, and testing, respectively. FreeSurfer~(\cite{fischl2012freesurfer}) was used to pre-process the brain MRI images, and label maps for 35 anatomical structures were provided for evaluation.

\subsubsection{XCAT-to-CT Registration}
Computerized phantoms have been widely used in the medical imaging field for algorithm optimization and imaging system validation~(\cite{Christoffersen2013, chen2019incorporating, zhang2017new}). The four-dimensional extended cardiac-torso (XCAT) phantom~(\cite{segars20104d}) was developed based on anatomical images from the Visible Human Project data. While the current XCAT phantom\footnote{as of October, 2021} can model anatomical variations through organ and phantom scaling, it cannot completely replicate the anatomical variations seen in humans. As a result, XCAT-to-CT registration (which can be thought of as atlas-to-image registration) has become a key method for creating anatomically variable phantoms~(\cite{chen2020generating, fu2021iphantom, segars2013population}). This research used a CT dataset from~(\cite{segars2013population}) that includes 50 non-contrast chest-abdomen-pelvis (CAP) CT scans that are part of the Duke University imaging database. Selected organs and structures were manually segmented in each patient's CT scan. The structures segmented included the following: the body outline, the bone structures, lungs, heart, liver, spleen, kidneys, stomach, pancreas, large intestine, prostate, bladder, gall bladder, and thyroid. The manual segmentation was done by several medical students, and the results were subsequently corrected by an experienced radiologist at Duke University. The CT volumes have voxel sizes ranging from $0.625\times0.625\times5$mm to $0.926\times0.926\times5$mm. We used trilinear interpolation to resample all volumes to an identical voxel spacing of $2.5\times2.5\times5$mm. The volumes were all cropped and zero-padded to have a size of $160\times160\times160$ voxels. The intensity values were first clipped in the range of $[-1000, 700]$ Hounsfield Units and then normalized to the range of $[0, 1]$. The XCAT attenuation map was generated with a resolution of $1.1\times1.1\times1.1$mm using the material compositions and attenuation coefficients of the constituents at 120 keV. It was then resampled, cropped, and padded so that the resulting volume matched the size of the CT volumes. The XCAT attenuation map's intensity values were also normalized to be within a range of $[0, 1]$. The XCAT and CT images were rigidly registered using the proposed affine network. The dataset was split into 35, 5, and 10 (7:1:2) volumes for training, validation, and testing. We conducted five-fold cross-validation on the fifty image volumes, resulting in 50 testing volumes in total.

\subsection{Baseline Methods}
We compared \texttt{TransMorph} to various registration methods that have previously demonstrated state-of-the-art registration performance. We begin by comparing \texttt{TransMorph} with four non-deep-learning-based methods. The hyper-parameters of these methods, unless otherwise specified, were empirically set to balance the trade-off between registration accuracy and running time. The methods and their hyperparameter settings are described below: 
\begin{itemize}[noitemsep]
    \item \texttt{SyN}\footnote{\url{https://github.com/ANTsX/ANTsPy}}(\cite{avants2008symmetric}): For both inter-patient and atlas-to-patient brain MR registration tasks, we used the mean squared difference (MSQ) as the objective function, along with a default Gaussian smoothing of 3 and three scales with 180, 80, 40 iterations, respectively. For XCAT-to-CT registration, we used cross-correlation (CC) as the objective function, a Gaussian smoothing of 5 and three scales with 160, 100, 40 iterations, respectively.
    \item \texttt{NiftyReg}\footnote{\url{https://www.ucl.ac.uk/medical-image-computing}}(\cite{modat2010fast}): We used the sum of squared differences (SSD) as the objective function and bending energy as a regularizer for all registration tasks. For inter-patient brain MR registration, we empirically used a regularization weighting of 0.0002 and three scales with 300 iterations each. For atlas-to-patient brain MR registration, the regularization weighting was set to 0.0006, and we used three scales with 500 iterations each. For XCAT-to-CT registration, we used a regularization weight of 0.0005 and five scales with 500 iterations each.
    \item \texttt{deedsBCV}\footnote{\url{https://github.com/mattiaspaul/deedsBCV}}~(\cite{heinrich2015multi}): The objective function was self-similarity context (SSC)~(\cite{heinrich2013towards}) by default. For both inter-patient and atlas-to-patient brain MR registration, we used the hyperparameter values suggested in~(\cite{hoffmann2020learning}) for neuroimaging, in which the grid spacing, search radius, and quantization step were set to $6\times5\times4\times3\times2$, $6\times5\times4\times3\times2$, and $5\times4\times3\times2\times1$, respectively. For XCAT-to-CT registration, we used the default parameters suggested for abdominal CT registration~(\cite{heinrich2015multi}), where the grid spacing, search radius, and quantization step were $8\times7\times6\times5\times4$, $8\times7\times6\times5\times4$, and $5\times4\times3\times2\times1$, respectively.
    \item \texttt{LDDMM}\footnote{\url{https://github.com/brianlee324/torch-lddmm}}~(\cite{beg2005computing}): MSE was used as the objective function by default. For both inter-patient and atlas-to-patient brain MR registration, we used the smoothing kernel size of 5, the smoothing kernel power of 2, the matching term coefficient of 4, the regularization term coefficient of 10, and the iteration number of 500. For XCAT-to-CT registration, we used the same kernel size, kernel power, the matching term coefficient, and the number of iteration. However, the regularization term coefficient was empirically set to 3.
\end{itemize}

Next, we compared the proposed method with several existing deep-learning-based methods. For a fair comparison, unless otherwise indicated, the loss function (Eqn. \ref{eqn:loss}) that consists of MSE (Eqn. \ref{eqn:MSE}) and diffusion regularization (Eqn. \ref{Eqn:diffusion}) was used for inter-patient brain MR registration, while we instead used LNCC (Eqn. \ref{eqn:lncc}) for atlas-to-patient MRI registration. For XCAT-to-CT registration, we used the loss function (Eqn. \ref{eqn:loss_w_seg}) that consists of LNCC (Eqn. \ref{eqn:lncc}), bending energy (Eqn. \ref{eqn:bending}), and Dice loss (Eqn. \ref{eqn:dice}). Auxiliary data (organ segmentation) was used for XCAT-to-CT registration only. Recall that the hyperparameters $\lambda$ and $\gamma$ define, respectively, the weight for deformation field regularization and Dice loss. The detailed parameter settings used for each method were as follows: 
\begin{itemize}[noitemsep]
    \item \texttt{VoxelMorph}\footnote{\url{http://voxelmorph.csail.mit.edu}}~(\cite{balakrishnan2018unsupervised,balakrishnan2019voxelmorph}): We employed two variants of \texttt{VoxelMorph}, the second variant doubles the number of convolution filters in the first variant; they are designated as \texttt{VoxelMorph}-1 and -2, respectively. For inter-patient and atlas-to-patient brain MR registration, the regularization hyperparameter $\lambda$ was set, respectively, to 0.02 and 1, where these values were reported as the optimal values in~\cite{balakrishnan2019voxelmorph}. For XCAT-to-CT registration, we set $\lambda=\gamma=1$.
    \item \texttt{VoxelMorph-diff}\footnote{\url{http://voxelmorph.csail.mit.edu}}~(\cite{dalca2019unsupervised}): For both inter-patient and atlas-to-patient brain MR registration tasks, the loss function $\mathcal{L}_{prob.}$ (Eqn. \ref{eqn:L_prob}) was used with $\sigma$ set to 0.01 and $\lambda$ set to 20. For XCAT-to-CT registration, we used the loss function $\mathcal{L}_{prob. w/ aux.}$ (Eqn. \ref{eqn:l_prob_w_seg}) with $\sigma=\sigma_s=0.01$ and $\lambda=20$.
    \item \texttt{CycleMorph}\footnote{\url{https://github.com/boahK/MEDIA_CycleMorph}}~(\cite{kim2021cyclemorph}): In \texttt{CycleMorph}, the hyerparameters $\alpha$, $\beta$, and $\lambda$, correspond to the weights for cycle loss, identity loss, and deformation field regularization. For inter-patient brain MR registration, we set $\alpha=0.1$, $\beta=0.5$, and $\lambda=0.02$. Whereas for atlas-to-patient brain MR registration, we set $\alpha=0.1$, $\beta=0.5$, and $\lambda=1$. These values were recommended in~(\cite{kim2021cyclemorph}) as the optimal values for neuroimaging. For XCAT-to-CT registration, we modified the \texttt{CycleMorph} by adding a Dice loss with a weighting of 1 to incorporate organ segmentation during training, and we set $\alpha=0.1$ and $\beta=1$. We observed that the $\lambda$ value of 1 suggested in~(\cite{kim2021cyclemorph}) yielded over-smoothed deformation field in our application. Therefore, the value of $\lambda$ was decreased to 0.1. 
    \item \texttt{MIDIR}\footnote{\url{https://github.com/qiuhuaqi/midir}}~(\cite{qiu2021learning}): The same loss function and $\lambda$ value as VoxelMorph were used. In addition, the control point spacing $\delta$ for B-spline transformation was set to 2 for all tasks, which was shown to be an optimal value in~\cite{qiu2021learning}. 
\end{itemize}
To evaluate the proposed Swin-Transformer-based network architecture, we compared its performance to existing Transformer-based networks that achieved state-of-the-art performance in other applications (e.g., image segmentation, object detection, etc.). We customized these models to make them suitable for image registration. They were modified to produce 3-dimensional deformation fields that warp the given moving image. Note that the only change between the methods below and \texttt{VoxelMorph} is the network architecture, with the spatial transformation function, loss function, and network training procedures remaining the same. The first three models used the hybrid Transformer-ConvNet architecture (i.e., \texttt{ViT-V-Net}, \texttt{PVT}, and \texttt{CoTr}), while the last model used a pure Transformer-based architecture (i.e., \texttt{nnFormer}). Their network hyperparameter settings were as follows:
\begin{itemize}[noitemsep]
    \item \texttt{ViT-V-Net}\footnote{\url{ https://bit.ly/3bWDynR}}~(\cite{chen2021vit}): This registration network was developed based on ViT~(\cite{dosovitskiy2020image}). We applied the default network hyperparameter settings suggested in~(\cite{chen2021vit}).
    \item \texttt{PVT}\footnote{\url{https://github.com/whai362/PVT}}~(\cite{wang2021pyramid}): The default settings were applied, except that the embedding dimensions were to be $\{20, 40, 200, 320\}$, the number of heads was set to $\{2, 4, 8, 16\}$, and the depth was increased to $\{3, 10, 60, 3\}$ to achieve a comparable number of parameters to that of \texttt{TransMorph}.
    \item \texttt{CoTr}\footnote{\url{https://github.com/YtongXie/CoTr}}~(\cite{xie2021cotr}): We used the default network settings for all registration tasks.
    \item \texttt{nnFormer}\footnote{\url{https://github.com/282857341/nnFormer}}~(\cite{zhou2021nnformer}): Because \texttt{nnFormer} was also developed on the basis of Swin Transformer, we applied the same Transformer hyperparameter values as in \texttt{TransMorph} to make a fair comparison.
\end{itemize}

\subsection{Implementation Details}
The proposed \texttt{TransMorph} was implemented using PyTorch~(\cite{paszke2019pytorch}) on a PC with an NVIDIA TITAN RTX GPU and an NVIDIA RTX3090 GPU. All models were trained for 500 epochs using the Adam optimization algorithm, with a learning rate of $1\times 10^{-4}$ and a batch size of 1. The brain MR dataset was augmented with flipping in random directions during training, while no data augmentation was applied to the CT dataset. Restricted by the sizes of the image volumes, the window sizes (i.e., $\{M_x, M_y, M_z\}$) used in Swin Transformer were set to $\{5, 6, 7\}$ for MR brain registration, $\{5, 5, 5\}$ for XCAT-to-CT registration, and $\{\}$respectively. The Transformer hyperparameter settings for \texttt{TransMorph} are listed in the first row of Table. \ref{table: TransMorph_sz_ablation}. Note that the variants of \texttt{TransMorph} (i.e., \texttt{TransMorph-Bayes}, \texttt{TransMorph-bspl}, and \texttt{TransMorph-diff}) share the same Transformer settings as \texttt{TransMorph}. The hyperparameter settings for each proposed variant are described as follows:
\begin{itemize}[noitemsep]
    \item \texttt{TransMorph}: The identical loss function parameters as \texttt{VoxelMorph} were used for all tasks.
    \item \texttt{TransMorph-Bayes}: The identical loss function parameters as \texttt{VoxelMorph} were applied here for all tasks. The dropout probability was set to 0.15.
    \item \texttt{TransMorph-bspl}: The loss function settings for all tasks were the same ones as those used in \texttt{VoxelMorph}. The control point spacing, $\delta$, for B-spline transformation was also set to 2, the same value used in \texttt{MIDIR}.
    \item \texttt{TransMorph-diff}: We applied the same loss function parameters as those used in \texttt{VoxelMorph-diff}.
\end{itemize}

The affine model presented in this work comprises of a compact Swin Transformer. The Transformer parameter settings were identical to \texttt{TransMorph} except that the embedding dimension was set to be 12, the numbers of Swin Transfomer block were set to be $\{1,1,2,2\}$, and the head numbers were set to be $\{1,1,2,2\}$. The resulting affine model has a total number of 19.55 millions of parameters and a computational complexity of 0.4 GMacs. Because the MRI datasets were affinely aligned as part of the preprocessing, the affine model was only used in the XCAT-to-CT registration.

\subsection{Additional Studies}
In this section, we present experiments designed to verify the effect of the various Transformer modules in \texttt{TransMorph} architecture. Specifically, we carried out two additional studies of network components and model complexity. They are performed using the validation datasets from the three registration tasks, and the system-level comparisons are reported on test datasets. The following subsections provide detailed descriptions of these studies.

\subsubsection{Ablation study on network components}
We begin by examining the effects of several network components on registration performance. Table \ref{table: TransMorph_skip_ablation} lists three variants of \texttt{TransMorph} that either keep or remove the network's long skip connections or the positional embeddings in the Transformer encoder. In ``w/o conv. skip.", the long skip connections from the two convolutional layers were removed (including two convolutional layers), which are the green arrows in Fig. \ref{fig:net_arch}. In ``w/o trans. skip.", the long skip connections coming from the Swin Transformer blocks were removed, which are the orange arrows in Fig. \ref{fig:net_arch}. We claimed in section \ref{sec:reg_net} that the positional embedding (i.e., $\mathbf{E}_{pos}$ in Eqn. \ref{eqn:embding}) was not a necessary element of \texttt{TransMorph}, because the positional information of tokens can be learned implicitly in the network via the consecutive up-sampling in the decoder and backpropagating the loss between output and target. Here, we conducted experiments to study the effectiveness of positional embeddings. Table \ref{table: TransMorph_skip_ablation} also lists five variants of \texttt{TransMorph} that either keep or remove the positional embeddings in the Transformer encoder. In the third variation, "w/o positional embedding", we did not employ any type of positional embedding. In the fourth variant, ``w/ shuffling", we did not employ any positional embedding but instead randomly shuffled the positions of the tokens (i.e., the dimension $N$ of $\mathbf{z}$ in Eqn. \ref{eqn:embding} and \ref{eqn:sw_block}) just before the self-attention calculation. Following the self-attention calculation, the positions are permuted back into their original order. This way, the self-attention modules in the Transformer encoder are truly invariant to the order of the tokens. In the fifth variant, ``w/ rel. positional bias", we used the relative positional bias in the self-attention computation (i.e. $B$ in Eqn. \ref{eqn:atten}) as used in the Swin Transformer~(\cite{liu2021swin}). In the second to last variant, ``w/ lrn. positional embedding", we added the same learnable positional embedding to the patch embeddings at the start of the Transformer encoder as used in the ViT~(\cite{dosovitskiy2020image}) while keeping the relative positional bias. In the last variant, ``w/ sin. positional embedding", we substituted the learnable positional embedding with a sinusoidal positional embedding, the same embedding used in the original Transformer~(\cite{vaswani2017attention}), which hardcodes the positional information in the tokens.

\begin{table*}[!t]
\fontsize{6.5}{8}\selectfont
\parbox{.5\linewidth}{
\centering
\caption{Quantitative evaluation results of the inter-patient (i.e., the JHU dataset) and the atlas-to-patient (i.e., the IXI dataset) brain MRI registration. Dice score and percentage of voxels with a non-positive Jacobian determinant (i.e., folded voxels) are evaluated for different methods. The \textbf{bolded} numbers denote the highest scores, while the \textit{italicized} ones indicate the second highest.}
\begin{tabular}{ c | c | c | c | c}
\hline
  & \multicolumn{2}{c|}{Inter-patient MRI} & \multicolumn{2}{c}{Atlas-to-patient MRI} \\
 \hline
 Model & DSC & \% of $\vert J_{\phi}\vert\leq0$ & DSC & \% of $\vert J_{\phi}\vert\leq0$\\
 \hline
 Affine & 0.572$\pm$0.166 & - & 0.386$\pm$0.195 & -\\
 \hline
 \texttt{SyN}& 0.729$\pm$0.127 & $<$0.0001& 0.645$\pm$0.152 & $<$0.0001\\
 \hline
 \texttt{NiftyReg}& 0.723$\pm$0.131 & 0.061$\pm$0.093& 0.645$\pm$0.167 & 0.020$\pm$0.046\\
 \hline
 \texttt{LDDMM} & 0.716$\pm$0.131 & $<$0.0001& 0.680$\pm$0.135 & $<$0.0001\\
 \hline
 \texttt{deedsBCV}& 0.719$\pm$0.130 & 0.253$\pm$0.110& 0.733$\pm$0.126 & 0.147$\pm$0.050\\
 \hline
 \texttt{VoxelMorph}-1 & 0.718$\pm$0.134 & 0.426$\pm$0.231& 0.729$\pm$0.129 & 1.590$\pm$0.339\\
 \hline
 \texttt{VoxelMorph}-2 & 0.723$\pm$0.132 & 0.389$\pm$0.222& 0.732$\pm$0.123 & 1.522$\pm$0.336\\
 \hline
 \texttt{VoxelMorph-diff} & 0.715$\pm$0.137 & $<$0.0001& 0.580$\pm$0.165 & $<$0.0001\\
 \hline
 \texttt{CycleMorph} & 0.719$\pm$0.134 & 0.231$\pm$0.168& 0.737$\pm$0.123 & 1.719$\pm$0.382\\
 \hline
 \texttt{MIDIR} & 0.710$\pm$0.132 & $<$0.0001& 0.742$\pm$0.128 & $<$0.0001\\
 \hline
 \texttt{ViT-V-Net} & 0.729$\pm$0.128 & 0.402$\pm$0.249& 0.734$\pm$0.124& 1.609$\pm$0.319\\
 \hline
 \texttt{PVT} & 0.729$\pm$0.130 & 0.427$\pm$0.254& 0.727$\pm$0.128& 1.858$\pm$0.314\\
 \hline
 \texttt{CoTr} & 0.725$\pm$0.131 & 0.415$\pm$0.258& 0.735$\pm$0.135& 1.292$\pm$0.342\\
  \hline
 \texttt{nnFormer} & 0.729$\pm$0.128 & 0.399$\pm$0.234& 0.747$\pm$0.135& 1.595$\pm$0.358\\
 \hline
 \hline
 \texttt{TransMorph-Bayes} & \textit{0.744$\pm$0.125} & 0.389$\pm$0.241& 0.753$\pm$0.123& 1.560$\pm$0.333\\
 \hline
 \texttt{TransMorph-diff} & 0.730$\pm$0.129 & $<$0.0001& 0.594$\pm$0.163 & $<$0.0001\\
 \hline
 \texttt{TransMorph-bspl} & 0.740$\pm$0.123 & $<$0.0001& \textbf{0.761$\pm$0.122}& $<$0.0001\\
 \hline
 \texttt{TransMorph} & \textbf{0.745$\pm$0.125}& 0.396$\pm$0.240& \textit{0.754$\pm$0.124}& 1.579$\pm$0.328\\
 \hline
\end{tabular}
\label{tab:table_res}}
\quad
\fontsize{7.5}{9}\selectfont
\parbox{0.5\linewidth}{
\centering
\caption{Quantitative evaluation results of XCAT-to-CT registration. Dice score of 16 organs, percentage of voxels with a non-positive Jacobian determinant (i.e., folded voxels), and SSIM are evaluated for different methods. The \textbf{bolded} numbers denote the highest scores, while the \textit{italicized} ones indicate the second highest.}
\label{tab:table_res_XCAT}
\begin{tabular}{ c | c | c | c }
 \hline
 Model & DSC & \% of $\vert J_{\phi}\vert\leq0$ & SSIM\\
 \hline
 w/o registration & 0.220$\pm$0.242 & - & 0.576$\pm$0.071 \\
 \hline
 Affine Transformer & 0.330$\pm$0.291 & - & 0.751$\pm$0.018 \\
 \hline
 \texttt{SyN}& 0.498$\pm$0.342 & 0.001$\pm$0.002 & 0.894$\pm$0.021 \\
 \hline
 \texttt{NiftyReg}& 0.488$\pm$0.333 & 0.025$\pm$0.046 & 0.886$\pm$0.027\\
 \hline
 \texttt{LDDMM} & 0.519$\pm$0.265 & 0.006$\pm$0.007 & 0.874$\pm$0.031\\
 \hline
 \texttt{deedsBCV}& 0.568$\pm$0.306 & 0.126$\pm$0.123 & 0.863$\pm$0.029\\
 \hline
 \texttt{VoxelMorph}-1 & 0.532$\pm$0.313 & 2.275$\pm$1.283 & 0.899$\pm$0.027\\
 \hline
 \texttt{VoxelMorph}-2 & 0.548$\pm$0.317 & 1.696$\pm$0.909  & 0.910$\pm$0.027\\
 \hline
 \texttt{VoxelMorph-diff} & 0.526$\pm$0.330 & $<$0.0001  & 0.911$\pm$0.020\\
 \hline
 \texttt{CycleMorph} & 0.528$\pm$0.321 & 3.263$\pm$1.188  & 0.909$\pm$0.024\\
 \hline
 \texttt{MIDIR} & 0.551$\pm$0.303 & $<$0.0001 & 0.896$\pm$0.022\\
 \hline
 \texttt{ViT-V-Net} & 0.582$\pm$0.311 & 2.109$\pm$1.032 & 0.915$\pm$0.020\\
 \hline
 \texttt{PVT} & 0.516$\pm$0.321 & 2.939$\pm$1.162 & 0.900$\pm$0.027\\
 \hline
 \texttt{CoTr} & 0.550$\pm$0.313 & 1.530$\pm$1.052 & 0.905$\pm$0.029\\
  \hline
 \texttt{nnFormer} & 0.536$\pm$0.315 & 1.371$\pm$0.620 & 0.902$\pm$0.024\\
 \hline
 \hline
 \texttt{TransMorph-Bayes} & \textit{0.594$\pm$0.313} & 1.475$\pm$0.857 & \textbf{0.919}$\pm$\textbf{0.024}\\
 \hline
 \texttt{TransMorph-diff} & 0.541$\pm$0.324 & $<$0.0001 &0.910$\pm$0.025\\
 \hline
 \texttt{TransMorph-bspl} & 0.575$\pm$0.311 & $<$0.0001 & 0.908$\pm$0.025\\
 \hline
 \texttt{TransMorph} & \textbf{0.604$\pm$0.314} & 1.679$\pm$0.772 & \textit{0.918}$\pm$\textit{0.023}\\
 \hline
\end{tabular}}
\end{table*}

\begin{table}[!t]
\fontsize{8}{9.5}\selectfont
\parbox{\linewidth}{
\centering
\caption{Quantitative evaluation results for brain MRI registration of the OASIS dataset from the 2021 Learn2Reg challenge task 3. Dice score of 35 cortical and subcortical brain structures, the $95^{th}$ percentile percentage of the Hausdorff distance, and the standard deviation of the logarithm of the Jacobian determinant (SDlogJ) of the displacement field are evaluated for different methods. The validation results came from the challenge's leaderboard, whereas the test results came directly from the challenge's organizers. The \textbf{bolded} numbers denote the highest scores, while the \textit{italicized} ones indicate the second highest.}
\label{tab:table_res_OASIS}

\begin{tabular}{ c | c | c | c }
\hline
 \multicolumn{4}{c}{\textbf{Validation}}\\
 \hline
 Model & DSC & HdDist95 & SDlogJ\\
 \hline
 \cite{lv2022joint} & 0.827$\pm$0.013 & 1.722$\pm$0.318 & 0.121$\pm$0.015\\
 \hline
 \cite{siebert2021fast} &  0.846$\pm$0.016 & 1.500$\pm$0.304 & \textbf{0.067$\pm$0.005} \\
 \hline
 \cite{mok2021conditional} & \textit{0.861$\pm$0.015} & 1.514$\pm$0.337 & \textit{0.072$\pm$0.007}\\
 \hline
 \texttt{VoxelMorph}-huge & 0.847$\pm$0.014& 1.546$\pm$0.306 & 0.133$\pm$0.021\\
 \hline
 \texttt{TransMorph} & 0.858$\pm$0.014 & \textit{1.494$\pm$0.288} & 0.118$\pm$0.019 \\
 \hline
 \texttt{TransMorph-Large} & \textbf{0.862$\pm$0.014} & \textbf{1.431$\pm$0.282} & 0.128$\pm$0.021 \\
 \hline
 \hline
 \multicolumn{4}{c}{\textbf{Test}}\\
 \hline
 Model & DSC & HdDist95 & SDlogJ\\
 \hline
 Initial & 0.56 & 3.86 & -\\
 \hline
 \cite{lv2022joint} & 0.80 & 1.77 & \textit{0.08}\\
 \hline
 \cite{siebert2021fast} & 0.81 & \textbf{1.63} & \textbf{0.07}\\
 \hline
 \cite{mok2021conditional} & \textit{0.82}  & 1.67 & \textbf{0.07} \\
 \hline
 \texttt{TransMorph} & 0.816 & 1.692 & 0.124\\
 \hline
 \texttt{TransMorph-Large} & \textbf{0.820} & \textit{1.656} & 0.124 \\
 \hline
\end{tabular}}
\end{table}

\begin{figure}[!t]
\centering
\includegraphics[width=.45\textwidth]{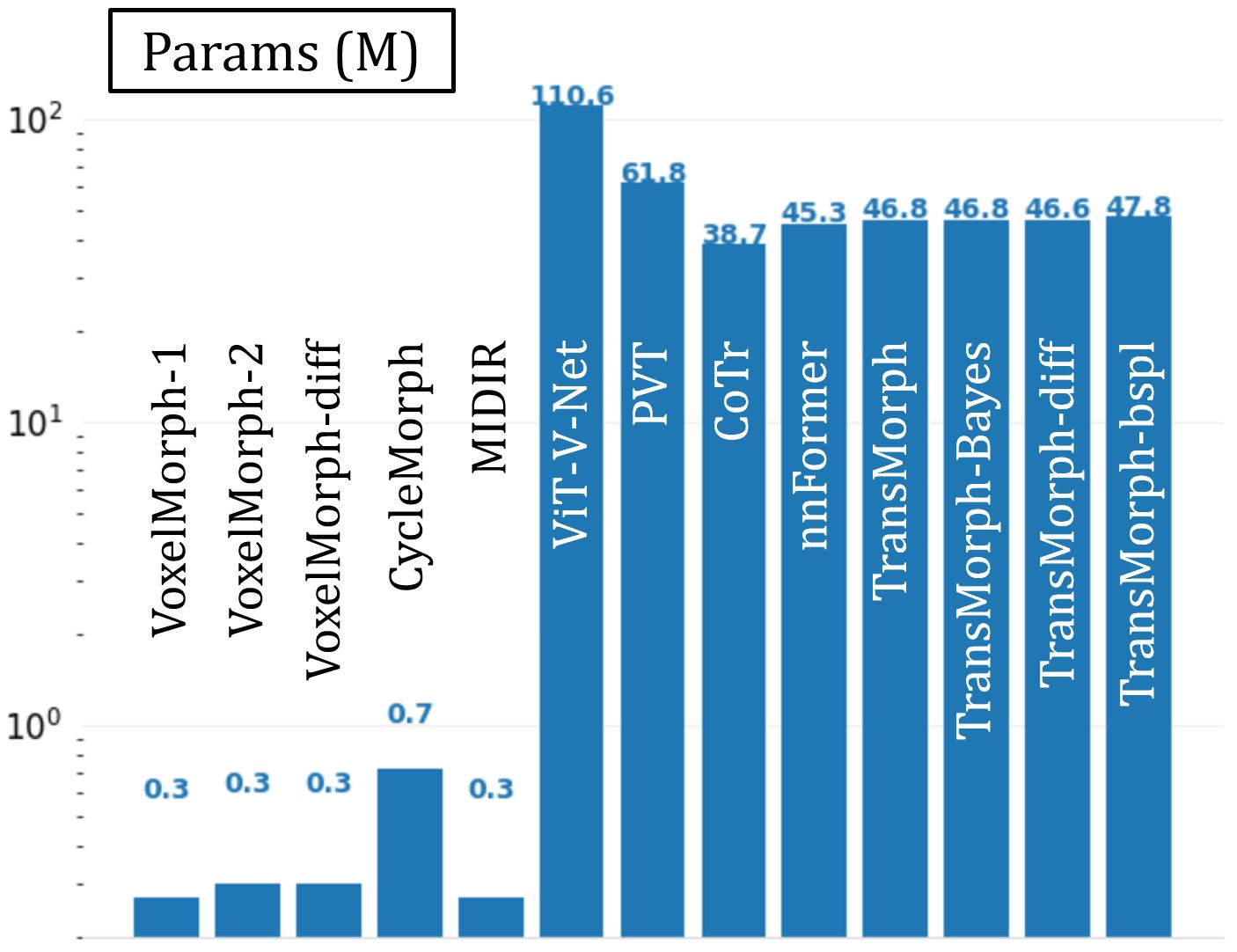}
\caption{The number of parameters in each deep-learning-based model. The values are in units of millions of parameters.\label{fig:model_params}}
\end{figure}
\subsubsection{Model complexity study}
The impact of model complexity on registration performance was also investigated in this paper. Table \ref{table: TransMorph_sz_ablation} listed the parameter settings and the number of trainable parameters of four variants of the proposed \texttt{TransMorph} model. In the base model, \texttt{TransMorph}, the embedding dimension $C$ was set to 96, and the number of Swin Transformer blocks in the four stages of the encoder was set to 2, 2, 4, and 2, respectively. Additionally, we introduced \texttt{TransMorph}-tiny, \texttt{TransMorph}-small, and \texttt{TransMorph}-large, which are about $1/200\times$, $1/4\times$, and $2\times$ the model size of \texttt{TransMorph}. Finally, we compared our model to a customized \texttt{VoxelMorph} (denoted \texttt{VoxelMorph}-huge), which has a comparable parameter size to that of \texttt{TransMorph} w/ lrn. positional embedding. Specifically, we maintained the same number of layers in \texttt{VoxelMorph}-huge as in \texttt{VoxelMorph}, but increased the number of convolution kernels in each layer. As a result, \texttt{VoxelMorph}-huge has 63.25 million trainable parameters.
\begin{figure*}[!t]
\centering
\includegraphics[width=.98\textwidth]{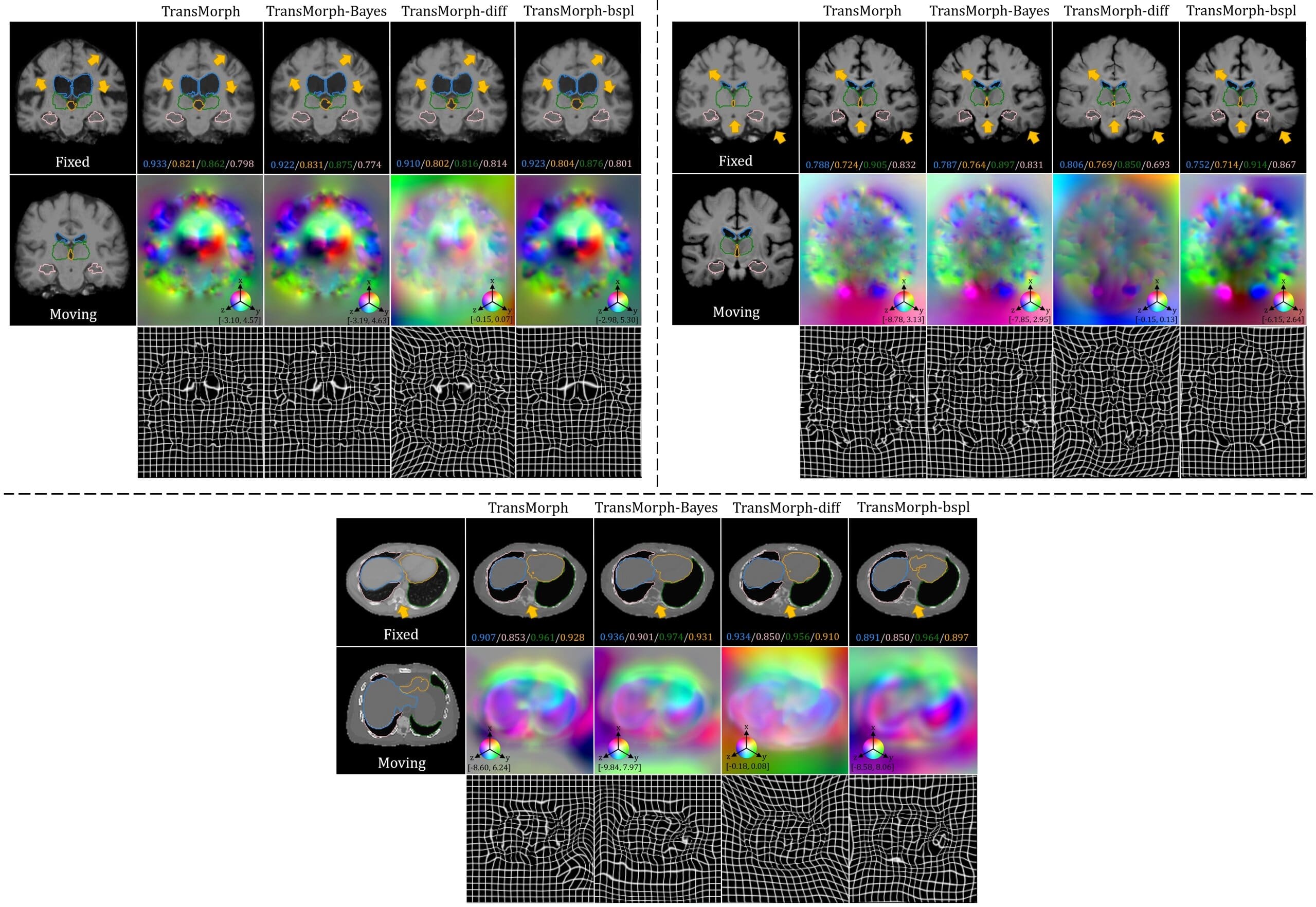}
\caption{Qualitative results of \texttt{TransMorph} ($2^{nd}$ column) and its Bayesian- ($3^{rd}$ column), probabilistic- ($4^{th}$ column), and B-spline ($5^{th}$ column) variants. Top-left \& Top-right panels: Results of inter-patient and atlas-to-patient brain MRI registration. The blue, orange, green, and pink contours define, respectively, the ventricles, third ventricle, thalami, and hippocampi. Bottom panel: Results of XCAT-to-CT registration. The blue, orange, green, and pink contours define, respectively, the liver, heart, left lung, and right lung. The second row in both panels exhibits the displacement fields $\mathbf{u}$, where spatial dimension $x$, $y$, and $z$ is mapped to each of the RGB color channels, respectively. The [$p$, $q$] in color bars denotes the magnitude range of the fields. \label{fig:brain_xcat_transmorph}}
\end{figure*}
\subsection{Evaluation Metrics}
The registration performance of each model was evaluated based on the volume overlap between anatomical/organ segmentation, which was quantified using the Dice score~(\cite{dice1945measures}). We averaged the Dice scores of all anatomical/organ structures for all patients. The mean and standard deviation of the averaged scores were compared across various registration methods. 

To quantify the regularity of the deformation fields, we also reported the percentages of non-positive values in the determinant of the Jacobian matrix on the deformation fields (i.e., $\vert J_{\phi}\vert\leq0$).

Additionally, for XCAT-to-CT registration, we used the structural similarity index (SSIM)~(\cite{wang2004image}) to quantify the structural difference between the deformed XCAT and the target CT images. The mean and standard deviation of the SSIM values of all patients were reported and compared.
\section{Results}
\label{sec:results}

\begin{figure*}[!t]
\centering
\includegraphics[width=.98\textwidth]{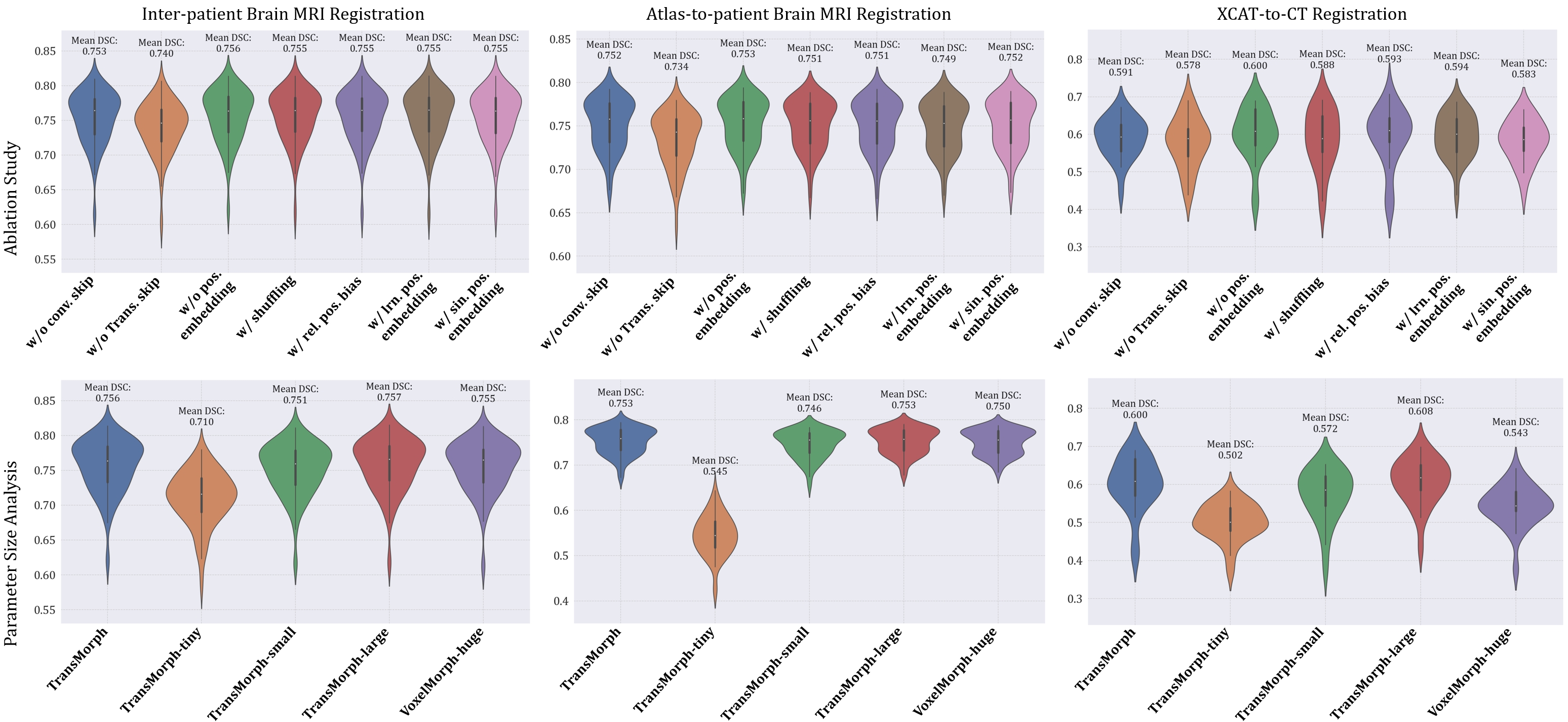}
\caption{Quantitative evaluation results of the additional studies performed on the validation datasets of the two brain MRI and XCAT-to-CT registration tasks.\label{fig:ablation_res}}
\end{figure*}

\subsection{Inter-patient Brain MRI Registration}
The top-left panel of Fig. \ref{fig:brain_xcat_transmorph} shows the qualitative results of a sample slice for inter-patient brain MRI registration. The scores in blue, orange, green, and pink correspond to ventricles, third ventricle, thalami, and hippocampi, respectively. Additional qualitative comparisons across all methods are shown in Fig. \ref{fig:jhu_brain} in \ref{sec:add_inter_MR}. Among the proposed models, diffeomorphic variants (i.e., \texttt{TransMorph-diff} and \texttt{TransMorph-bspl}) generated smoother displacement fields, with \texttt{TransMorph-bspl} producing the smoothest deformations inside the brain area. On the other hand, \texttt{TransMorph} and \texttt{TransMorph-Bayes} showed better qualitative results (highlighted by the yellow arrows) with higher Dice scores for the delineated structures. 

The quantitative evaluations are shown in Table \ref{tab:table_res}. The results presented in the table show that the proposed method, \texttt{TransMorph}, achieved the highest mean Dice score of 0.745. Although the diffeomorphic variants produced slightly lower Dice scores than \texttt{TransMorph}, they still outperformed the existing registration methods and generated almost no foldings (i.e., $\sim0\%$ of $\vert J_{\phi}\vert\leq0$) in the deformation fields. By comparison, \texttt{TransMorph} improved Dice score by $>$0.2 when compared to \texttt{VoxelMorph} and \texttt{CycleMorph}. We found that the Transformer-based models (i.e., \texttt{TransMorph}, \texttt{ViT-V-Net}, \texttt{PVT}, \texttt{CoTr}, and \texttt{nnFormer}) generally produced better Dice scores than the ConvNet-based models. Note that even though \texttt{ViT-V-Net} had almost twice the number of the trainable parameters (as shown in Fig. \ref{fig:model_params}), \texttt{TransMorph} still outperformed all the Transformer-based models (including \texttt{ViT-V-Net}) by at least 0.1 in the Dice score, demonstrating Swin-Transformer's superiority over other Transformer architectures. When we conducted hypothesis testing on the results using the paired $t$-test with Bonferroni correction~\cite{armstrong2014use} (i.e., dividing the $p$-values by 13, the total number of the paired $t$-tests performed), the $p$-values between the best performing \texttt{TransMorph} variant (i.e., \texttt{TransMorph}) and all other methods were $p\ll0.0005$.

Figs. \ref{fig:jhu_brain_conventional} and \ref{fig:jhu_brain_transformer} show additional Dice results for a variety of anatomical structures, with Fig. \ref{fig:jhu_brain_conventional} comparing \texttt{TransMorph} to current registration techniques (both optimization- and learning-based methods), and Fig. \ref{fig:jhu_brain_transformer} comparing the Dice scores between the Transformer-based models.

\subsection{Atlas-to-patient Brain MRI Registration}
The top-right panel of Fig. \ref{fig:brain_xcat_transmorph} shows the qualitative results of the \texttt{TransMorph} variants on a sample MRI slice for atlas-to-patient brain MRI registration. As highlighted by the yellow arrows, the diffeomorphic variants resulted in the deformed images that were less comparable to the fixed image in terms of visual appearance. In contrast, the variants without diffeomorphic deformations (i.e., \texttt{TransMorph} and \texttt{TransMorph-Bayes}) produced better qualitative results, with the sulci in the deformed atlas images more closely matching those in the fixed image. Additional qualitative comparisons are shown in Fig. \ref{fig:IXI_brain} in \ref{sec:add_atlas_MR}, where we observed that all the learning-based methods yielded more detailed and precise deformation fields than the conventional methods. This might be owing to the high parameterization of the DNNs, which enables the modeling of more complicated deformations.

Table. \ref{tab:table_res} shows the quantitative evaluation results of the atlas-to-patient registration. The highest mean Dice score of 0.761 was achieved by the proposed \texttt{TransMorph-bspl} with nearly no folded voxels. The second best Dice score of 0.754 was achieved by both \texttt{TransMorph} and \texttt{TransMorph-Bayes}, while \texttt{TransMorph-Bayes} yielded a smaller standard deviation. In comparison to these \texttt{TransMorph} variants, \texttt{TransMorph-diff} produced a lower Dice score of 0.594. However, note that this score is still higher ($\sim$0.02) than the one produced by \texttt{VoxelMorph-diff}, which is the base model of \texttt{TransMorph-diff}. Additionally, we observed that the registration methods that used MSE for training or optimization resulted in lower Dice scores (i.e., \texttt{SyN}, \texttt{NiftyReg}, \texttt{LDDMM}, \texttt{VoxelMorph-diff}, and \texttt{TransMorph-diff}). This was most likely due to the significant disparity in the intensity values of brain sulci between the atlas and the patient MRI images. As seen in the top-right panel of Fig \ref{fig:brain_xcat_transmorph}, the sulci in the atlas image (i.e., the moving image) exhibited low-intensity values comparable to the background, but the sulci in the patient MRI image had intensity values more comparable to the neighboring gyri. Thus, the discrepancies in the sulci intensity values may account for the majority of the MSE loss during training, compelling the registration models to fill the sulci in the atlas image with other brain structures (as shown in Fig. \ref{fig:IXI_brain}, these models produced significantly smaller sulci than models trained with LNCC), thereby limiting registration performance. The paired $t$-tests with Bonferroni correction~(\cite{armstrong2014use}) revealed the $p$-values of $p\ll0.0005$ between the best performing model (i.e., \texttt{TransMorph}-bspl) and all other methods. This indicates that the proposed method outperformed the comparative registration methods and network architectures.

A detailed breakdown of Dice scores for a variety of anatomical structures is shown in Figs. \ref{fig:IXI_brain_conventional} and \ref{fig:IXI_brain_transformer} in \ref{sec:add_atlas_MR}.
\begin{table*}[!t]
\fontsize{10}{11}\selectfont
\caption{System-level comparison of various \texttt{TransMorph} designs and the customized \texttt{VoxelMorph} on the validation and test datasets of inter-patient MRI, atlas-to-patient MRI, and XCAT-to-CT registration tasks. ”Val. DSC” denotes the Dice scores on the validation dataset; ``Test DSC" denotes the system-level comparison of the Dice scores on the test dataset. The \textbf{bolded} numbers denote the highest scores, while the \textit{italicized} ones indicate the second highest.}
\centering
    \begin{tabular}{ c || c | c | c | c | c | c }
 \hline
  & \multicolumn{2}{c|}{Inter-patient MRI} & \multicolumn{2}{c|}{Atlas-to-patient MRI} & \multicolumn{2}{c}{XCAT-to-CT} \\
 \hline
 Model & Val. DSC & Test DSC & Val. DSC & Test DSC & Val. DSC & Test DSC\\
 \hline
 w/o conv. skip.& 0.753$\pm$0.119 & 0.743$\pm$0.124 & 0.752$\pm$0.129 & 0.754$\pm$0.125 & 0.591$\pm$0.319 & 0.586$\pm$0.314\\
 \hline
 w/o Trans. skip. & 0.740$\pm$0.124 & 0.727$\pm$0.130  & 0.734$\pm$0.127 & 0.736$\pm$0.125 & 0.578$\pm$0.315 & 0.588$\pm$0.314\\
 \hline
 w/ shuffling & 0.755$\pm$0.119 & 0.744$\pm$0.125 & 0.751$\pm$0.127 & \textbf{0.754$\pm$0.123} & 0.588$\pm$0.314 & 0.597$\pm$0.310\\
 \hline
 w/ rel. positional bias & 0.755$\pm$0.120 & 0.742$\pm$0.125& 0.751$\pm$0.131 & 0.753$\pm$0.127 & 0.593$\pm$0.315 & 0.592$\pm$0.319\\
 \hline
 w/ lrn. positional embedding & 0.755$\pm$0.120 & 0.744$\pm$0.125 & 0.749$\pm$0.131 & 0.751$\pm$0.129 & 0.594$\pm$0.315 & 0.586$\pm$0.315\\
 \hline
 w/ sin. positional embedding & 0.755$\pm$0.120 & 0.744$\pm$0.125  & \textit{0.752$\pm$0.126} & \textbf{0.754$\pm$0.123} & 0.583$\pm$0.320 & 0.572$\pm$0.317\\
 \hline
 \hline
 \texttt{TransMorph}& \textit{0.756$\pm$0.119} & \textit{0.745$\pm$0.125}& \textbf{0.753$\pm$0.127} & \textit{0.754$\pm$0.124} & \textit{0.600$\pm$0.317} & \textit{0.604$\pm$0.314}\\
 \hline
 \texttt{TransMorph}-tiny & 0.710$\pm$0.132 & 0.696$\pm$0.140 &0.545$\pm$0.180 & 0.543$\pm$0.180 & 0.502$\pm$0.311 & 0.501$\pm$0.312\\
 \hline
 \texttt{TransMorph}-small & 0.751$\pm$0.121 & 0.740$\pm$0.126  & 0.746$\pm$0.128 & 0.747$\pm$0.125 & 0.572$\pm$0.320 & 0.570$\pm$0.318\\
 \hline
 \texttt{TransMorph}-large & \textbf{0.757$\pm$0.119}& \textbf{0.746$\pm$0.124}  & \textit{0.753$\pm$0.130} & 0.754$\pm$0.128 & \textbf{0.608$\pm$0.305} & \textbf{0.611$\pm$0.311}\\
 \hline
 \texttt{VoxelMorph}-huge & 0.755$\pm$0.119& 0.744$\pm$0.124  & 0.750$\pm$0.133 & 0.751$\pm$0.130 & 0.543$\pm$0.320 &  0.550$\pm$0.319 \\
 \hline
\end{tabular}
\label{tab:table_ablation_sys}
\end{table*}

\begin{figure*}[!t]
\centering
\includegraphics[width=.85\textwidth]{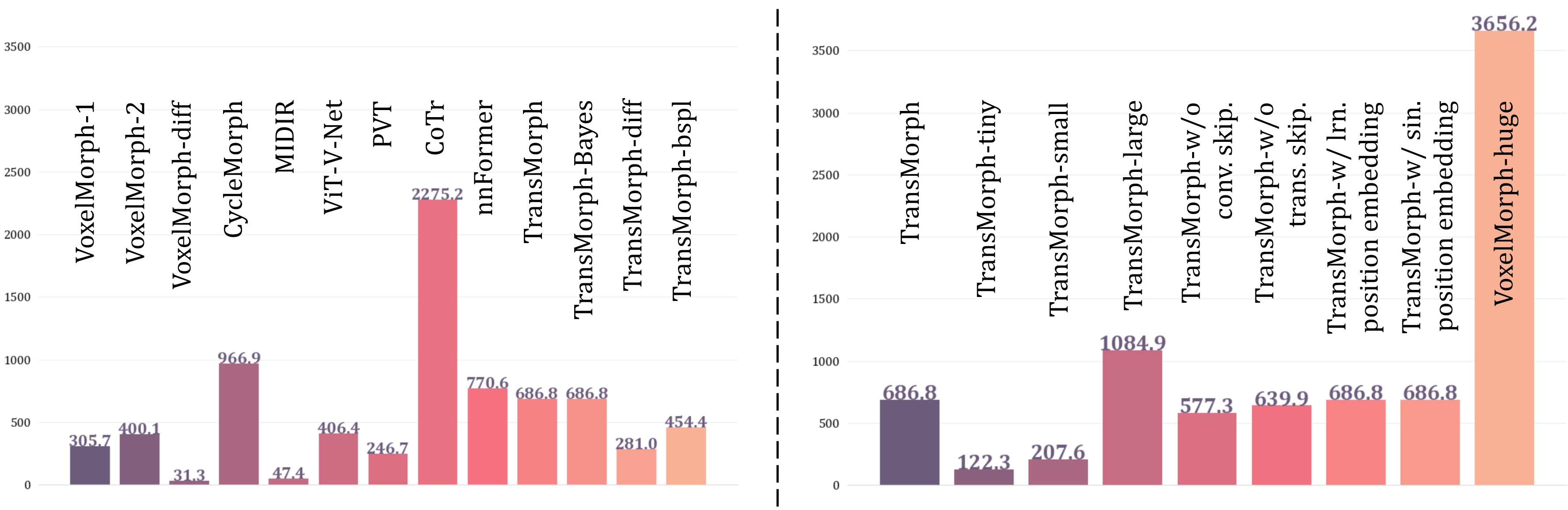}
\caption{Model computational complexity comparisons represented in Giga multiply–accumulate operations (GMACs). Greater values imply a greater degree of computational complexity. These values were obtained using an input image of size $160\times 192\times 224$.  \label{fig:MACs_compare}}
\end{figure*}

\subsection{Learn2Reg OASIS Brain MRI Registration}
Table \ref{tab:table_res_OASIS} shows the quantitative results of the validation and test sets of the challenge. The validation scores of the various methods were obtained from the leaderboard of the challenge, whilst the test scores were obtained directly from the organizers. \texttt{TransMorph} performed similarly to the best-performing method (\texttt{LapIRN} (\cite{mok2021conditional})) of the challenge on the validation set, where \texttt{TransMorph}-large achieved the best mean Dice score of 0.862 and mean HdDist95 of 1.431. \texttt{VoxelMorph}-huge performed significantly poor than \texttt{TransMorph}, with a $p$-value less than 0.01 from paired $t$-test. This reveals the superiority of Transformer-based architecture over ConvNet despite having a comparable number of parameters. On the test set, the \texttt{TransMorph} and \texttt{TransMorph}-large achieved comparable mean Dice score to that of \texttt{LapIRN}. Despite the comparable performance, \texttt{LapIRN} produced much more uniform deformation fields as measured by SDlogJ. In a separate study, we presented a simple extension of \texttt{TransMorph} that significantly outperformed \texttt{LapIRN} while maintaining smooth deformation fields. We direct interested readers to (\cite{chen2022unsupervised}) for further details. Moreover, \texttt{LapIRN} employed a multiresolution framework in which three ConvNet registration backbones were involved in generating deformation fields at three different scales. \texttt{TransMorph}, however, operated on a single resolution. We underline that \texttt{TransMorph} is a registration backbone, and that it may be easily adapted to \texttt{LapIRN} or any advanced registration frameworks.

\subsection{XCAT-to-CT Registration}
The bottom panel of Fig. \ref{fig:brain_xcat_transmorph} shows the qualitative results for a representative CT slice. The blue, orange, green, and pink lines denote the liver, heart, left lung, and right lung, respectively, while the bottom values show the corresponding Dice scores. Similar to the findings in the previous sections, \texttt{TransMorph} and \texttt{TransMorph-Bayes} gave more accurate registration results (highlighted by the yellow arrows and the delineated structures), while the diffeomorphic variants produced smoother deformations. Additional qualitative comparisons are shown in Fig. \ref{fig:xcat_CT} in \ref{sec:add_xcat_ct}. It is possible to see certain artifacts in the displacement field created by \texttt{nnFormer} (as shown in Fig. \ref{fig:xcat_CT}); these were most likely caused by the patch operations of the Transformers used in its architecture. \texttt{nnFormer} is a near-convolution-free model (convolutional layers are employed only to form displacement fields). In contrast to the relatively small displacements in brain MRI registration, displacements in XCAT-to-CT registration may exceed the patch size. Consequently, the lack of convolutional layers to refine the stitched displacement field patches may have resulted in artifacts. Four example coronal slices of the deformed XCAT phantoms generated by various registration methods are shown in Fig. \ref{fig:xcat_deform} in \ref{sec:add_xcat_ct}.

The quantitative evaluation results are presented in Table \ref{tab:table_res_XCAT}. They include Dice scores for all organs and scans, the percentage of non-positive Jacobian determinants, and the structural similarity index (SSIM)~(\cite{wang2004image}) between the deformed XCAT phantom and the target CT scan. The window size used in SSIM was set to 7. Without registration or affine transformation, a Dice score of 0.22 and an SSIM of 0.576 demonstrate the large dissimilarity between the original XCAT phantom and patient CT scans. The Dice score and SSIM increased to 0.33 and 0.751, respectively, after aligning the XCAT and patient CT using the proposed affine Transformer. Among the traditional registration methods, \texttt{deedsBCV}, which was initially designed for abdominal CT registration-based segmentation~(\cite{heinrich2015multi}), achieved the highest Dice score of 0.568, which is even higher than most of the learning-based methods. Among the learning-based methods, Transformer-based models outperformed ConvNet-based models on average, which is consistent with the findings from the brain MR registration tasks. The $p$-values from the paired $t$-tests with Bonferroni correction~(\cite{armstrong2014use}) between \texttt{TransMorph} and all non-\texttt{TransMorph} methods were $p\ll0.05$. The proposed \texttt{TransMorph} models yielded the highest Dice and SSIM scores of all methods in general, with the best Dice of 0.604 given by \texttt{TransMorph} and the best SSIM of 0.919 given by \texttt{TransMorph-Bayes}. The diffeomorphic variants produced lower Dice and SSIM scores as a consequence of not having any folded voxels in the deformation.

Figs. \ref{fig:xcat_brain_conventional} and \ref{fig:xcat_brain_transformer} show additional boxplots of Dice scores on the various abdominal organs, with Fig. \ref{fig:xcat_brain_conventional} comparing \texttt{TransMorph} to current registration techniques (both optimization- and learning-based methods), and Fig. \ref{fig:xcat_brain_transformer} comparing the Dice scores between the Transformer-based models.
\begin{figure*}[!t]
\centering
\includegraphics[width=.98\textwidth]{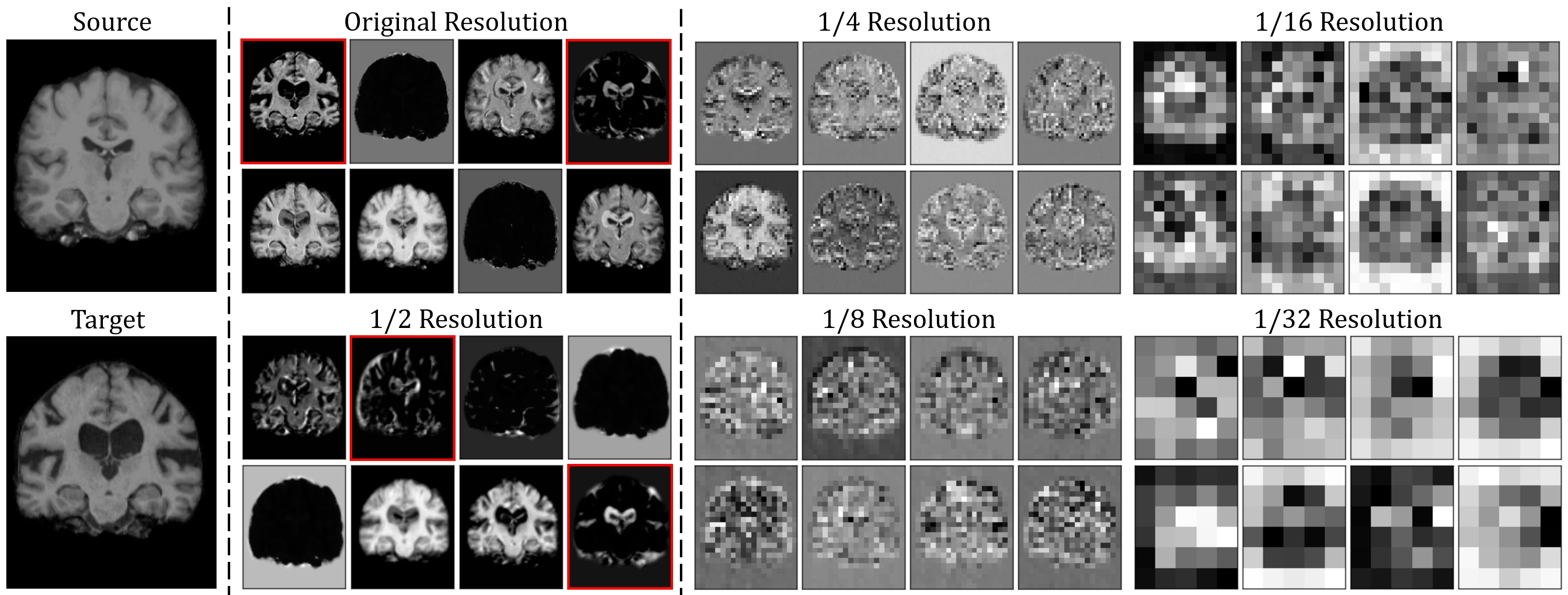}
\caption{Examples of feature maps in \texttt{TransMorph}'s skip connections. Eight feature maps are \textit{randomly} selected from the feature maps associated with each skip connection. Left panel: Example 2D slices of source and target images (i.e., $I_m$ and $I_f$), which are used as inputs to \texttt{TransMorph}. Middle panel: Feature maps in the skip connections of the two convolutional layers (denoted by the green arrows in Fig. \ref{fig:net_arch}). Right panel: Feature maps in the skip connections of the Swin Transformer blocks (denoted by the orange arrows in Fig. \ref{fig:net_arch}). \label{fig:skip_feats_8}}
\end{figure*}

\subsection{Ablation Studies}
\label{sec:ablation_res}
\paragraph{Inter-patient Registration} The first figure in the first row of Fig. \ref{fig:ablation_res} shows the violin plots of Dice scores from the ablation study on the validation dataset of inter-patient brain MR registration. When evaluating the effectiveness of skip connections, we observed that the skip connections from both the convolution and Transformer layers improved registration performance. \texttt{TransMorph} scored a mean Dice of 0.753 after the skip connections from the convolutional layers were removed. However, the score decreased to 0.740 when the skip connections from the Transformer blocks were removed. In comparison, the skip connections from convolutional layers were less effective, with a mean Dice improvement of 0.003. Note that the \texttt{TransMorph} with shuffling, and with and without positional embeddings all generated comparable mean Dice scores and violin plots, suggesting that positional embedding may not be necessary.

\paragraph{Atlas-to-patient Registration} The violin plots from the ablation study on the atlas-to-patient registration task are shown in the second figure in the first row of Fig. \ref{fig:ablation_res}. Comparable violin plots with similar mean Dice scores around 0.752 were observed with and without the skip connections from the convolutional layers. When the skip connections from the Transformer blocks were removed, the Dice score decreased by 0.019, reflecting the effectiveness of these skip connections. Comparable violin plots and mean Dice scores around 0.750 were observed with shuffling, and with and without various positional embeddings, confirming that \texttt{TransMorph}'s performance is unaffected by whether or not positional embedding was used.

\paragraph{XCAT-to-CT Registration} The second to last figure in the first row of Fig. \ref{fig:ablation_res} shows the violin plots from the validation dataset of XCAT-to-CT registration task. Without the skip connections from the convolution and Transformer layers, the Dice scores dropped by 0.013 and 0.016, respectively, when compared to \texttt{TransMorph}, further supporting the observation that skip connections can improve performance. Learnable and relative positional embeddings yielded comparable mean Dice scores for XCAT-to-CT registration in the range of 0.593. When sinusoidal positional embedding was employed, a score of 0.583 was attained, whereas a score of 0.588 was produced when the positions were shuffled. With a score of 0.600, without using positional embeddings yielded a slight improvement among other variants. The effect of each component is addressed in depth in the Discussion section (section \ref{sec:net_ele_discuss}).

In conclusion, the results from all three tasks indicate that using skip connections improves performance. The results of the three tasks (i.e., inter-patient, atlas-to-patient, and XCAT-to-CT registration tasks) reveal that with and without using positional embedding or even randomly shuffling the token positions produced similar results. Additionally, we applied the \texttt{TranMorph} models to the test datasets of the three registration tasks for system-level comparisons, and the results are shown in the upper panel of Table. \ref{tab:table_ablation_sys}. The scores on the test datasets followed the same trend as those on the validation datasets, where the positional embeddings had an insignificant influence on registration performance.

\subsection{Computational Complexity}
The barplot in the left panel of Fig. \ref{fig:MACs_compare} shows the computational complexity comparisons between the deep-learning-based registration models. The plot was created using an input image with a resolution of $160\times192\times224$, the same size as the brain MRI images. The numbers were expressed in Giga multiply-accumulate operations (GMACs), with a higher value indicating a more computationally expensive model that may also be more memory intensive. The proposed model, \texttt{TransMorph}, and its Bayesian variant, \texttt{TransMorph-Bayes}, had a moderate computational complexity with 687 GMACs which is much less than \texttt{CoTr} and \texttt{CycleMorph}. In practice, the GPU memory occupied during training for \texttt{TransMorph} was about 15 GiB with a batch size of 1 and an input image size of $160\times192\times224$. The diffeomorphic variants, \texttt{TransMorph-diff} and \texttt{TransMorph-bspl}, had 281 and 454 GMACs, which are comparable to that of the conventional ConvNet-based registration models, \texttt{VoxelMorph}-1 and -2. In practice, they occupied approximately 11 GiB of GPU memory during training, which is a size that can be readily accommodated by the majority of modern GPUs. In terms of the number of parameters, all ConvNet-based models had fewer than 1M network parameters (as shown in Fig. \ref{fig:model_params}); yet their GMACs (i.e., computational complexity) were comparable to \texttt{TransMorph}, but their registration performances were significantly inferior. Transformer-based models were all of large scale, with more than 30M parameters. Notably, \texttt{ViT-V-Net} and \texttt{PVT} had around 2$\times$ and 1.5$\times$ more parameters than \texttt{TransMorph}, nevertheless \texttt{TransMorph} outperformed them by a significant margin on all of the evaluated registration tasks. This demonstrates that the success of \texttt{TransMorph} owes not just to the large model size but also to the architecture itself.

Fig. \ref{fig:ablation_res} shows the quantitative results of \texttt{TransMorph} models with various architectural settings and the customized ConvNet-based model \texttt{VoxelMorph}-huge on the validation datasets of the three registration datasets. When parameter size is the only variable in \texttt{TransMorph} models, there is a strong correlation between model complexity (as shown in the right panel of Fig. \ref{fig:MACs_compare}) and registration performance. \texttt{TransMorph}-tiny produced the lowest mean Dice of 0.710, 0.545, and 0.502 on the validation set of the three registration tasks, respectively. The Dice score steadily improves as the complexity of the model increases. Note that for inter-patient and atlas-to-patient brain MRI registration (the first and second figures in the bottom row of Fig. \ref{fig:ablation_res}), the improvement in mean Dice score from \texttt{TransMorph} to \texttt{TransMorph}-large were mostly under 0.01 but the latter was almost twice as computationally costly (as shown in the right panel of Fig. \ref{fig:MACs_compare}). The customized ConvNet-based model, \texttt{VoxelMorph}-huge, had the comparable number of parameters as \texttt{TransMorph}. However, it achieved slightly lower mean Dice scores than those of \texttt{TransMorph} for the JHU and IXI brain MR registration tasks, and significantly lower scores for OASIS brain MR and the XCAT-to-CT registration task. This further indicates the architectural advantages of \texttt{TransMorph} for image registration. A significant disadvantage of \texttt{VoxelMorph}-huge was its computational complexity, with 3656 GMACs (as seen in the right panel of Fig. \ref{fig:MACs_compare}), it was nearly five times as computationally expensive as \texttt{TransMorph}, making it memory-intensive ($\sim$ 22 GiB for a patch size of 1 during training) and slow to train in practice. However, \texttt{TransMorph} was able to accommodate a larger number of parameters without significantly increasing computational complexity. The promising performances brought by the larger scale of parameters demonstrate the superior scaling property of Transformer-based models as described in (\cite{zhai2022scaling, liu2022convnet}). The \texttt{TranMorph} models with different model parameter settings and \texttt{VoxelMorph}-huge were applied to the test datasets for system-level comparisons, and the results are shown in the bottom panel of Table. \ref{tab:table_ablation_sys}.

\begin{figure}[!h]
\centering
\includegraphics[width=.48\textwidth]{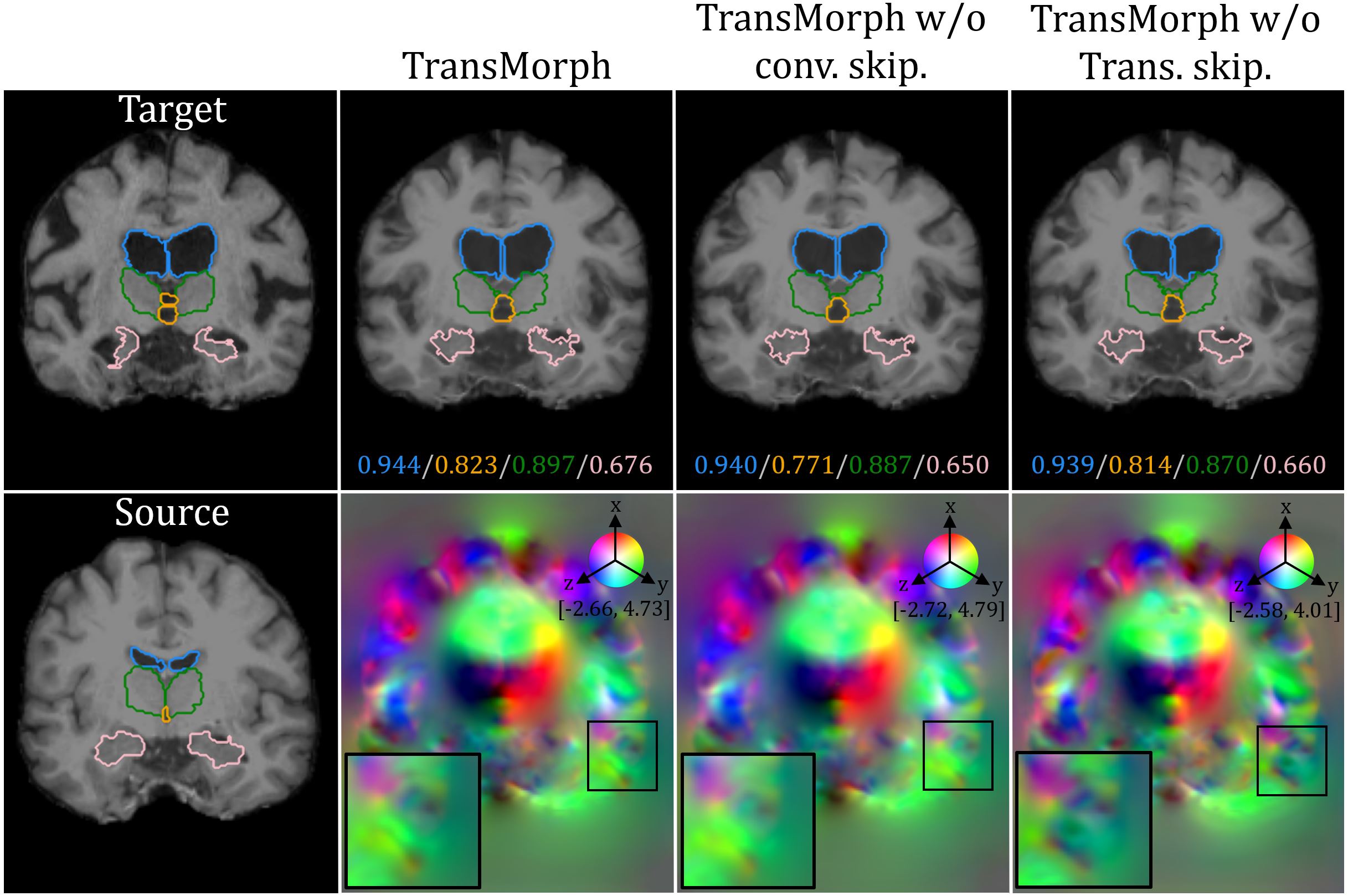}
\caption{Qualitative impact of skip connections on the deformation fields. The spatial dimension $x$, $y$, and $z$ in the displacement field is mapped to each of the RGB color channels, respectively. The [$p$, $q$] in color bars denotes the magnitude range of the fields.\label{fig:skip_res}}
\end{figure}
\section{Discussion}
\label{sec:discussion}

\subsection{Network Components in TransMorph}
\label{sec:net_ele_discuss}
\subsubsection{Skip Connections}

As previously shown in section \ref{sec:ablation_res}, skip connections may aid in enhancing registration accuracy. In this section, we give further insight into the skip connections' functionality. 

Fig. \ref{fig:skip_feats_8} shows some example feature maps in each skip connection (a full feature map visualization is shown in Fig. \ref{fig:skip_feat} in Appendix). Specifically, the left panel shows sample slices of the input volumes; the center panel illustrates selected feature maps in the skip connections of the convolutional layers, and the right panel illustrates selected feature maps in the skip connections of the Swin Transformer blocks. As seen from these feature maps that the Swin Transformer blocks provided more abstract information (right panel in Fig. \ref{fig:skip_feats_8}), in comparison to the convolutional layers (middle panel in Fig. \ref{fig:skip_feats_8}). Since a Transformer divides an input image volume into patches to create tokens for self-attention operations (as described in section \ref{sec:reg_net}), it can only deliver information up to a certain resolution, which is often a factor of the patch size lower than the original resolution (i.e., $\frac{H}{P}\times\frac{W}{P}\times\frac{L}{P}$, and $P=4$ in our case). On the other hand, the convolutional layers resulted in higher resolution feature maps with more detailed and human-readable information (e.g., edge and boundary information). Certain feature maps even revealed distinctions between the moving and fixed images (highlighted by the red boxes). Fig. \ref{fig:skip_res} shows the qualitative comparisons between the proposed model with and without a specific type of skip connection. As seen by the magnified areas, \texttt{TransMorph} with both skip connection types provided a more detailed and accurate displacement field. Therefore, adding the skip connections from the convolutional layers is still recommended, although the actual Dice improvement were subtle on the validation datasets ($0.003$ for inter-patient brain MRI, $0.001$ for atlas-to-patient brain MRI, and $0.009$ for XCAT-to-CT registration).
\begin{figure}[!h]
\centering
\includegraphics[width=.48\textwidth]{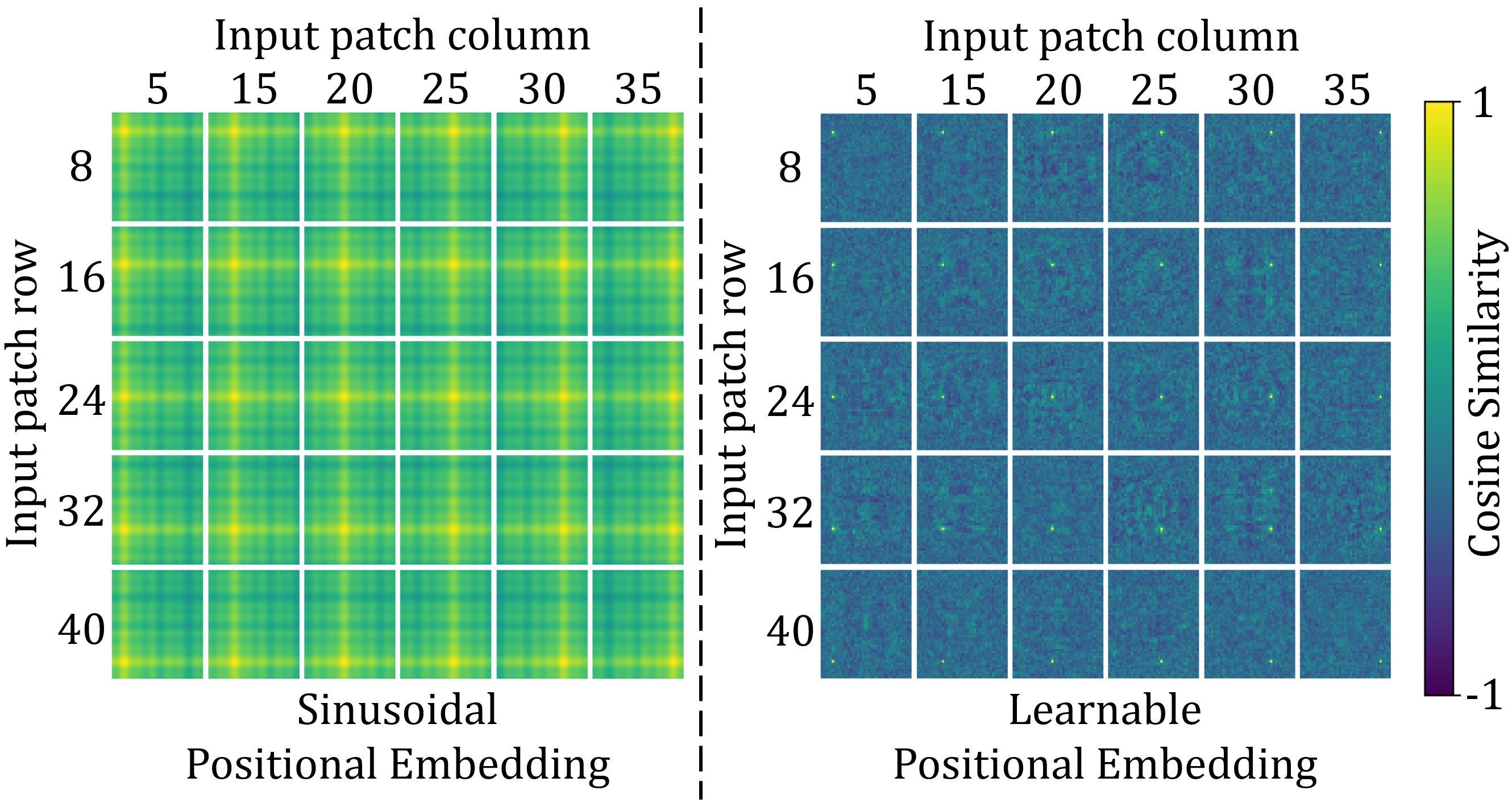}
\caption{Example slice of the positional embeddings used in \texttt{TransMorph}. Left panel: Sinusoidal positional embedding. Right panel: Learnable positional embedding. Tiles in both panels show the cosine similarities between the position embedding of the token with the indicated row and column and the position embeddings of all other tokens.\label{fig:pos_embed}}
\end{figure}

\begin{figure*}[!t]
\centering
\includegraphics[width=.95\textwidth]{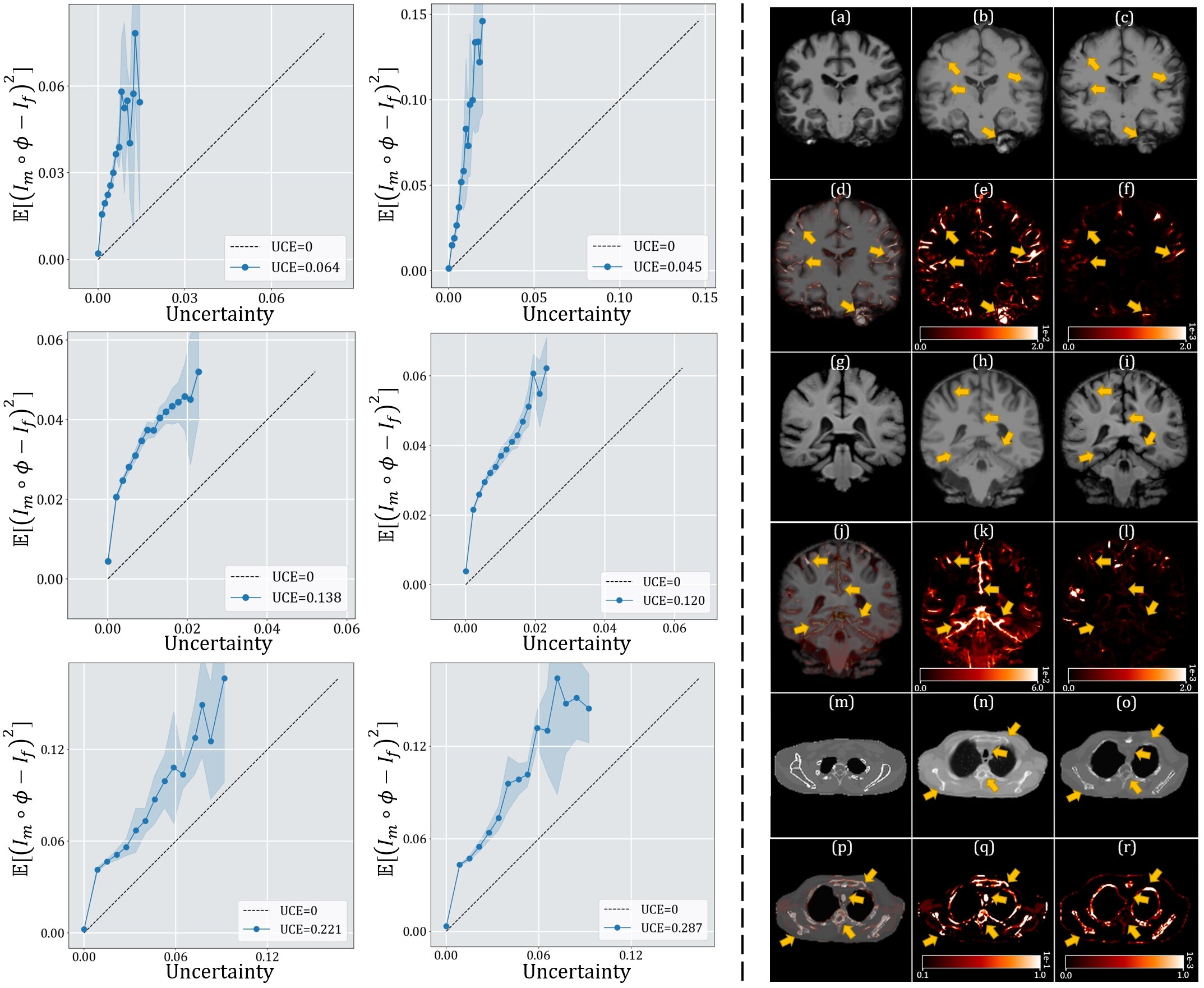}
\caption{Comparisons of the appearance uncertainty estimates derived from the predictive variance and the predicted model error. Left panel: Calibration plots and uncertainty calibration error (UCE) for \texttt{TransMorph-Bayes} on two inter-patient brain MR test sets (top), two atlas-to-patient brain MR test sets (middle), and two XCAT-to-CT test sets (bottom). The blue lines represent the results obtained using the uncertainty estimate $\hat{\Sigma}_f^2$. The dashed lines represent the perfect calibration, which are the results achieved when the uncertainty estimate is $\Sigma_f^2$ or $err(I_m\circ\phi)$ (i.e., the expected model error). The values are obtained from 10 repeated runs, and the shaded regions represent the standard deviation. Right panel: Visualization of the registration uncertainty on an inter-patient brain MRI test set (i.e., a-f), an atlas-to-patient brain MRI test set (i.e., g-l), and a CT test set (i.e., m-r). (a), (g), \& (m): Moving image. (b), (h), \& (n): Fixed image. (c), (i), \& (o): Deformed moving image. (d), (j), \& (p): Per-pixel uncertainty, represented by $\Sigma_f^2$, overlays the deformed image. (e), (k), \& (q): Per-pixel uncertainty given by $\Sigma_f^2$ (i.e., the proposed method). (f), (l), \& (r): Per-pixel uncertainty given by $\hat{\Sigma}_f^2$. The yellow arrows highlight sites where $\Sigma_f^2$ identifies registration failures but $\hat{\Sigma}_f^2$ does not. \label{fig:cal_plots}}
\end{figure*}

\subsubsection{Positional Embedding}
\label{sec:pos_embd}
Transformers in computer vision were initially designed for image classification tasks~(\cite{dosovitskiy2020image, liu2021swin, dong2021cswin, wang2021pyramid}). Such a Transformer produces a condensed probability vector that is not in the image domain but instead a description of the likelihood of being a certain class. The loss calculated based on this vector does not backpropagate any spatial information into the network. Thus, it is critical to encode positional information on the patched tokens; otherwise, as the network gets deeper, Transformer would lose track of the tokens' locations relative to the input image, resulting in unstable training and inferior predictions. However, for pixel-level tasks like image registration, the condensed features generated by Transformers are often subsequently expanded using a decoder whose output is an image with the same resolution as the input and target images. Any spatial mismatching between the output and target contributes to the loss, which is then backpropagated throughout the network. As a result, the Transformer implicitly learns the positional information of tokens, thus obviating the need for positional embedding. In this work, we compared the registration performance of \texttt{TransMorph} and \texttt{TransMorph} with positional embedding on brain MRI and XCAT-to-CT registration. The results shown in section \ref{sec:ablation_res} indicated that positional embedding did not improve registration performance; rather, it introduced more parameters into the network. In this section, we discuss the positional embeddings in further detail.

Three positional embeddings were studied in this paper: sinusoidal~(\cite{vaswani2017attention}), learnable~(\cite{dosovitskiy2020image}), and relative~(\cite{liu2021swin}) embeddings, which are also the major types of positional embedding. In sinusoidal positional embedding, the position of each patched token is represented by a value drawn from a predetermined sinusoidal signal according to the token's position relative to the input image. Whereas with learnable positional embedding, the network learns the representation of the token's location from the training dataset rather than giving a hardcoded value. The relative positional bias hardcodes the relative position relations between any two tokens in the dot product of the query and key representations (i.e., $B$ in Eqn. \ref{eqn:atten}). To validate that the network learned the positional information,~\cite{dosovitskiy2020image} computed the cosine similarities between a learned embedding of a token and that of all other tokens. The obtained similarity values were then used to form an image. If positional information is learned, the image should reflect increased similarities at the token's and nearby tokens' positions.
\begin{figure*}[!t]
\centering
\includegraphics[width=.98\textwidth]{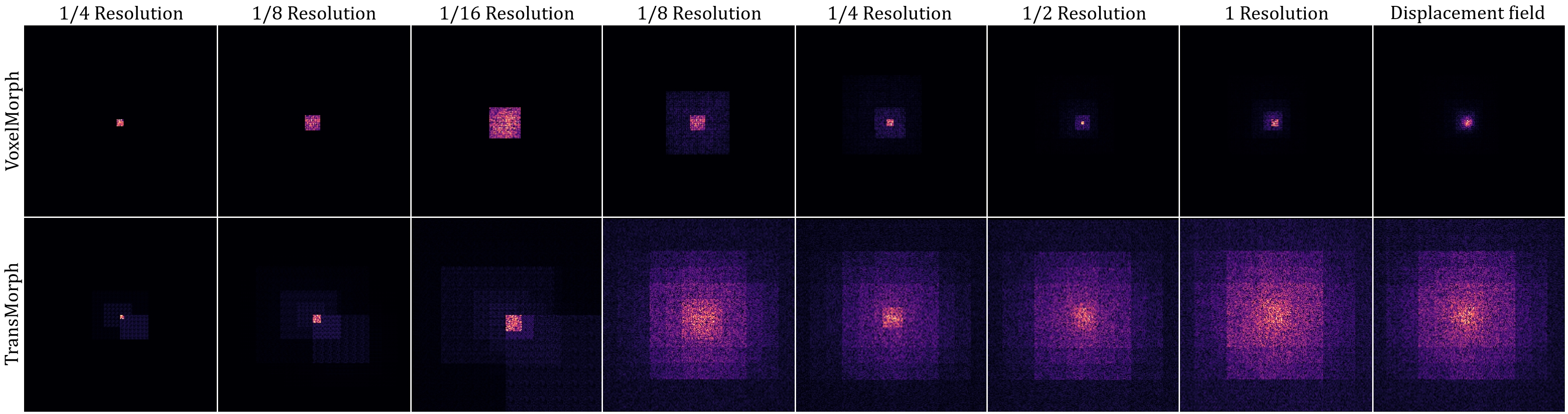}
\caption{Example ERFs of \texttt{VoxelMorph} and the proposed Transformer-based model \texttt{TransMorph}. The top row shows the ERF slices (i.e., $y=80$) at each stage of the network on an input image size of $160\times160\times160$. For a consistent comparison of ERFs between \texttt{VoxelMorph} and \texttt{TransMorph}, the ERFs at 1/2 of \texttt{VoxelMorph} and 1/32 resolution of \texttt{TransMorph} were omitted.\label{fig:ERF_example}}
\end{figure*}
Here, we computed the images of cosine similarities for both sinusoidal and learnable positional embeddings used in this work. The left and right panels in Fig. \ref{fig:pos_embed} show the images of cosine similarities. These images were generated based on an input image size of $160\times192\times224$ and a patch size of $4\times4\times4$ (resulting in $40\times48\times56$ patches). Each image has a size of $40\times48$ representing an image of cosine similarities in the plane of $z=28$ (i.e., the middle slice). There should have been a total of $40\times48$ images in each panel. However, for better visualization, just a few images were shown here. The images were chosen with step sizes of 5 and 8 in $x$ and $y$ direction, respectively, resulting in $6\times5$ images in each panel. As seen from the left panel, the images of sinusoidal embeddings exhibit a structured pattern, showing a high degree of correlation between tokens' relative locations and image intensity values. Note that the brightest pixel in each image represents the cosine similarity between a token's positional embedding and itself, which reflects the token's actual location relative to all other tokens. The similarity then gradually decreases as it gets farther away from the token. On the other hand, images generated with learnable embeddings (right panel of Fig. \ref{fig:pos_embed}) lack such structured patterns, implying that the network did not learn the positional information associated with the tokens in the learnable embeddings. To further demonstrate that the network implicitly learned the positional information, we randomly shuffled the token positions when computing self-attention during training and testing. As a result, the self-attention modules could not explicitly perceive input tokens' positional information. However, as seen from the Dice scores in Fig. \ref{fig:ablation_res}, regardless of shuffling and which positional embedding was employed, the mean Dice scores and violin plots were quite comparable to those produced without positional embedding. Thus, the findings confirmed that \texttt{TransMorph} learned the positional information of the tokens implicitly and that the learnable, sinusoidal, and relative positional embeddings were redundant in the model and had a negligible effect on registration performance.

\subsection{Uncertainty Quantification of TransMorph-Bayes}
As previously mentioned in section \ref{sec:bayes_variant}, the appearance uncertainty estimates produced by the predictive variance (Eqn. \ref{eqn:pred_var_0}) were actually miscalibrated, meaning that the uncertainty values did not properly correlate to predicted model errors since variance was computed using the predictive mean instead of target image $I_f$. We proposed to directly use the expected model error to express appearance uncertainty since the target image is available at all times in image registration. Thus, the resulting appearance uncertainty estimate is perfectly calibrated. In this section, we examine how the proposed and existing methods differ in their estimates of appearance uncertainty.

To quantify the calibration error, we used the Uncertainty Calibration Error (UCE) introduced in~(\cite{pmlr-v121-laves20a}), which is calculated on the basis of the binned difference between the expected model error (i.e., $\mathbb{E}\left[(I_m\circ\phi-I_f)^2\right]$) and the uncertainty estimation (e.g., $\hat{\Sigma}^2_f$ in Eqn. \ref{eqn:pred_var_0} or $\Sigma^2_f$ in Eqn. \ref{eqn:pred_var_1}). We refer the interested reader to the corresponding references for further details about UCE. The plots in the left panel of Fig. \ref{fig:cal_plots} exhibit the calibration plots and UCE obtained on four representative test sets. All results were based on a sample size of 25 (i.e., $T=25$ in Eqn. \ref{eqn:pred_mean}, \ref{eqn:pred_var_0}, and \ref{eqn:pred_var_1}) from 10 repeated runs. The blue lines show the results produced with the $\hat{\Sigma}^2_f$ and the shaded regions represent the standard deviation from the 10 runs, while the dashed black lines indicate the perfect calibration achieved with the proposed method. Notice that the uncertainty values obtained using $\hat{\Sigma}^2_f$ did not match well to the expected model error; in fact, they were consistently being underestimated (for reasons described in section \ref{sec:uncert_cal}). In comparison, the proposed method enabled perfect calibration with UCE $=0$ since its uncertainty estimate equaled the expected model error. In the right panel of Fig. \ref{fig:cal_plots}, we show the visual comparisons of the uncertainty derived from $\Sigma^2_f$ and $\hat{\Sigma}^2_f$. When we compare either (e) to (f) or (k) to (l), we see that the former (i.e., (e) and (k)) captured more registration failures than the latter (as highlighted by the yellow arrows), indicating a stronger correlation between deformation uncertainty and registration failures. This is thus further evidence that the proposed method provides the perfect uncertainty calibration.

Despite the promising results, there are some limitations of using $\sigma_f$ to estimate appearance uncertainty. In this work, we modeled $\sigma_f$ as $\mathbb{E}\left[(I_m\circ\phi-I_f)^2\right]$, which is the MSE of the Monte Carlo sampled registration outputs relative to the fixed image. MSE, on the other hand, is not necessarily the optimal metric for expressing the expected error. In multi-modal registration instances like PET to CT or MRI to CT registration, MSE is anticipated to be high, given the vast difference in image appearance and voxel values across modalities. Thus, if MSE is employed to quantify the appearance uncertainty in these instances, the uncertainty values will be dominated by the squared bias (i.e., $(\hat{I}_f-I_f)^2$ in Eqn. \ref{eqn:exp_err_var}), resulting in an ineffective uncertainty estimate. In these instances, the predicted variance may be a more appropriate choice for appearance uncertainty quantification.

Additional results for both appearance and transformation uncertainty estimations are shown in Fig. \ref{fig:jhu_brain_uncert} in Appendix. Observably, the two uncertainty measures provide estimates that are substantially different, with appearance uncertainty values being high in locations with substantial appearance mismatches and transformation uncertainty values being high in regions with large deformations and generally constant intensity values.

\begin{figure*}[!h]
\centering
\includegraphics[width=.98\textwidth]{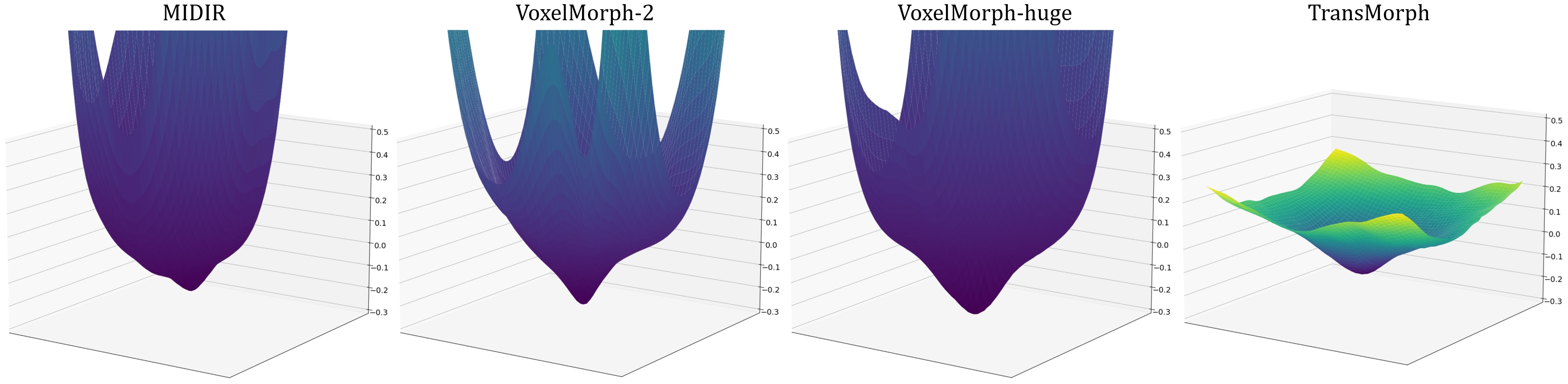}
\caption{The loss landscapes of \texttt{MIDIR}, \texttt{VoxelMorph}-2, \texttt{VoxelMorph}-huge, and \texttt{TransMorph}, where the loss function is composed of LNCC and diffusion regularizer. \texttt{TransMorph} yielded a much flatter landscape that those of ConvNet-based models. \label{fig:loss_landscape}}
\end{figure*}
\subsection{Comparison of Effective Receptive Fields}
\label{sec:comp_ERF}
We demonstrate in this section that the \textit{effective receptive fields} (ERFs) of Transformer-based models are larger than that of ConvNet-based models and spans the whole spatial domain of an image. We used the definition of ERF introduced in~(\cite{luo2016understanding}), which quantifies the amount of influence that each input voxel has on the output of a neural network. In the next paragraph, we briefly discuss the computation of ERF and recommend interested readers to the reference for further information.

Assume the voxels in the input image $I_m$ and the output displacement field $\mathbf{u}$ are indexed by $(i, j, k)$, with an image size of $160\times160\times160$ (i.e., the size of CT scans used in this work), the center voxel is located at $(80, 80, 80)$. ERF quantifies how much each $I_m(i,j,k)$ contributes to the center voxel of the displacement field, i.e. $\mathbf{u}(80,80,80)$. This is accomplished using the partial derivative $\partial\mathbf{u}(80,80,80)/\partial I_m(i,j,k)$, which indicates the relative relevance of $I_m(i,j,k)$ to $\mathbf{u}(80,80,80)$. To obtain this partial derivative, we set the error gradient to:
\begin{linenomath}
\begin{equation}
   \frac{\partial\ell}{\partial\mathbf{u}(i,j,k)} =
  \begin{cases}
    1, & \text{for } (i,j,k)=(80,80,80) \\
    0, & \text{otherwise}\\
  \end{cases},
\end{equation}
\end{linenomath}
where $\ell$ denotes an arbitrary loss function. Then this gradient is propagated downward from $\mathbf{u}$ to the input $I_m$, where the resulting gradient of $I_m$ represents the desired partial derivative $\partial\mathbf{u}(80,80,80)/\partial I_m(i,j,k)$. This partial derivative is independent of the input and loss function and is only a function of the network architecture and the index $(i,j,k)$, which adequately describes the distribution of the effective receptive field.

A comparison of the ERFs of \texttt{VoxelMorph} and \texttt{TransMorph} is shown in Fig. \ref{fig:ERF_example}. Note that the other ConvNet-based models were omitted because they adopted a similar network architecture as \texttt{VoxelMorph} (e.g., CycleMorph and MIDIR). Due to the locality of convolution operations, \texttt{VoxelMorph}'s ERF at each stage (top row in Fig. \ref{fig:ERF_example}) was highly localized, particularly in the encoding stages (i.e., 1/4, 1/8, and 1/16 resolution). Even at the end of the network, the theoretical receptive field of \texttt{VoxelMorph} encompassed the entire image; yet, its ERF emphasized only a small portion of the image. In contrast, the ERFs of the proposed \texttt{TransMorph} were substantially larger than those of \texttt{VoxelMorph} at each stage, and the ERFs in the decoding stage covered the entire image (bottom row in Fig. \ref{fig:ERF_example}). The ERFs reveal that ConvNet-based architectures can only perceive a portion of the input image, particularly during the encoding stages, indicating that they cannot explicitly comprehend the spatial relationships between distant voxels. For tasks that require large deformations, ConvNets may fall short of establishing accurate voxel correspondences between the moving and fixed images, which is essential for image registration. On the other hand, \texttt{TransMorph} adopts substantially large kernels at the encoding stages leading to substantially large ERFs throughout the network thanks to the self-attention mechanism of the Transformer.

\begin{figure*}[!h]
\centering
\includegraphics[width=.9\textwidth]{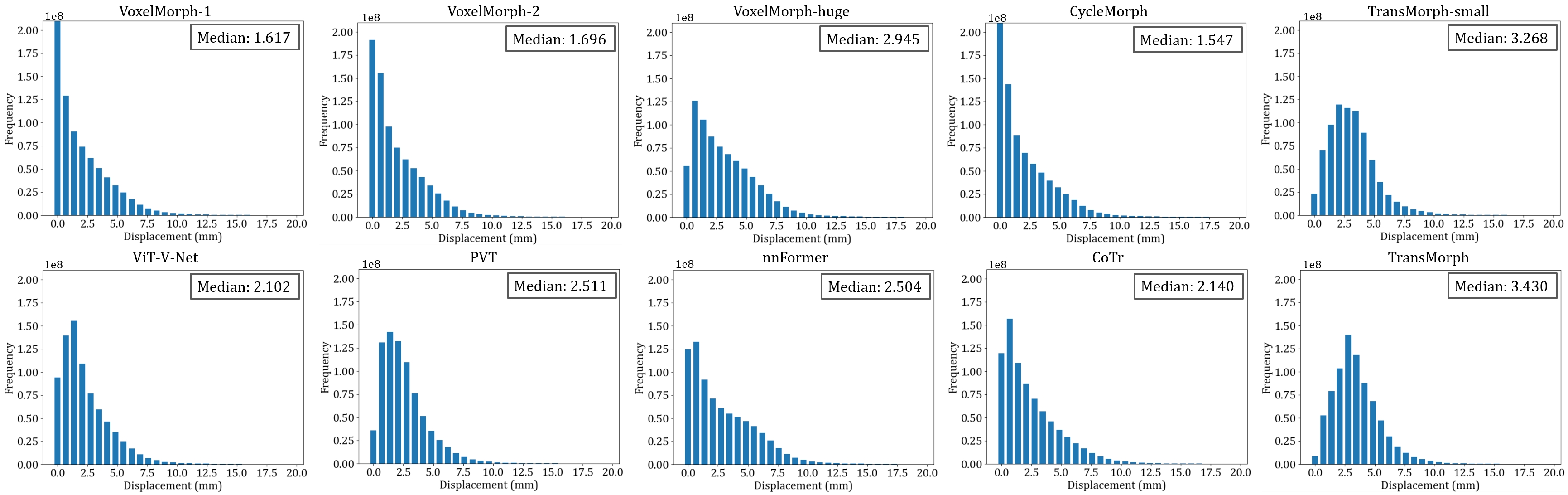}
\caption{Histograms of the displacement magnitudes in millimeters. These histograms were generated using 115 test volumes from the IXI dataset. The displacement magnitude is computed as $\sqrt{d_x^2+d_y^2+d_z^2}$, where $d_\{x,y,z\}$ denotes the displacement in $x$, $y$, and $z$ directions. The median displacement magnitude is shown in the upper right corner of each plot. To provide fair comparisons, only models that produce dense displacements are shown here. \texttt{VoxelMorph} and \texttt{CycleMorph} are ConvNet-based models, whereas the other models are Transformer-based. \label{fig:disp_hist}}
\end{figure*}

\begin{figure*}[!h]
\centering
\includegraphics[width=.9\textwidth]{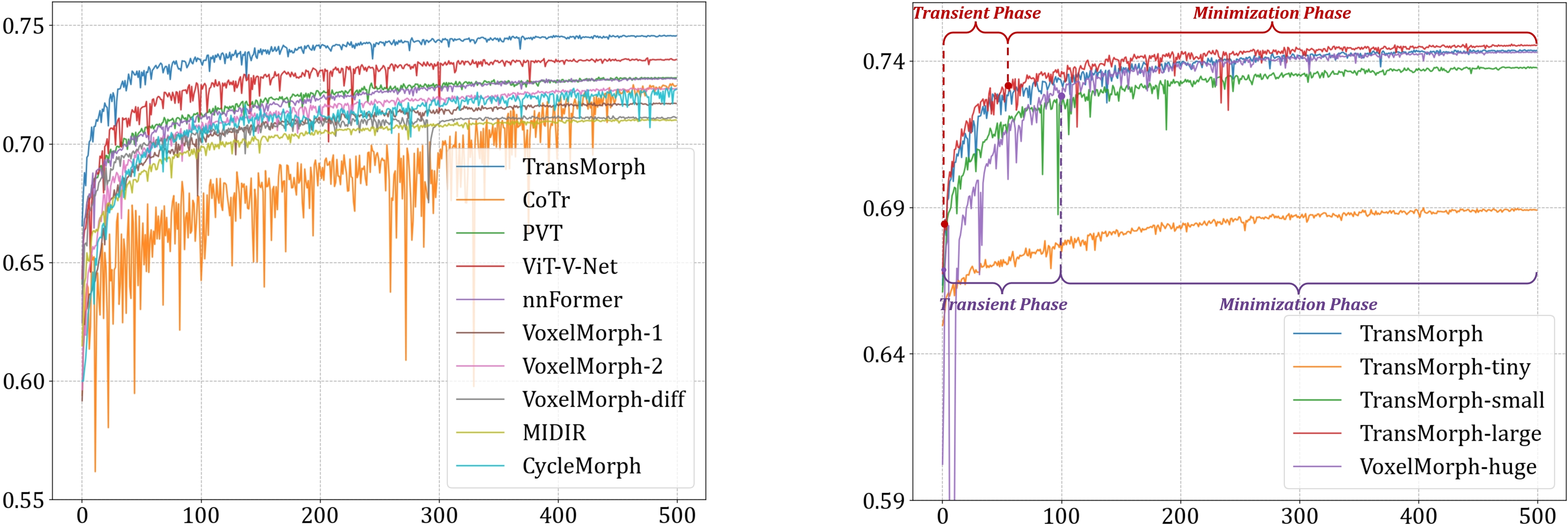}
\caption{Validation Dice scores for inter-patient brain MRI registration during training. The validation dataset comprises 104 image pairings that were not included in the training or testing set.\label{fig:convergence}}
\end{figure*}

\subsection{Comparison of Displacement Magnitudes}
As demonstrated in section \ref{sec:comp_ERF}, \texttt{TransMorph} had substantially larger effective receptive fields than \texttt{VoxelMorph}, which might be beneficial for capturing semantic information that is necessary for coping with large deformations (\cite{ha2020semantically}). In this section, we provide more evidence that Transformer-based models are more capable of producing larger deformations. We used 115 test volumes from the IXI dataset to generate histograms of displacement magnitudes in millimeters. Fig. \ref{fig:disp_hist} shows histograms of the displacement magnitudes for the various methods. The models that produced dense displacement fields are shown for fair comparisons.  Note that \texttt{VoxelMorph} and \texttt{CylceMorph} are ConvNet-based models, whereas the other models are Transformer-based. All models were trained under the identical setting (e.g., loss functions, number of epochs, optimizers, etc.), where the only variable was the network architecture. As indicated by the histograms, all Transformer-based models had much more larger displacements than ConvNet-based models. The displacement distributions of ConvNet-based models had a mode near 0 and had more smaller displacements. We additionally showed the histograms of \texttt{VoxelMorph}-huge and \texttt{TransMorph}-small, the former of which had 63.25M parameters and the latter of which had 11.76M parameters. Despite having around 6$\times$ more parameters, \texttt{VoxelMorph}-huge still exhibited smaller displacements than \texttt{TransMorph}-small. This further indicates that the larger displacements produced by \texttt{TransMorph} were not a consequence of an increase in the number of parameters but rather the network architecture. Given the above-demonstrated improved registration performance of the Transformer-based models, these histograms indicate that in cases where larger displacements are required, the Transformer-based models will likely provide better registration.

\subsection{Comparison of Loss Landscapes}
In this section, the loss landscapes of \texttt{TransMorph} and ConvNet-based models are compared. We adopted the loss landscape visualization method described in (\cite{li2018visualizing, goodfellow2014qualitatively, im2016empirical}), in which a set of pre-trained model parameters (denoted as $\theta$) are perturbed in two random directions (denoted as $\delta$ and $\eta$) with step sizes of $\alpha$ and $\beta$ to acquire loss values at different locations. The loss landscape was plotted based on the function of the form:
\begin{linenomath}
\begin{equation}
    f(\alpha, \beta) = \mathcal{L}(\theta+\alpha\delta+\beta\eta),
\end{equation}
\end{linenomath}
where $\mathcal{L}$ denotes the loss function made up of LNCC and diffusion regularizer. We averaged the loss landscapes of ten samples from the validation set of the atlas-to-patient registration task to obtain the final 3D contour plot for each model. For comparison between ConvNet-based models and \texttt{TransMorph}, the loss landscapes of \texttt{VoxelMorph}, \texttt{MIDIR}, and \texttt{TransMorph} were created as shown in Fig. \ref{fig:loss_landscape}. \texttt{TransMorph} produced a substantially flatter loss landscape than that of the ConvNet-based models. This observation is consistent with the findings given in (\cite{park2022vision}), which suggest that Transformers tend to promote flatter loss landscapes. Many studies have demonstrated that a flatter landscape results in improved performance and better generalizability (\cite{park2022vision, keskar2016large, santurkar2018does, foret2020sharpness, li2018visualizing}). The flatter landscape of TransMorph further demonstrates the advantages of Transformer-based models for image registration.

\subsection{Convergence and Speed}
The left panel of Fig. \ref{fig:convergence} shows the validation dice scores of the learning-based methods during training. In comparison to other methods, the proposed \texttt{TransMorph} achieved $>0.7$ in Dice within the first 20 epochs, showing that it learned the spatial correspondence between image pairs quicker than the competing models. Notably, \texttt{TransMorph} consistently outperformed the other Transformer-based models while having a comparable number of parameters and computational complexity. This implied Swin Transformer architecture was more effective than other Transformers, resulting in a performance improvement for \texttt{TransMorph}. On average, Transformer-based models provided better validation scores than ConvNet-based models, with the exception of \texttt{CoTr}, whose validation results were volatile during training (as seen from the orange curve in Fig. \ref{fig:convergence}). The performance of \texttt{CoTr} may be limited by its architecture design, which substitutes a Transformer for the skip connections and bottleneck of a U-shaped CovnNet. As a result, it lacks the direct flow of features learned during the encoding stage to the layers creating the registration, making it difficult to converge. The right panel of Fig. \ref{fig:convergence} shows the training curves of the \texttt{TransMorph} variants and the customized \texttt{VoxelMorph}-huge. As described in (\cite{im2016empirical, sutskever2013importance, darken1991towards}), the training curve of a deep learning model consists of two phases: a ``transient" phase followed by a ``minimization" phase, where the former identifies the neighborhood of local minima and the latter seeks the local minima inside that neighborhood. As seen in the figure, \texttt{TransMorph} variants had shorter ``transient" phases than that of \texttt{VoxelMorph}-huge, indicating that they identified the local minima neighborhood more quickly. A fast convergent algorithm is often preferred since it not only saves time but also computing resources and costs. There have been many efforts to accelerate the convergence rate of deep learning models (\cite{darken1991towards, looney1996stabilization, zeiler2013rectified, smith2019super}). \texttt{TransMorph} tends to accelerate convergence rate compared to ConvNet-based models, which promotes its potential of faster training using fewer epochs, saving time and reducing the carbon footprint.

\begin{table}[!t]
\centering
\fontsize{8}{9.5}\selectfont
\begin{tabular}{ c || c | c}
 \hline
 Model & Training (min/epoch) & Inference (sec/image)\\
 \hline
 \hline
 \texttt{SyN} & - & 192.140\\
 \hline
 \texttt{NiftyReg} & - &30.723\\
 \hline
 \texttt{LDDMM} & - & 66.829\\
 \hline
 \texttt{deedsBCV} & - & 31.857\\
 \hline
 \texttt{VoxelMorph}-1 & 8.75 & 0.380\\
 \hline
 \texttt{VoxelMorph}-2 & 9.40 & 0.430\\
 \hline
 \texttt{VoxelMorph}-diff & 4.20 & 0.049\\
 \hline
 \texttt{VoxelMorph}-huge & 28.50 & 1.107\\
 \hline
 \texttt{CycleMorph} & 41.90 & 0.281\\
 \hline
 \texttt{MIDIR} & 4.05 & 1.627\\
 \hline
 \texttt{ViT-V-Net} & 9.20 & 0.197 \\
 \hline
 \texttt{PVT} & 13.80 & 0.209\\
 \hline
 \texttt{CoTr} & 17.10 & 0.372\\
  \hline
 \texttt{nnFormer} & 6.35 & 0.105\\
 \hline
 \hline
 \texttt{TransMorph-Bayes} & 22.60 & 7.739\\
 \hline
 \texttt{TransMorph-diff} & 7.35 & 0.099\\
 \hline
 \texttt{TransMorph-bspl} & 10.50 & 1.739\\
 \hline
 \texttt{TransMorph} & 14.40 & 0.329\\
 \hline
\end{tabular}
\caption{Average training and inference time for methods used in this work. Note that \texttt{SyN}, \texttt{NiftyReg}, and \texttt{deedsBCV} were applied using CPUs, while \texttt{LDDMM} and the learning-based methods were implemented on GPU. Inference time was averaged based on 40 repeated runs.}\label{tab:table_time}
\end{table}

Table \ref{tab:table_time} compares the training time in min per epoch (min/epoch) and inference time in seconds per image (sec/image) among the methods used in this paper. Note that  \texttt{SyN}, \texttt{NiftyReg}, and \texttt{deedsBCV} packages are all CPU-based, while \texttt{LDDMM} and the deep-learning-based methods are all GPU-based. The speed was calculated using an input image size of $160\times192\times224$, which corresponds to the size of the brain MRI scans. The training time per epoch was computed based on 768 training image pairs. The most and second most time-consuming methods to train are two ConvNet-based methods, \texttt{CycleMorph} and the customized \texttt{VoxelMorph}-huge, which required approximately $(41.90\text{min}\times500)/(60\text{min}\times24 \text{hr})\approx15$ days and $(28.50\text{min}\times500)/(60\text{min}\times24 \text{hr})\approx10$ days for 500 epochs of training, respectively. \texttt{CycleMorph} was time-consuming because the cycle-consistent training virtually trains four networks simultaneously in a single epoch. Whereas the training of \texttt{VoxelMorph}-huge was slowed down by the extensive convolution operations. The proposed \texttt{TransMorph} has a moderate training speed, roughly $1.5\times$ that of \texttt{VoxelMorph}-2 but $0.5\times$ that of the customized \texttt{VoxelMorph}-huge. In terms of inference time, learning-based models undoubtedly operated orders of magnitudes faster than traditional registration methods. Note that \texttt{TransMorph} is about $3\times$ faster than \texttt{VoxelMorph}-huge during inference. These finds are proportional to the calculated computational complexity as shown in the barplot on the left in Fig. \ref{fig:MACs_compare}. Among the learning-based models, \texttt{TransMorph-Bayes} required the highest inference time. However, the time required is due to the sampling of $T=25$ images for a single prediction and uncertainty estimation.

\subsection{Limitations}
There are some limitations to our work. First, rather than doing extensive grid searches for optimal hyperparameters for the baseline methods, the hyperparameters are either determined empirically or based on the values suggested in the original paper. Due to the time required to train some of the baseline methods and the limited memory available on the GPU, we were unable to afford the intensive grid search. Moreover, because this study introduced a generic network architecture for image registration, we concentrated on architectural comparison rather than on selecting optimal hyperparameters for loss functions or complex training methods. However, the proposed \texttt{TransMorph} architecture is readily adaptable using either the cycle-consistent training method used by \texttt{CycleMorph}~(\cite{kim2021cyclemorph}) or the symmetric training method proposed in~(\cite{mok2020fast}). Additionally, the proposed network may be used in conjunction with any registration loss function. 


In the future, we will investigate alternative loss functions, such as mutual information, in an effort to expand the potential of the proposed method for multi-modal registration tasks.

\section{Conclusion}
\label{sec:conclusion}
In this paper, we introduced \texttt{TransMorph}, a novel model for unsupervised deformable image registration. \texttt{TransMorph} is built on Transformer, which is well-known for its capability to establish long-range spatial correspondence between image voxels, making \texttt{TransMorph} a strong candidate for image registration tasks.

Two variants of \texttt{TransMorph} are proposed, which provide topology-preserved deformations. Additionally, we introduced Bayesian deep learning to the Transformer encoder of \texttt{TransMorph}, enabling deformation uncertainty estimation without degrading registration performance.

We evaluated \texttt{TransMorph} on the task of inter-patient brain MR registration and a novel task of phantom-to-CT registration. The results revealed that \texttt{TransMorph} achieved superior registration accuracy than various traditional and learning-based methods, demonstrating its effectiveness for medical image registration.

\section*{Declaration of Competing Interest}
The authors declare that they have no competing interests.

This manuscript has not been submitted to, nor is under review
at, another journal or other publishing venue.

\section*{CRediT authorship contribution statement}
\textbf{Junyu Chen:} Conceptualization, Methodology, Software, Data curation, Investigation, Writing - original draft, Visualization. \textbf{Eric C. Frey}: Validation, Resources, Writing - Review and Editing, Supervision, Funding acquisition. \textbf{Yufan He:} Methodology, Validation, Investigation, Writing - Review and Editing. \textbf{William P. Segars}: Data curation. \textbf{Ye Li:} Validation. \textbf{Yong Du:} Validation, Resources, Data curation, Writing - Review and Editing, Supervision, Funding acquisition.

\section*{Acknowledgments}
This work was supported by grants from the National Institutes of Health, U01-CA140204, R01EB031023, and U01EB031798. The views expressed in written conference materials or publications and by speakers and moderators do not necessarily reflect the official policies of the NIH; nor does mention by trade names, commercial practices, or organizations imply endorsement by the U.S. Government.
\onecolumn
\appendix
\section*{Appendix}
\section{Affine Network Architecture}
\begin{figure*}[!h]
\centering
\includegraphics[width=.7\textwidth]{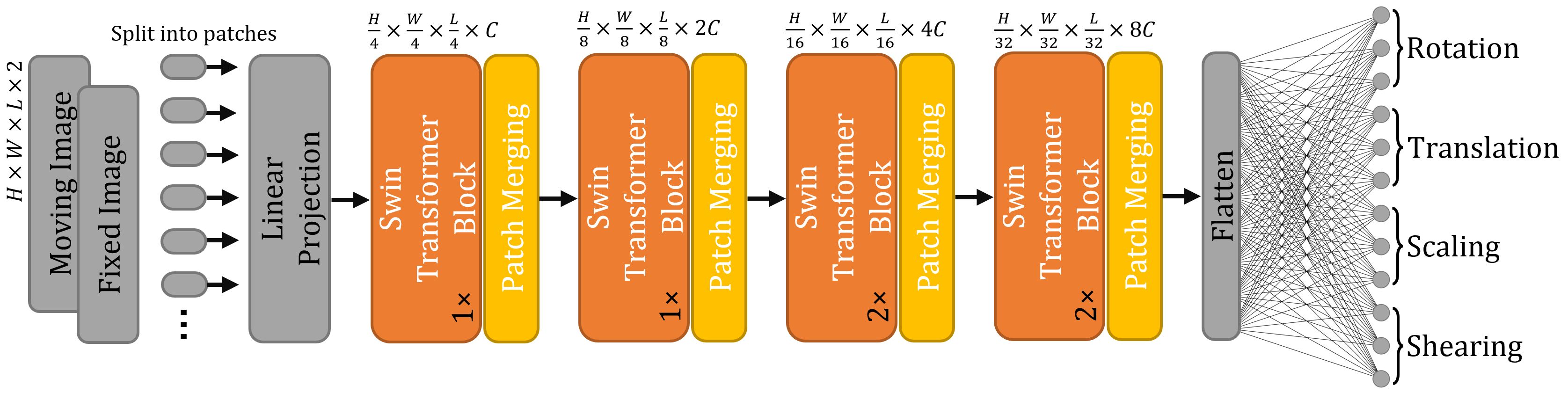}
\caption{Visualization of the proposed Swin-Transformer-based affine network. This network outputs three rotation, three translation, three scaling, and three shearing parameters for rigid registration. The embedding dimension $C$ in the network was set to 12. \label{fig:affine_arch}}
\end{figure*}
\section{Miscalibration in Predictive Variance}
\label{app:mis_cal}
The expected model error (characterized by MSE) is defined as:
\begin{linenomath}
\begin{equation}
\label{eqn:exp_err}
    err(I_m\circ\phi) = \mathbb{E}\left[(I_m\circ\phi-I_f)^2\right]=\frac{1}{T}\sum^T_{t=1}\left(I_m\circ\phi_t - I_f\right)^2,
\end{equation}
\end{linenomath}
where $t$ represents the $t^{th}$ sample from a total number of $T$ samples. We denote $I_d=I_m\circ\phi$ for convenience, and it can be shown that:
\begin{linenomath}
\begin{equation}
    \begin{split}
        \mathbb{E}\left[(I_d-I_f)^2\right]&=\mathbb{E}\left[(I_d-\mathbb{E}[I_d]+\mathbb{E}[I_d]-I_f)^2\right]\\
        &=\mathbb{E}\left[\left(I_d-\mathbb{E}[I_d]\right)^2\right]+\left(\mathbb{E}[I_d]-I_f\right)^2+ 2(\mathbb{E}[I_d]-I_f)\mathbb{E}\left[I_d-\mathbb{E}[I_d]\right]\\
        &=\mathbb{E}\left[\left(I_d-\mathbb{E}[I_d]\right)^2\right]+\left(\mathbb{E}[I_d]-I_f\right)^2.
    \end{split}
\end{equation}
\end{linenomath}
Therefore,
\begin{linenomath}
\begin{equation}
\label{eqn:exp_err_var}
\begin{split}
     err(I_m\circ\phi)&=\frac{1}{T}\sum^T_{t=1}\left(I_m\circ\phi_t - I_f\right)^2\\
     &=\frac{1}{T}\sum^T_{t=1}\left(I_m\circ\phi_t - \frac{1}{T}\sum^T_{t=1}I_m\circ\phi_t\right)^2+\left(\frac{1}{T}\sum^T_{t=1}I_m\circ\phi_t-I_f\right)^2\\
     &=\hat{\Sigma}^2_f+\left(\hat{I}_f-I_f\right)^2,\\
\end{split}
\end{equation}
\end{linenomath}
where $\hat{I}_f-I_f$ is referred to as the bias between the predictive mean $\hat{I}_f$ and the target image $I_f$. Due to the problem of overfitting the training set in supervised algorithms (e.g., deep learning)~(\cite{bishop2006pattern}), this bias may be less noticeable on training dataset but more noticeable on test images, which is a phenomenon referred to as the \textit{bias-variance tradeoff}~(\cite{friedman2017elements}). As a consequence, the predictive variance $\hat{\Sigma}^2_f$ is systematically smaller than the expected error $err(I_m\circ\phi)$, resulting in miscalibrated uncertainty estimations.
\newpage
\section{Additional Results for Inter-patient Brain MRI Registration}
\label{sec:add_inter_MR}
\begin{figure*}[!htp]
\centering
\includegraphics[width=.98\textwidth]{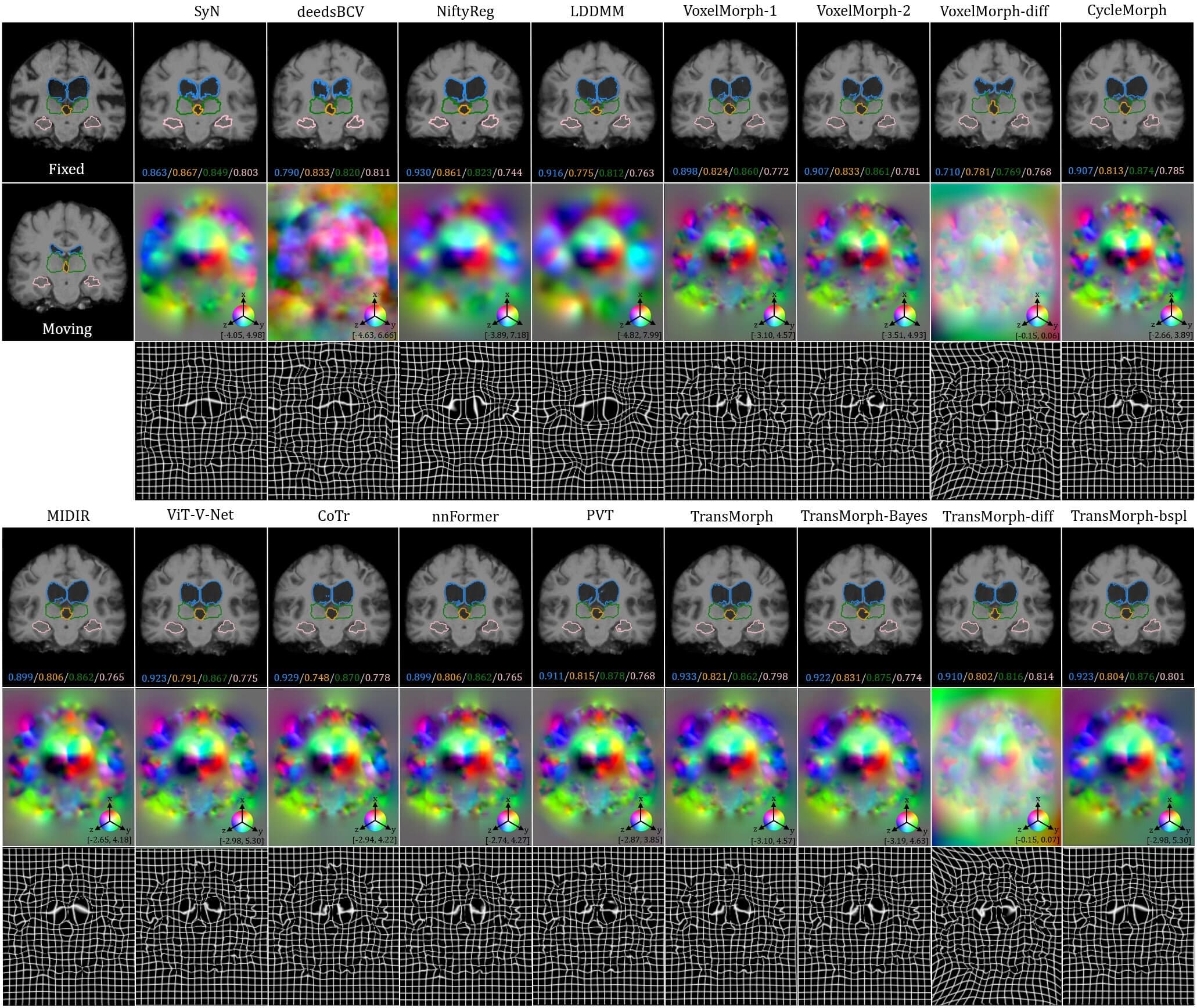}
\caption{Additional qualitative comparison of various registration methods on the inter-patient brain MR registration task. The first row shows the deformed moving images, the second row shows the deformation fields, and the last row shows the deformed grids. The spatial dimension $x$, $y$, and $z$ in the displacement field is mapped to each of the RGB color channels, respectively. The [$p$, $q$] in color bars denotes the magnitude range of the fields.\label{fig:jhu_brain}}
\end{figure*}

\begin{figure*}[!htp]
\centering
\includegraphics[width=.98\textwidth]{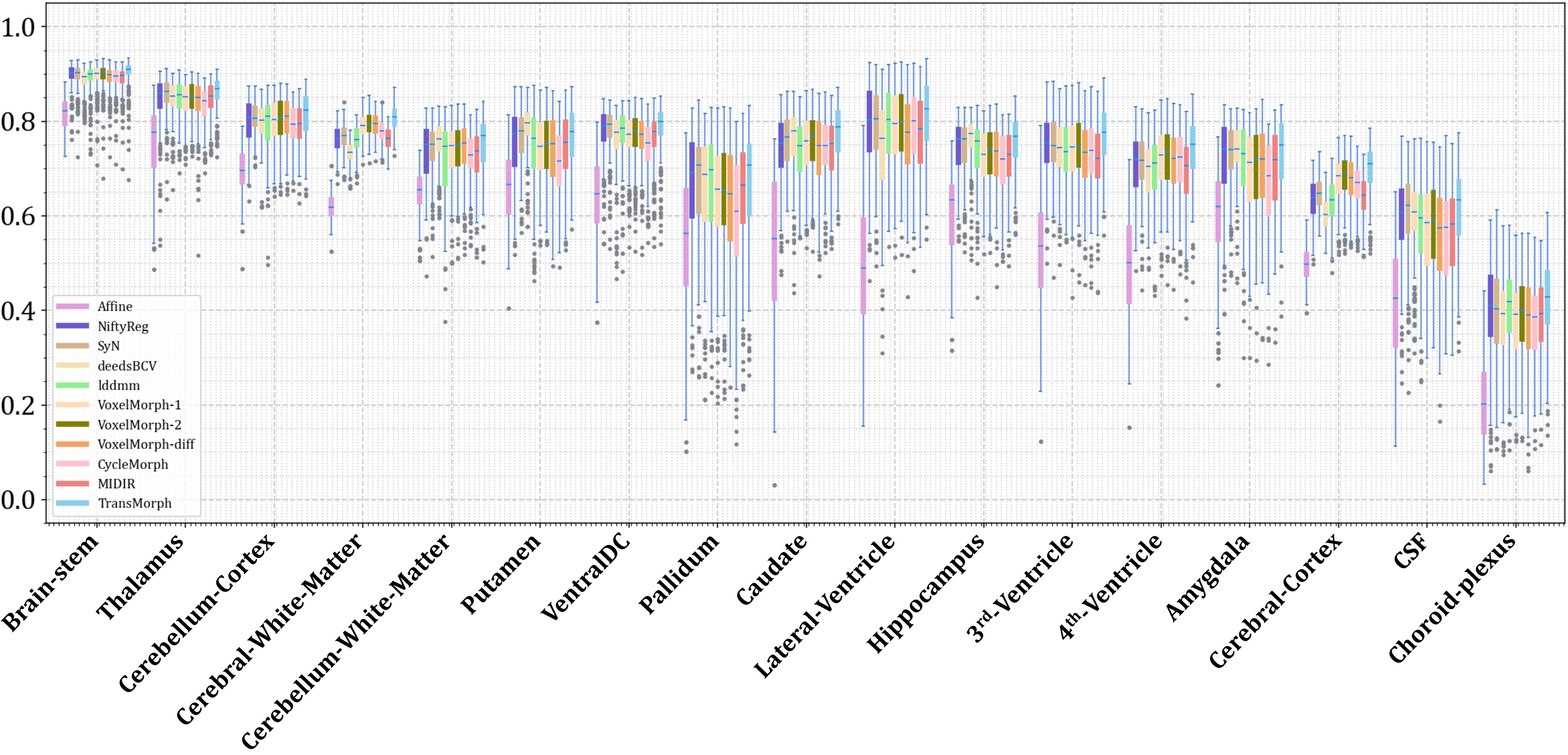}
\caption{Quantitative comparison of the various registration methods on the inter-patient brain MR registration task. Boxplots showing Dice scores for different brain MR substructures using the proposed \texttt{TransMorph} and existing image registration methods. \label{fig:jhu_brain_conventional}}
\end{figure*}

\begin{figure*}[!htp]
\centering
\includegraphics[width=.98\textwidth]{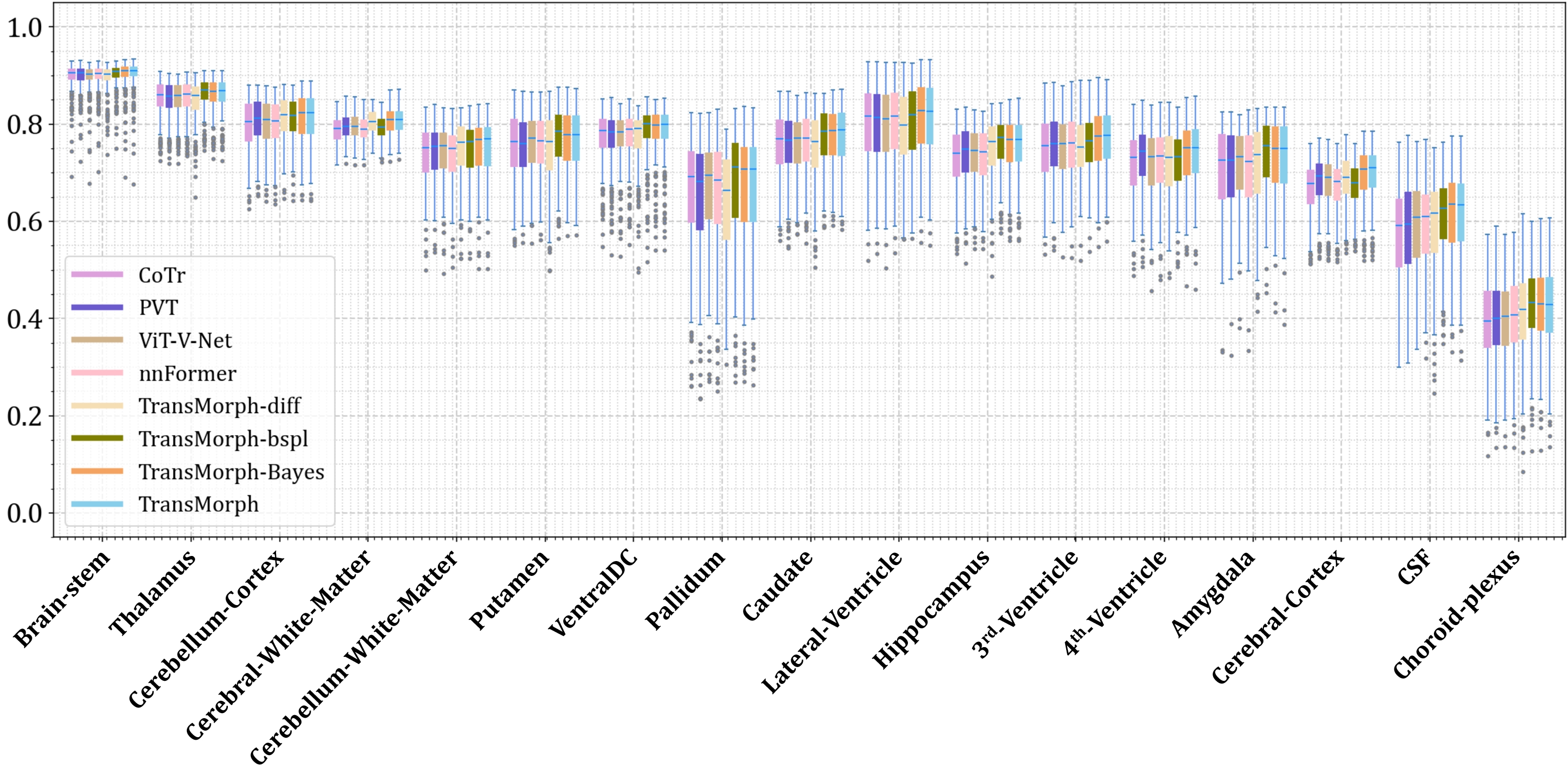}
\caption{Quantitative comparison of the Transformer-based models on the inter-patient brain MR registration task. Boxplots showing Dice scores for different brain MR substructures using the proposed \texttt{TransMorph}, the variants of \texttt{TransMorph}, and other Transformer architectures.\label{fig:jhu_brain_transformer}}
\end{figure*}

\newpage
\section{Additional Results for Atlas-to-patient Brain MRI Registration}
\label{sec:add_atlas_MR}
\begin{figure*}[!htp]
\centering
\includegraphics[width=.98\textwidth]{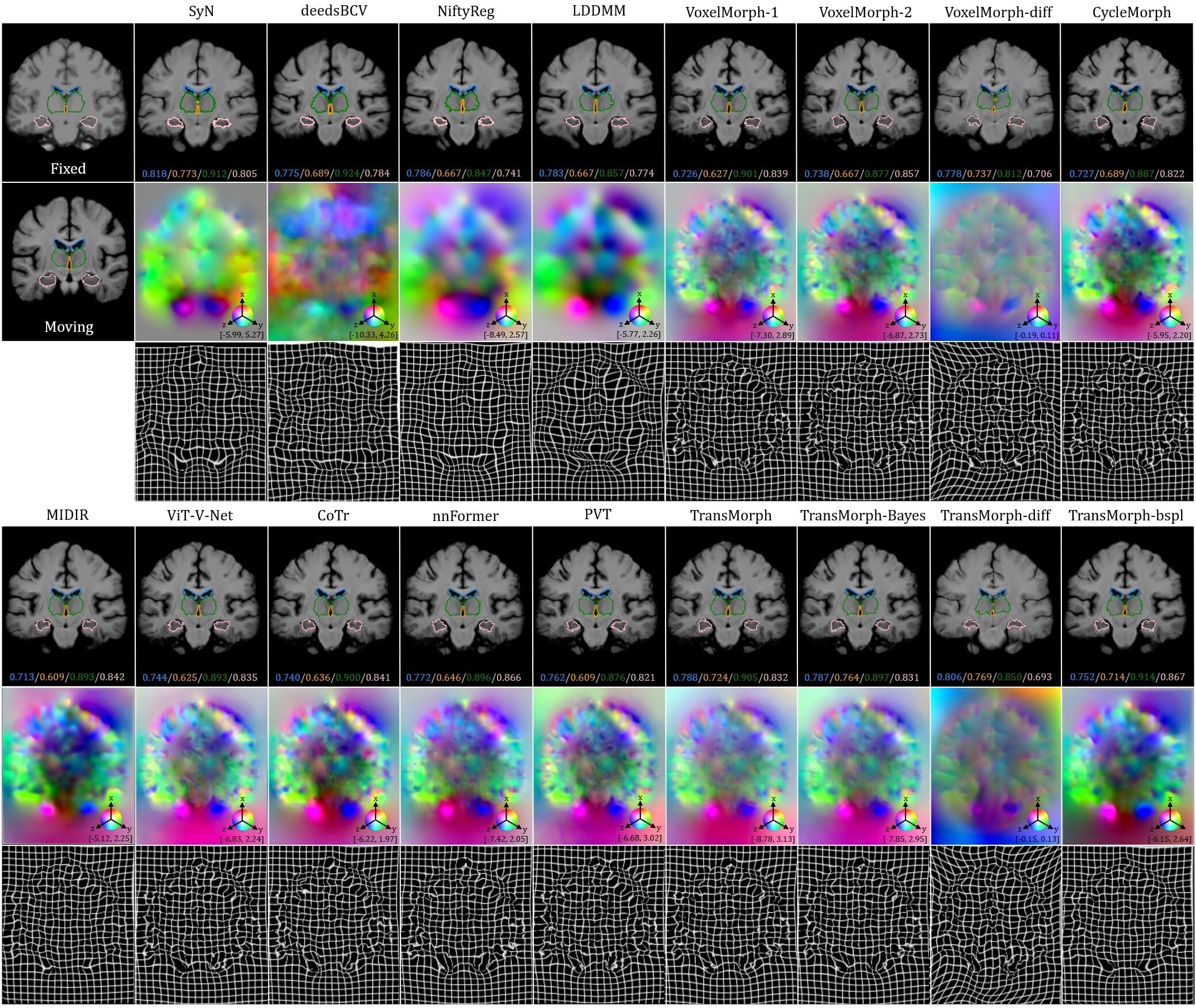}
\caption{Additional qualitative comparison of various registration methods on the atlas-to-patient brain MR registration task. The first row shows the deformed moving images, the second row shows the deformation fields, and the last row shows the deformed grids. The spatial dimension $x$, $y$, and $z$ in the displacement field is mapped to each of the RGB color channels, respectively. The [$p$, $q$] in color bars denotes the magnitude range of the fields.\label{fig:IXI_brain}}
\end{figure*}

\newpage
\begin{figure*}[!htp]
\centering
\includegraphics[width=.98\textwidth]{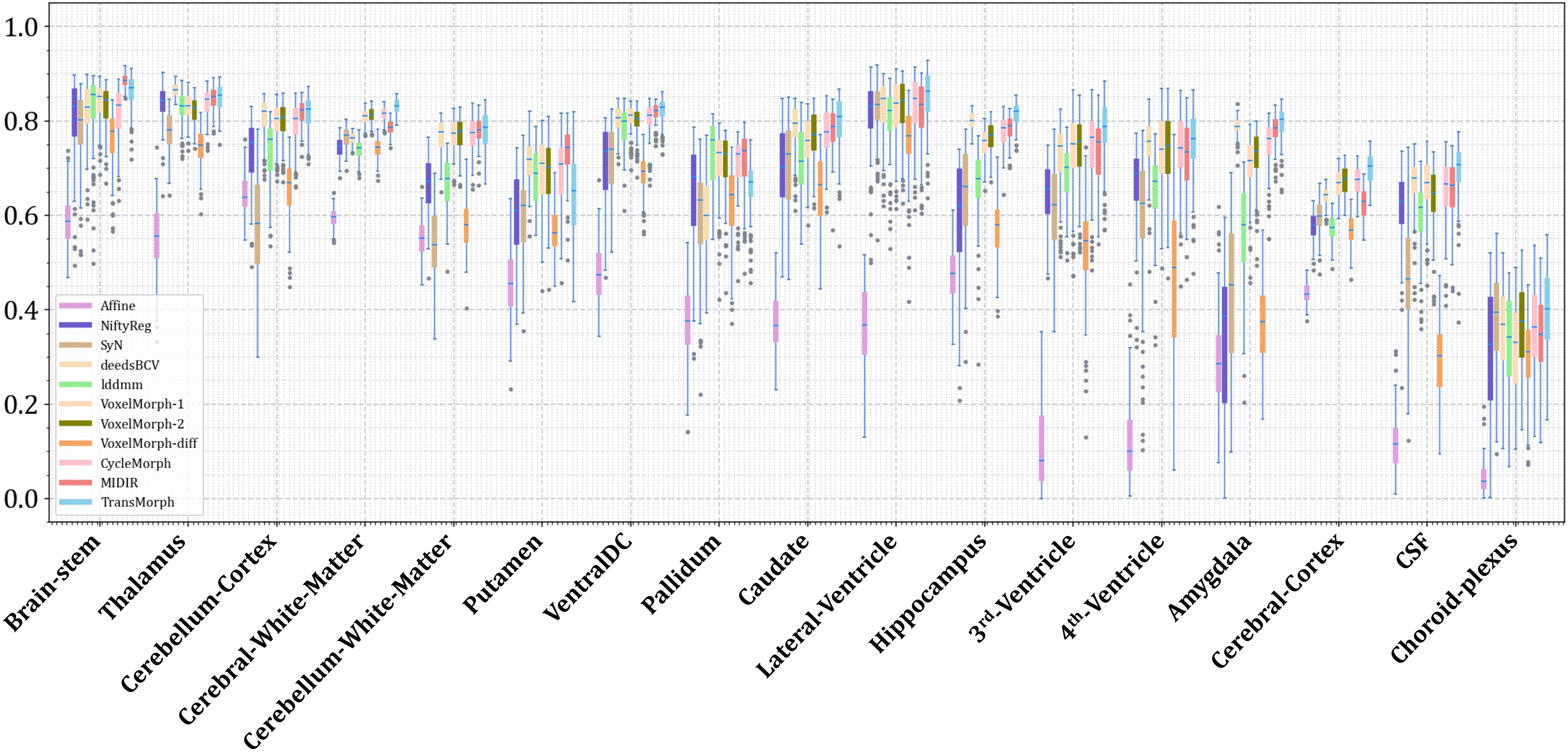}
\caption{Quantitative comparison of the various registration methods on the atlas-to-patient brain MR registration task. Boxplots showing Dice scores for different brain MR substructures using the proposed \texttt{TransMorph} and existing image registration methods. \label{fig:IXI_brain_conventional}}
\end{figure*}

\begin{figure*}[!htp]
\centering
\includegraphics[width=.98\textwidth]{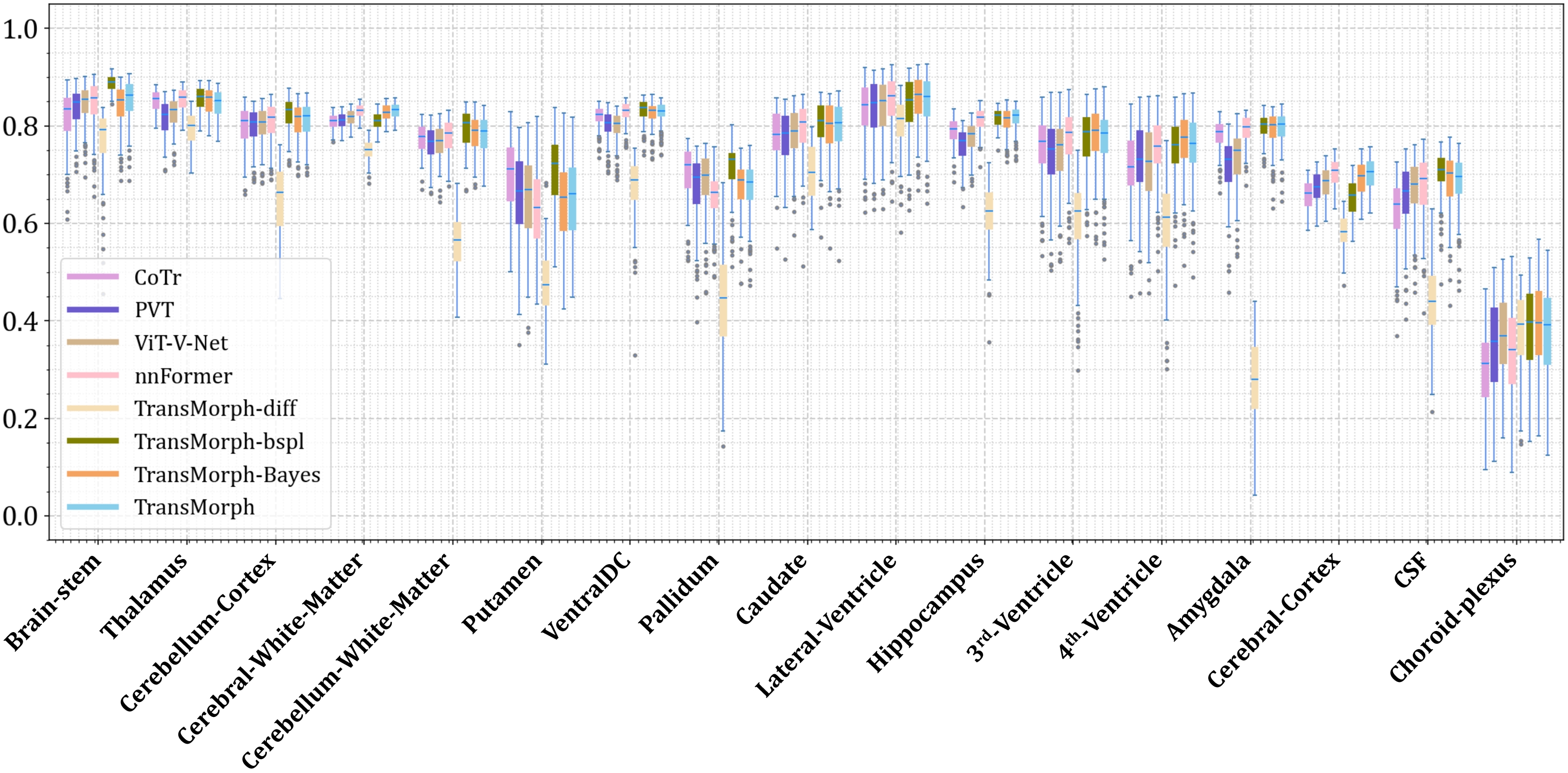}
\caption{Quantitative comparison of the Transformer-based models on the atlas-to-patient brain MR registration task. Boxplots showing Dice scores for different brain MR substructures using the proposed \texttt{TransMorph}, the variants of \texttt{TransMorph}, and other Transformer architectures.\label{fig:IXI_brain_transformer}}
\end{figure*}

\newpage
\section{Additional Results for XCAT-to-CT Registration}
\label{sec:add_xcat_ct}
\begin{figure*}[!htp]
\centering
\includegraphics[width=.98\textwidth]{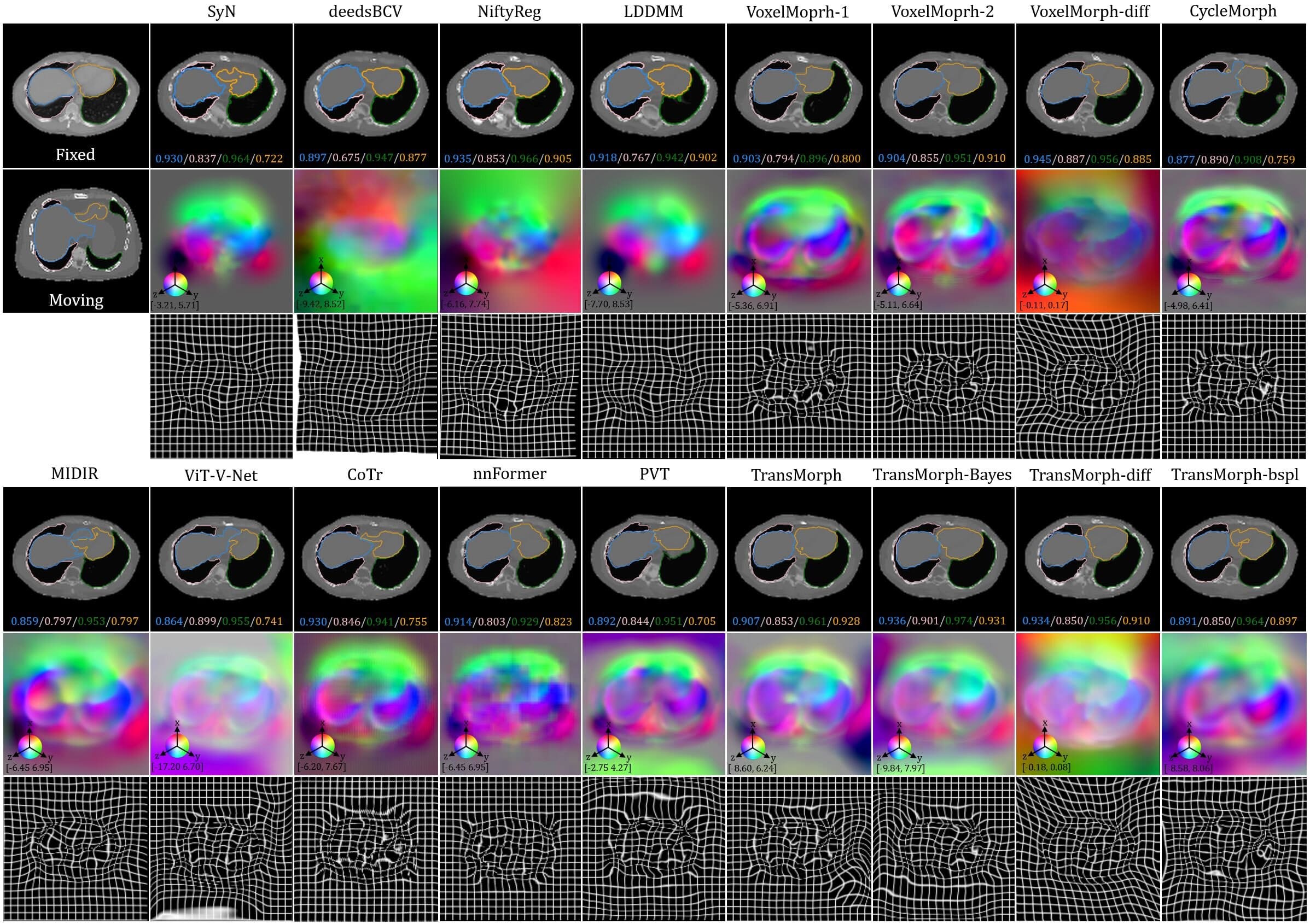}
\caption{Additional qualitative comparison of various registration methods on the XCAT-to-CT registration task. The first row shows the deformed moving images, the second row shows the deformation fields, and the last row shows the deformed grids. The spatial dimension $x$, $y$, and $z$ in the displacement field is mapped to each of the RGB color channels, respectively. The [$p$, $q$] in color bars denotes the magnitude range of the fields.\label{fig:xcat_CT}}
\end{figure*}

\newpage
\begin{figure*}[!htp]
\centering
\includegraphics[width=.98\textwidth]{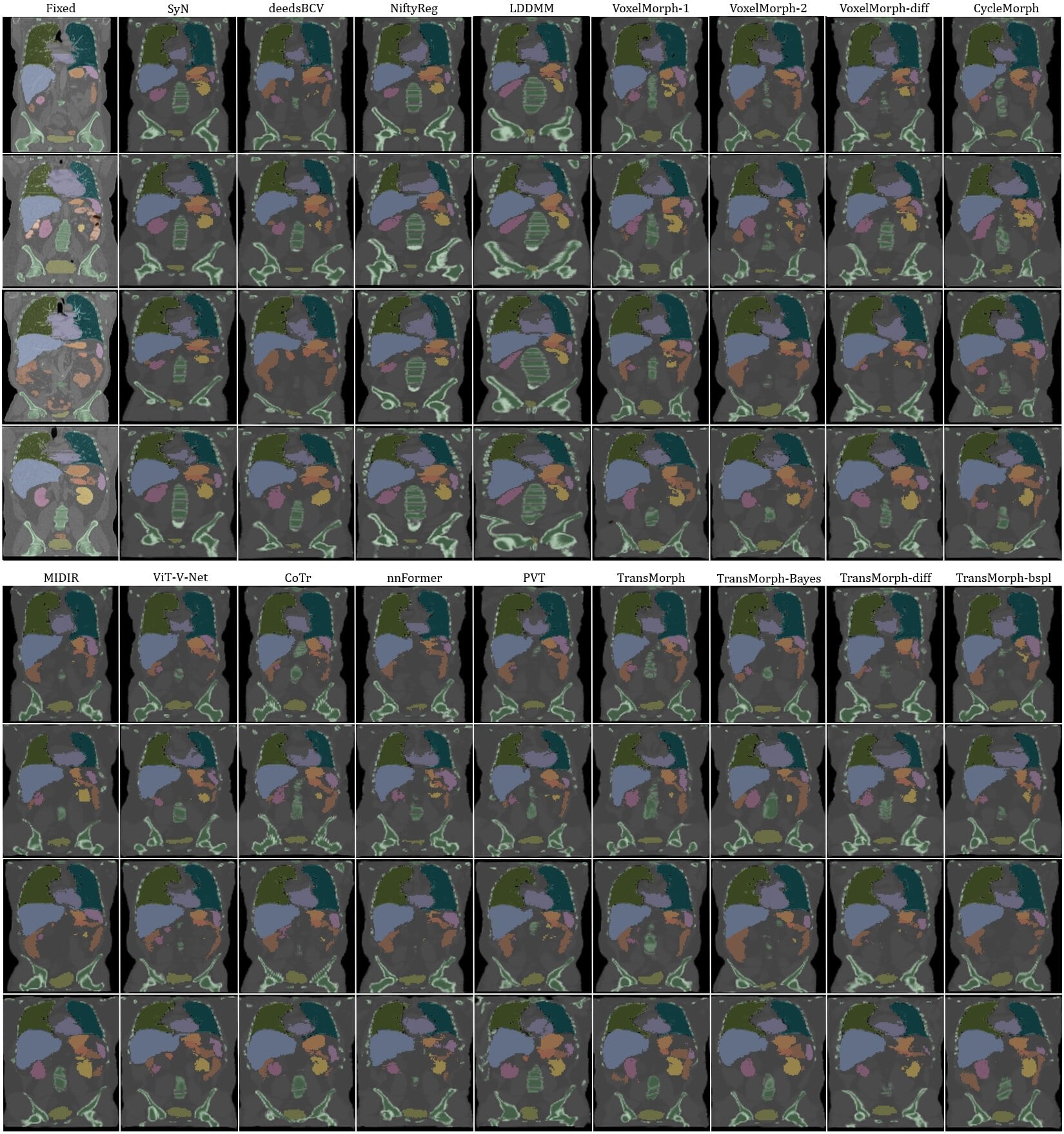}
\caption{Additional coronal slices of the deformed XCAT phantom generated by various registration methods. \label{fig:xcat_deform}}
\end{figure*}

\begin{figure*}[!htp]
\centering
\includegraphics[width=.98\textwidth]{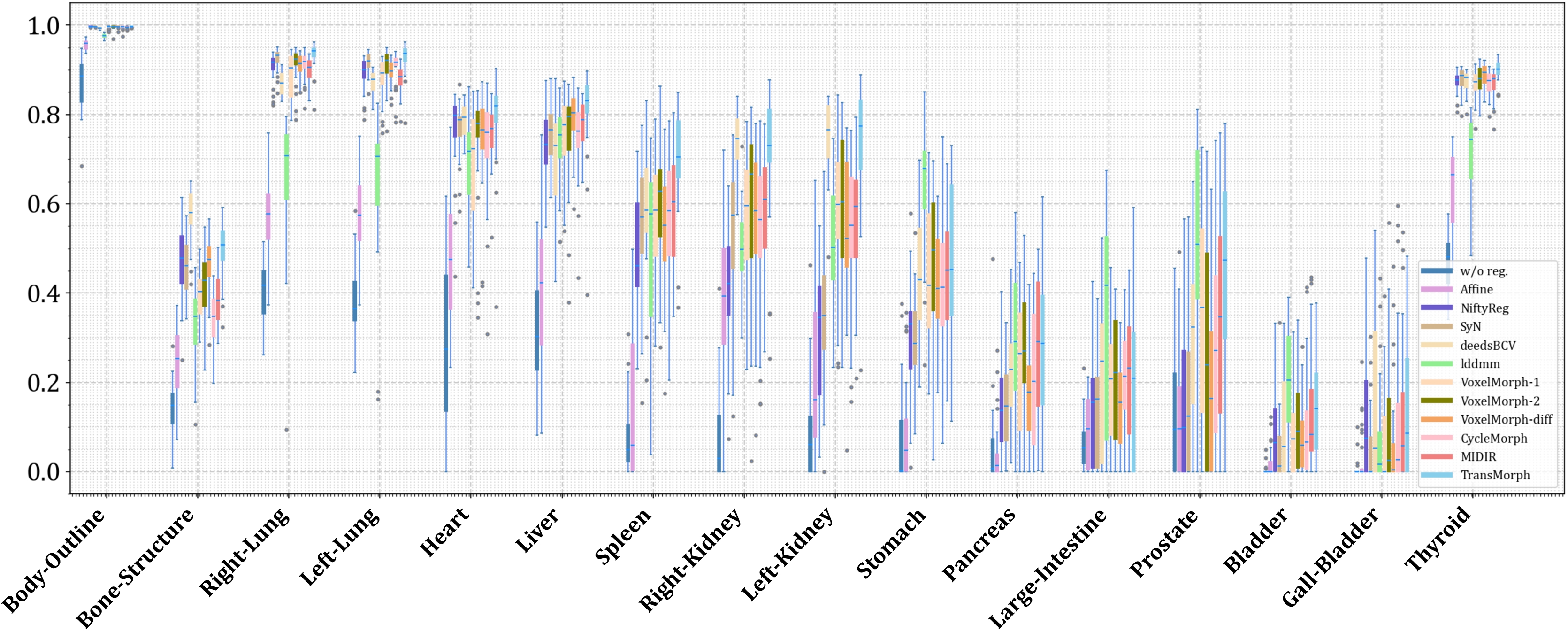}
\caption{Quantitative comparison of various registration methods on the XCAT-to-CT registration task. Boxplots showing Dice scores for different organs in CT obtained using the proposed \texttt{TransMorph} and existing image registration methods.\label{fig:xcat_brain_conventional}}
\end{figure*}

\begin{figure*}[!htp]
\centering
\includegraphics[width=.98\textwidth]{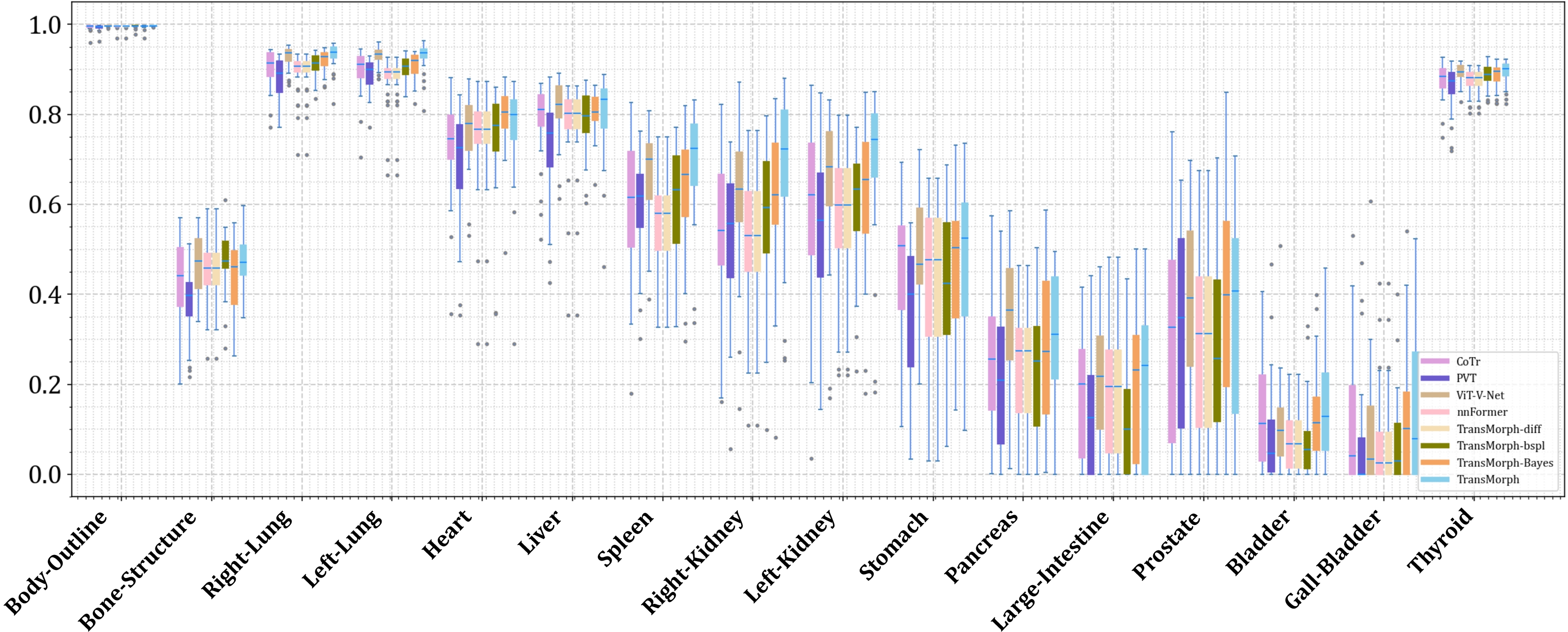}
\caption{Quantitative comparison of the Transformer-based models on the XCAT-to-CT registration task. Boxplots showing Dice scores for different organs in CT obtained using the proposed \texttt{TransMorph}, the variants of \texttt{TransMorph}, and other Transformer architectures.\label{fig:xcat_brain_transformer}}
\end{figure*}
\newpage
\section{Additional Qualitative Results for Uncertainty Quantification}
\begin{figure*}[!htp]
\centering
\includegraphics[width=.85\textwidth]{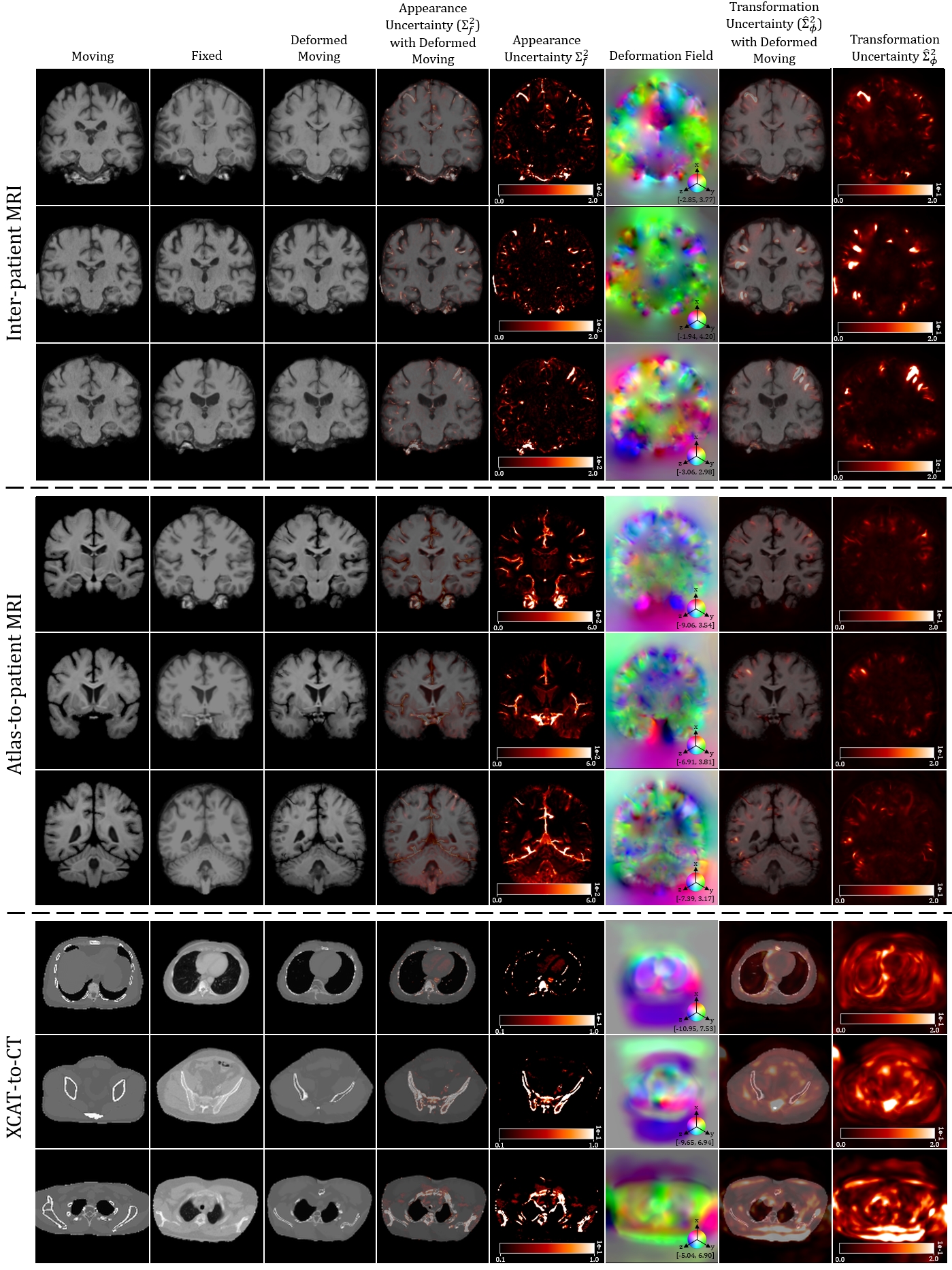}
\caption{Qualitative results and registration uncertainty estimate with \texttt{TransMorph-Bayes}. The fourth and the fifth columns exhibit the appearance uncertainties estimated using the proposed uncertainty estimation scheme (i.e., $\Sigma_f^2$). The last column shows the transformation uncertainties, i.e., $\hat{\Sigma}_\phi^2$, where the uncertainty maps were taken as square root of the sum of the variances of the deformation in $x$, $y$,
and $z$ direction. The spatial dimension $x$, $y$, and $z$ in the displacement field is mapped to each of the RGB color channels, respectively. The [$p$, $q$] in color bars denotes the magnitude range of the fields. \label{fig:jhu_brain_uncert}}
\end{figure*}
\newpage

\section{Visualization of Feature Maps in Skip Connections}
\begin{figure*}[!htp]
\centering
\includegraphics[width=.98\textwidth]{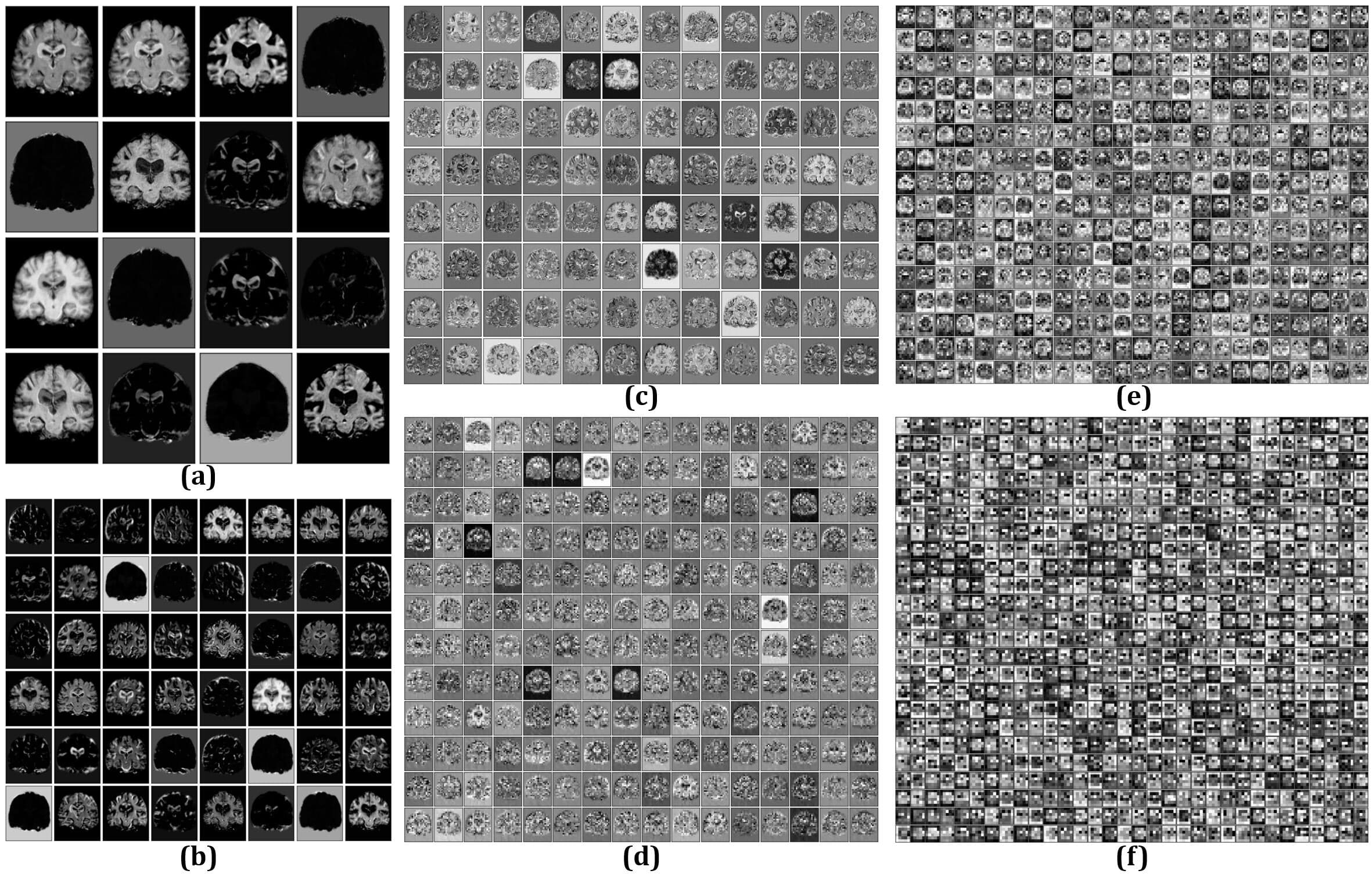}
\caption{Feature maps in \texttt{TransMorph}'s skip connections. (a) and (b) exhibit, respectively, the feature maps in the first and second skip connections from the convolutional layers in the encoder (i.e., the green arrows in Fig. \ref{fig:net_arch}); (c)-(f) exhibit the feature maps in the skip connections from the Transformer blocks (i.e., the orange arrows in Fig. \ref{fig:net_arch}).\label{fig:skip_feat}}
\end{figure*}

\section{Probabilistic diffeomorphic registration}
\label{sec:prob_diff}
As shown in section \ref{sec:prob_bspl}, we introduced a variational inference framework to the proposed \texttt{TransMorph} (which we denote as \texttt{TransMorph-diff}). A prior distribution 
\begin{linenomath}
\begin{equation}
    p(\mathbf{u})=\mathcal{N}(\mathbf{u};\mathbf{0}, \mathbf{\Sigma}_\mathbf{u})
\end{equation}
\end{linenomath} was placed over the dense displacement field $\mathbf{u}$, where $\mathbf{0}$ and $\mathbf{\Sigma}_\mathbf{u}$ are the mean and covariance of the multivariate Gaussian distribution. We followed~(\cite{dalca2019unsupervised}) and defined $\mathbf{\Sigma}_\mathbf{u}^{-1}=\Lambda_\mathbf{u}=\lambda\mathbf{L}$, where $\Lambda_\mathbf{u}$ denotes the precision matrix, $\lambda$ controls the scale of $\mathbf{u}$, $\mathbf{L}=\mathbf{D}-\mathbf{A}$ is the Laplacian matrix of a neighborhood graph formed on the voxel grid, $\mathbf{D}$ is the graph degree matrix, and $\mathbf{A}$ is a voxel neighborhood adjacency matrix. The probability $p(I_f|I_m)$ can be computed using the law of total probability:
\begin{linenomath}
\begin{equation}
\label{eqn:prob_IF_IM}
    p(I_f|I_m)=\int_{\mathbf{u}}p(I_f|\mathbf{u}, I_m)p(\mathbf{u})d\mathbf{u}.
\end{equation}
\end{linenomath}
The likelihood $p(I_f|\mathbf{u}, I_m)$ was also assumed to be Gaussian
\begin{linenomath}
\begin{equation}
    p(I_f|\mathbf{u}, I_m)=\mathcal{N}(I_f;I_m\circ\phi_\mathbf{u}, \sigma^2_\mathbf{I}\mathbb{I}),
\end{equation}
\end{linenomath}
where $\sigma^2_\mathbf{I}$ captures the variance of the image noise, and $\phi_{\mathbf{u}}$ is the group exponential of the time-stationary velocity field $\mathbf{u}$, i.e. $\phi=\exp(\mathbf{u})$, and was computed using a \textit{scaling-and-squaring} approach (section \ref{sec:ss_diff}). 

Our goal is to estimate the posterior probability $p(\mathbf{u}|I_f, I_m)$. Due to the intractable nature of the integral over $\mathbf{u}$ in Eqn. \ref{eqn:prob_IF_IM}, $p(I f|I m)$ is usually calculated using just the $\mathbf{u}$'s that are most likely to have generated $I_f$~(\cite{krebs2019learning}). Since computing the posterior $p(\mathbf{u}|I_f, I_m)$ analytically is also intractable, we instead assumed a variational posterior $q_\psi(\mathbf{u}|I_f, I_m)$ learned by the network with parameters $\psi$. The Kullback-Leibler divergence (KL) was used to relate the variational posterior to the actual posterior, which results in the evidence lower limit (ELBO)~(\cite{kingma2013auto}):
\begin{linenomath}
\begin{equation}
\label{eqn:ELBO}
    \begin{split}
        \log &p(I_f|I_m)-\text{KL}\left[q_\psi(\mathbf{u}|I_f, I_m)\Vert p(\mathbf{u}|I_f, I_m)\right]=\\
        &\mathbb{E}_{\mathbf{u}\sim q_{\psi}}\left[\log p(I_f|\mathbf{u}, I_m)\right]-\text{KL}\left[q_\psi(\mathbf{u}|I_f, I_m)\Vert p(\mathbf{u})\right],
    \end{split}
\end{equation}
\end{linenomath}
where the KL-divergence on the left hand side vanishes if the variational posterior is identical to the actual posterior. Therefore, maximizing $\log p(I_f|I_m)$ is equivalent to minimizing the negative of ELBO on the right hand side of Eqn. \ref{eqn:ELBO}. Since the prior distribution $p(\mathbf{u})$ was assumed to be a multivariate Gaussian, the variational posterior is likewise a multivariate Gaussian, defined as:
\begin{linenomath}
\begin{equation}
    q_\psi(\mathbf{u}|I_f, I_m)=\mathcal{N}(\mathbf{u}; \mathbf{\mu}_\psi(\mathbf{u}|I_f, I_m), \mathbf{\Sigma}_\psi(\mathbf{u}|I_f, I_m)),
\end{equation}
\end{linenomath}
where $\mathbf{\mu}_\psi$ and $\mathbf{\Sigma}_\psi$ are the voxel-wise mean and variance generated by the network with parameters $\psi$. In each forward pass, the dense displacement field $\mathbf{u}$ is sampled using reparameterization $\mathbf{u}=\mathbf{\mu}_\psi+\mathbf{\Sigma}_\psi\odot\mathbf{\epsilon}$ with $\mathbf{\epsilon}\sim\mathcal{N}(\mathbf{0}, \mathbf{I})$. The variational parameters $\mathbf{\mu}_\psi$ and $\mathbf{\Sigma}_\psi$ are learned by minimizing the loss
(\cite{dalca2019unsupervised}):
\begin{linenomath}
\begin{equation}
\label{eqn:L_prob_append}
    \begin{split}
        &\mathcal{L}_{prob.}(I_f, I_m, \phi_\mathbf{u}; \psi)\\&=-\mathbb{E}_{\mathbf{u}\sim q_{\psi}}\left[\log p(I_f|\mathbf{u}, I_m)\right]+\text{KL}\left[q_\psi(\mathbf{u}|I_f, I_m)\Vert p(\mathbf{u})\right]\\
        &=\frac{1}{2\sigma^2}\Vert I_f-I_m\circ\phi_\mathbf{u}\Vert^2+\frac{1}{2}\Bigg[\text{tr}(\lambda\mathbf{D}\mathbf{\Sigma}_\psi-\log \mathbf{\Sigma}_\psi)+\mathbf{\mu}^\top_\psi\Lambda_\mathbf{u}\mathbf{\mu}_\psi\Bigg],
    \end{split}
\end{equation}
\end{linenomath}
where $\mathbf{\mu}^\top_\psi\Lambda_{\mathbf{u}}\mathbf{\mu}_\psi$ can be thought of as a diffusion regluarization (Eqn. \ref{Eqn:diffusion}) placed over the mean displacement field $\mathbf{\mu}_\psi$, that is  $\mathbf{\mu}^\top_\psi\Lambda_\mathbf{u}\mathbf{\mu}_\psi=\frac{\lambda}{2}\sum_\mathbf{p}\sum_{i\in N_{(\mathbf{p})}}(\mathbf{\mu}(\mathbf{p})-\mathbf{\mu}(i))^2$, where $N_{(\mathbf{p})}$ represents the neighboring voxels of the $\mathbf{p}^{th}$ voxel. 

As discussed in section \ref{sec:aux_seg}, when the auxiliary segmentation information is available (i.e., the label maps of $I_f$ and $I_m$, denoted as $s_f$ and $s_m$), Dice loss can be used for training the network to further enhance registration performance. Dice loss, however, does not preserve a Gaussian approximation of the deformation fields. Instead, we follow~(\cite{dalca2019unsupervised}) and replace the KL divergence in Eqn. \ref{eqn:ELBO} with:
\begin{linenomath}
\begin{equation}
    \begin{split}
         \text{KL}\left[q_\psi(\mathbf{u}|I_f, I_m)\Vert p(\mathbf{u}|I_f, s_f; I_m, s_m)\right],
    \end{split}
\end{equation}
\end{linenomath}
which yields a loss function of the form:
\begin{linenomath}
\begin{equation}
\label{eqn:l_prob_w_seg_append}
    \begin{split}
        \mathcal{L}_{prob.\ w/\ aux.}&(I_f, s_f, I_m, s_m, \phi_\mathbf{u}; \psi)\\
        &=\frac{1}{2\sigma^2}\Vert I_f-I_m\circ\phi_\mathbf{u}\Vert^2+\frac{1}{2\sigma_s^2}\Vert s_f-s_m\circ\phi_\mathbf{u}\Vert^2\\
        &\ \ \ \ +\frac{1}{2}\Bigg[\text{tr}(\lambda\mathbf{D}\mathbf{\Sigma}_\psi-\log \mathbf{\Sigma}_\psi)+\mathbf{\mu}^\top_\psi\Lambda_\mathbf{u}\mathbf{\mu}_\psi\Bigg].
    \end{split}
\end{equation}
\end{linenomath}
In~(\cite{dalca2019unsupervised}), $s_f$ and $s_m$ represent \textit{anatomical surfaces} obtained from label maps. In contrast, we directly used the \textit{label maps} as $s_f$ and $s_m$ in this work. They were image volumes with multiple channels, each channel contained a binary mask defining the segmentation of a certain structure/organ.

\section{B-splines diffeomorphic registration}
\label{sec:bspline_diff}
As demonstrated in section \ref{sec:prob_bspl}, we incorporated a cubic B-spline model~(\cite{qiu2021learning}) into \texttt{TransMorph} (which we denote as \texttt{TransMorph-bspl}). This network produces a lattice of low-dimensional control points instead of producing a dense displacement field at the original resolution, which might be computationally costly. As shown in Fig. \ref{fig:transmorph_var}, we denote the displacements of the B-spline control points generated by the network as $\mathbf{u}_B$ and the spacing between the control points as $\mathbf{\delta}$. Then, a weighted combination of cubic B-spline basis functions (i.e., $\beta_d$)~(\cite{rueckert1999nonrigid}) is used to generate the dense displacement field (i.e., the B-spline tensor product in Fig. \ref{fig:transmorph_var}):
\begin{linenomath}
\begin{equation}
    \hat{\mathbf{u}}(\mathbf{p})=\sum_{\mathbf{c}\in C}\mathbf{u}_{B}(\mathbf{c})\prod_{d\in\{x,y,z\}}\beta_d(\mathbf{p}_d-k(\mathbf{c}_d)),
\end{equation}
\end{linenomath}
where $\mathbf{c}$ is the index of the control points on the lattice $C$, and $k$ denotes the coordinates of the control points $\mathbf{u}_{B}(\mathbf{c})$ in image space. Then the final time-stationary displacement $\mathbf{u}$ is obtained using the same \textit{scaling-and-squaring} approach described in section \ref{sec:ss_diff}.

 \clearpage
\twocolumn
\typeout{}
\bibliographystyle{model2-names.bst}\biboptions{authoryear}
\bibliography{references}


\clearpage
\end{document}

%% file: introduction.tex
Deformable image registration (DIR) is fundamental for many medical imaging analysis tasks. It functions by establishing spatial correspondence in order to minimize the differences between a pair of fixed and moving images. Traditional methods formulate image registration as a variational problem for estimating a smooth mapping between the points in one image and those in another (\cite{avants2008symmetric, beg2005computing, vercauteren2009diffeomorphic, heinrich2013mrf, modat2010fast}). However, such methods are computationally expensive and usually slow in practice because the optimization problem needs to be solved de novo for each pair of unseen images.

Recently, deep neural networks (DNNs), especially convolutional neural networks (ConvNets), have demonstrated state-of-the-art performance in many computer vision tasks, including object detection (\cite{redmon2016you}), image classification (\cite{he2016deep}), and segmentation (\cite{long2015fully}). Ever since the success of U-Net in the ISBI cell tracking challenge of 2015 (\cite{ronneberger2015u}), ConvNet-based methods have become a major focus of attention in medical image analysis fields, such as tumor segmentation (\cite{isensee2021nnu, zhou2019unet++}), image reconstruction (\cite{zhu2018image}), and disease diagnostics (\cite{lian2018hierarchical}). In medical image registration, ConvNet-based methods can produce significantly improved registration performance while operating orders of magnitudes faster (after training) compared to traditional methods. ConvNet-based methods replace the costly per-image optimization seen in traditional methods with a single global function optimization during a training phase. The ConvNets learn the common representation of image registration from training images, enabling rapid alignment of an unseen image pair after training. Initially, the supervision of ground-truth deformation fields (which are usually generated using traditional registration methods) is needed for training the neural networks (\cite{onofrey2013semi, yang2017quicksilver, rohe2017svf}). Recently, the focus has been shifted towards developing unsupervised methods that do not depend on ground-truth deformation fields (\cite{balakrishnan2019voxelmorph, dalca2019unsupervised, kim2021cyclemorph, de2019deep, de2017end, lei20204d, chen2020generating, zhang2018inverse}). Nearly all of the existing deep-learning-based methods mentioned above used U-Net (\cite{ronneberger2015u}) or the simply modified versions of U-Net (e.g., tweaking the number of layers or changing down- and up-sampling schemes) as their ConvNet designs.

ConvNet architectures generally have limitations in modeling explicit long-range spatial relations (i.e., relations between two voxels that are far away from each other) present in an image due to the intrinsic locality (i.e., the limited effective receptive field) of convolution operations (\cite{luo2016understanding}). The U-Net (or V-Net (\cite{milletari2016v})) was proposed to overcome this limitation by introducing down- and up-sampling operations into a ConvNet, which theoretically enlarges the receptive field of the ConvNet and, thus, encourages the network to consider long-range relationships between points in images. However, several problems remain: first, the receptive fields of the first several layers are still restricted by the convolution-kernel size, and the global information of an image can only be viewed at the deeper layers of the network; second, it has been shown that as the convolutional layers deepen, the impact from far-away voxels decays quickly (\cite{li2021medical}). Therefore, the effective receptive field of a U-Net is, in practice, much smaller than its theoretical receptive field, and it is only a portion of the typical size of a medical image. This limits the U-Net's ability to perceive semantic information and model long-range relationships between points. Yet, it is believed that the ability to comprehend semantic scene information is of great importance in coping large deformations (\cite{ha2020semantically}). Many works in other fields (e.g., image segmentation) have addressed this limitation of U-Net (\cite{zhou2019unet++, jha2019resunet++, devalla2018drunet, alom2018recurrent}). To allow for a better flow of multi-scale contextual information throughout the network, Zhou et al. (\cite{zhou2019unet++}) proposed a nested U-Net (i.e., U-Net++), in which the complex up- and down-samplings along with multiple skip connections were used. Devalla et al. (\cite{devalla2018drunet}) introduced dilated convolution to the U-Net architecture that enlarges the network's effective receptive field. A similar idea was proposed by Alom et al. (\cite{alom2018recurrent}), where the network's effective receptive field was increased by deploying recurrent convolutional operations. Jha et al. proposed ResUNet++ (\cite{jha2019resunet++}) that incorporates the attention mechanisms into U-Net for modeling long-range spatial information. Despite these methods' promising performance in other medical imaging fields, there has been limiting work on using advanced network architectures for medical image registration.

Transformer, which originated from natural language processing tasks (\cite{vaswani2017attention}), has shown its potential in computer vision tasks. A Transformer deploys self-attention mechanisms to determine which parts of the input sequence (e.g., an image) are essential based on contextual information. Unlike convolution operations, whose effective receptive fields are limited by the size of convolution kernels, the self-attention mechanisms in a Transformer have large size effective receptive fields, making a Transformer capable of capturing long-range spatial information (\cite{li2021medical}). Dosovitskiy et al. (\cite{dosovitskiy2020image}) proposed Vision Transformer (ViT) that applies the Transformer encoder from NLP directly to images. It was the first purely self-attention-based network for computer vision and achieved state-of-the-art performance in image recognition. Subsequent to their success, Swin Transformer (\cite{liu2021swin}) and its variants (\cite{dai2021dynamic, dong2021cswin}) have demonstrated their superior performances in object detection, and semantic segmentation. Recently, Transformer-related methods have gained increased attention in medical imaging (\cite{chen2021transunet, xie2021cotr, wang2021transbts, li2021medical, wang2021ted, zhang2021transct}); the major application has been the task of image segmentation. 

Transformer can be a strong candidate for image registration because it can better comprehend the spatial correspondence between the moving and fixed images. Registration is the process of establishing such correspondence, and intuitively, by comparing different parts of the moving to the fixed image. A ConvNet has a narrow field of view: it performs convolution locally, and its field of view grows in proportion to the ConvNet's depth; hence, the shallow layers have a relatively small receptive field, limiting the ConvNet's ability to associate the distant parts between two images. For example, if the left part of the moving image matches the right part of the fixed image, ConvNet will be unable to establish the proper spatial correspondence between the two parts if it cannot see both parts concurrently (i.e., when one of the parts falls outside of the ConvNet's field of view). However, Transformer is capable of handling such circumstances and rapidly focusing on the parts that need deformation, owing to its large receptive field and self-attention mechanism.

Our group has previously shown preliminary results that demonstrated the bridging of ViT and V-Net provided good performance in image registration (\cite{chen2021vit}). In this work, we extended that preliminary work and investigated various Transformer models from other tasks (i.e., computer vision and medical imaging tasks). We present a hybrid Transformer-ConvNet framework, \texttt{TransMorph}, for volumetric medical image registration. In this method, the Swin Transformer (\cite{liu2021swin}) was employed as the encoder to capture the spatial correspondence between the input moving and fixed images. Then, a ConvNet decoder processed the information provided by the Transformer encoder into a dense displacement field. Long skip connections were deployed to maintain the flow of localization information between the encoder and decoder stages. We also introduced diffeomorphic variations of \texttt{TransMorph} to ensure a smooth and topology-preserving deformation. Additionally, we applied variational inference on the parameters of \texttt{TransMorph}, resulting in a Bayesian model that predicts registration uncertainty based on the given image pair. Qualitative and quantitative evaluation of the experimental results demonstrate the robustness of the proposed method and confirm the efficacy of Transformers for image registration.

The main contributions of this work are summarized as follows:
\begin{itemize}[noitemsep]
    \item \textit{Transformer-based model:} This paper presents the pioneering work on using Transformers for image registration. A novel Transformer-based neural network, \texttt{TransMorph}, was proposed for affine and deformable image registration.
    \item \textit{Architecture analysis:} Experiments in this paper demonstrate that positional embedding, which is a commonly used element in Transformer by convention, is not required for the proposed hybrid Transformer-ConvNet model. Secondly, we show that Transformer-based models have larger effective receptive fields than ConvNets. Moreover, we demonstrated that \texttt{TransMorph} promotes a flatter registration loss landscape.
    \item \textit{Diffeomorphic registration:} We demonstrate that \texttt{TransMorph} can be easily integrated into two existing frameworks as a registration backbone to provide diffeomorphic registration.
    \item \textit{Uncertainty quantification:} This paper also provides a Bayesian uncertainty variant of \texttt{TransMorph} that yields transformer uncertainty and perfectly calibrated appearance uncertainty estimates.
    \item \textit{State-of-the-art results:} We extensively validate the proposed registration models on two brain MRI registration applications (inter-patient and atlas-to-patient registration) and on a novel application of XCAT-to-CT registration with an aim to create a population of anatomically variable XCAT phantom. The datasets used in this study (which include a publicly available dataset, the IXI dataset\footnote{\url{https://brain-development.org/ixi-dataset/}}) contain over 1000 image pairs for training and testing. The proposed models were compared with various registration methods and demonstrated state-of-the-art performance. \textbf{Eight} registration approaches were employed as baselines, including learning-based methods and widely used conventional methods. The performances of \textbf{four} recently proposed Transformer architectures from other tasks (e.g., semantic segmentation, classification, etc.) were also evaluated on the task of image registration.
    \item \textit{Open source:} We provide the community with a fast and accurate tool for deformable registration. The source code, the pre-trained models, and our preprocessed IXI dataset are publicly available at \url{https://bit.ly/37eJS6N}.
\end{itemize}

The paper is organized as follows. Section \ref{sec:rel_work} discusses related work. Section \ref{sec:methods} explains the proposed methodology. Section \ref{sec:experiments} discusses experimental setup, implementation details, and datasets used in this study. Section \ref{sec:results} presents experimental results. Section \ref{sec:discussion} discusses the findings based on the results, and Section \ref{sec:conclusion} concludes the paper.

%% file: rel_works.tex
This section reviews the relevant literature and provides fundamental knowledge for the proposed method.

\subsection{Image Registration}
Deformable image registration (DIR) establishes spatial correspondence between two images by optimizing an energy function:
\begin{linenomath}
\begin{equation}
\label{eqn:E}
    E(I_m, I_f, \phi) = E_{sim}(I_m\circ\phi, I_f) + \lambda R(\phi), 
\end{equation}
\end{linenomath}
where $I_m$ and $I_f$ denote, respectively, the moving and fixed image, $\phi$ denotes the deformation field that warps the moving image (i.e., $I_m\circ\phi$), $R(\phi)$ imposes smoothness of the deformation field, and $\lambda$ is the regularization hyper-parameter that determines the trade-off between image similarity and deformation field regularity. The optimal warping, $\hat{\phi}$ is given by minimizing this energy function:
\begin{linenomath}
\begin{equation}
    \hat{\phi} = \argmin_\phi E(I_m, I_f, \phi).
\end{equation}
\end{linenomath}

In the energy function, $E_{sim}$ measures the level of alignment between the deformed moving image, $I_m\circ\phi$, and the fixed image, $I_f$. Some common choices for $E_{sim}$ are mean squared error (MSE) (\cite{beg2005computing, wolberg2000robust}), normalized cross-correlation (NCC) (\cite{avants2008symmetric}), structural similarity index (SSIM) (\cite{chen2020generating}), and mututal information (MI) (\cite{viola1997alignment}). The regularization term, $R(\phi)$, imposes spatial smoothness on the deformation field. A common assumption in most applications is that similar structures exist in both moving and fixed images. As a result, a continuous and invertible deformation field (i.e., a diffeomorphism) is needed to preserve topology, and the regularization, $R(\phi)$ is meant to enforce or encourage this. Isotropic diffusion (equivalent to Gaussian smoothing) (\cite{balakrishnan2019voxelmorph}), anisotropic diffusion (\cite{pace2013locally}), total variation (\cite{vishnevskiy2016isotropic}), and bending energy (\cite{johnson2002consistent}) are popular options for $R(\phi)$.

\subsubsection{Image registration via deep neural networks}
While traditional image registration methods iteratively minimize the energy function in (\ref{eqn:E}) for each pair of moving and fixed images, DNN-based methods optimize the energy function for a training dataset, thereby learning a global representation of image registration that enables alignment of an unseen pair of volumes. DNN methods are often categorized as supervised or unsupervised, with the former requiring a ground truth deformation field for training and the latter relying only on the image datasets. 

In supervised DNN methods, the ground-truth deformation fields are either produced synthetically or generated by traditional registration methods (\cite{yang2017quicksilver, sokooti2017nonrigid, cao2018deep}). \cite{yang2017quicksilver} proposed a supervised ConvNet that predicts the LDDMM (\cite{beg2005computing}) momentum from image patches. \cite{sokooti2017nonrigid} trained a registration ConvNet with synthetic displacement fields. The ground-truth deformation fields are often computationally expensive to generate, and the registration accuracy of these methods is highly dependent on the quality of the ground truth.

Due to the limitations of supervised methods, the focus of research has switched to unsupervised DNN methods that do not need ground-truth deformation fields. Unsupervised DNNs optimize an energy function on the input images, similar to traditional methods. However, DNN-based methods learn a common registration representation from a training set and then apply it to unseen images. Note that the term ``unsupervised" refers to the absence of ground-truth deformation fields, but the network still needs training (this is also known as ``self-supervised"). \cite{de2019deep, balakrishnan2018unsupervised, balakrishnan2019voxelmorph} are representative of unsupervised DNN-based methods.

More recently, diffeomorphic deformation representations have been developed to address the issue of non-smooth deformations in DNN-based methods. We briefly introduce its concepts in the next subsection.
\begin{figure*}[!t]
\centering
\includegraphics[width=.85\textwidth]{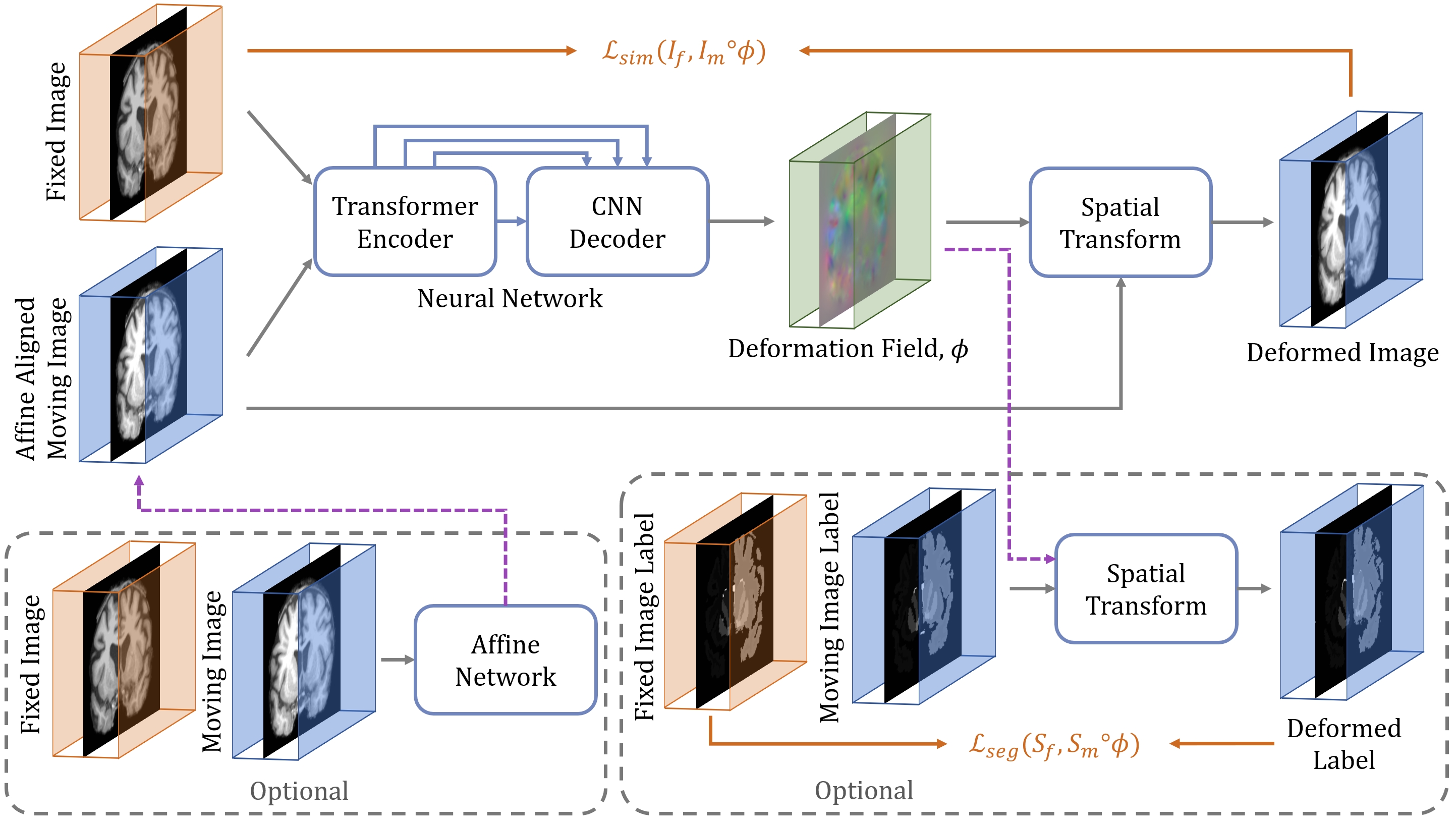}
\caption{The overall framework of the proposed Transformer-based image registration model, \texttt{TransMorph}. The proposed hybrid Transformer-ConvNet network takes two inputs: a fixed image and a moving image that is affinely aligned with the fixed image. The network generates a nonlinear warping function, which is then applied to the moving image through a spatial transformation function. If an image pair has not been affinely aligned, an affine Transformer may be used prior to the deformable registration (left dashed box). Additionally, auxiliary anatomical segmentations may be leveraged during training the proposed network (right dashed box).}\label{fig:overall}
\end{figure*}
\subsubsection{Diffeomorphic image registration}
\label{sec:ss_diff}
Diffeomorphic deformable image registration is important in many medical image applications, owing to its special properties including topology preservation and transformation invertibility. A diffeomorphic transformation is a smooth and continuous one-to-one mapping with invertible derivatives (i.e., non-zero Jacobian determinant). Such a transformation can be achieved via the time-integration of time-dependent (\cite{beg2005computing, avants2008symmetric}) or time-stationary velocity fields (SVFs) (\cite{arsigny2006log, ashburner2007fast, vercauteren2009diffeomorphic, hernandez2009registration}). In the time-dependent setting (e.g., LDDMM (\cite{beg2005computing}) and SyN (\cite{avants2008symmetric})), a diffeomorphic transformation $\phi$ is obtained via integrating the sufficiently smooth time-varying velocity fields $\nu^{(t)}$, i.e., $\frac{d}{dt}\phi^{(t)}=\nu^{(t)}(\phi^{(t)})$, where $\phi^{(t)} = id$ is the identity transform. On the other hand, in the stationary velocity fields (SVFs) setting (e.g., DARTEL \cite{ashburner2007fast} and diffeomorphic Demons (\cite{vercauteren2009diffeomorphic})), the velocity fields are assumed to be stationary over time, i.e., $\frac{d}{dt}\phi^{(t)}=\nu(\phi^{(t)})$. Dalca et al. (\cite{dalca2019unsupervised}) first adopt the diffeomorphism formulation in a deep learning model, using the SVFs setting with an efficient \textit{scaling-and-squaring} approach (\cite{arsigny2006log}). In the \textit{scaling-and-squaring} approach, the deformation field is represented as a Lie algebra member that is exponentiated to generate a time 1 deformation $\phi^{(1)}$, which is a member of the Lie group: $\phi^{(1)} = \exp(\nu)$. This means that the exponentiated flow field compels the mapping to be diffeomorphic and invertible using the same flow field. Starting from an initial deformation field:
\begin{linenomath}
\begin{equation}
   \phi^{(1/2^T)} = p + \frac{\nu(p)}{2^T},
\end{equation}
\end{linenomath}
where $p$ denotes the spatial locations. The $\phi^{(1)}$ can be obtained using the recurrence:
\begin{linenomath}
\begin{equation}
\label{eqn:sqring}
    \phi^{(1/2^{t-1})} = \phi^{(1/2^{t})} \circ \phi^{(1/2^{t})}.
\end{equation}
\end{linenomath}
Thus, $\phi^{(1)} = \phi^{(1/2)} \circ \phi^{(1/2)}$.

In practice, a neural network first generates a displacement field, which is then scaled by $1/2^T$ to produce an initial deformation field $\phi^{(1/2^T)}$. Subsequently, the squaring technique (i.e., Eqn. \ref{eqn:sqring}) is applied recursively to $\phi^{(1/2^T)}$ $T$ times via a spatial transformation function, resulting in a final diffeomorphic deformation field $\phi^{(1)}$. Despite the fact that diffeomorphisms are theoretically guaranteed to be invertible, interpolation errors can lead to invertibility errors that increase linearly with the number of interpolation steps (\cite{avants2008symmetric, mok2020fast}).

\subsection{Self-attention Mechanism and Transformer}
Transformer makes use of a self-attention mechanism that estimates the relevance of one input sequence to another via the Query-Key-Value (QKV) model (\cite{vaswani2017attention, dosovitskiy2020image}). The input sequences often originate from the flattened patches of an image. Let $\mathbf{x}$ be an image volume defined over a 3D spatial domain (i.e., $\mathbf{x} \in \mathbb{R}^{H\times W\times L}$). The image is first divided into $N$ flattened 3D patches $\mathbf{x}_p\in\mathbb{R}^{N\times P^3}$ , where $(H, W, L)$ is the size of the original image, $(P,P,P)$ is the size of each image patch, and $N=\frac{HWL}{P^3}$. Then, a learnable linear embedding $\mathbf{E}$ is applied to $\mathbf{x}_p$, which projects each patch into a $D\times1$ vector representation:
\begin{linenomath}
\begin{equation}
    \hat{\mathbf{x}}_e=[\mathbf{x}^1_p\mathbf{E};\mathbf{x}^2_p\mathbf{E};...;\mathbf{x}^N_p\mathbf{E}],\ \ \ \mathbf{E}\in\mathbb{R}^{P^3\times D}
\end{equation}
\end{linenomath}
where the dimension $D$ is a user-defined hyperparemeter. Then, a learnable positional embedding is added to $\hat{\mathbf{x}}_e$ so that the patches can retain their positional information, i.e., $\mathbf{x}_e=\hat{\mathbf{x}}_e+\mathbf{E}_{pos}$, where $\mathbf{E}_{pos}\in\mathbb{R}^{N\times D}$. These vector representations, often known as tokens, are subsequently used as inputs for self-attention computations.

\paragraph{Self-attention} To compute self-attention (SA), $\mathbf{x}_e\in\mathbb{R}^{N\times D}$ is encoded by $\mathbf{U}$ (i.e., a linear layer) to three matrix representations: Queries $\mathbf{Q}\in\mathbb{R}^{N\times D_k}$, Keys $\mathbf{K}\in\mathbb{R}^{N\times D_k}$, and Values $\mathbf{V}\in\mathbb{R}^{N\times D_v}$. The scaled dot-product attention is given by:
\begin{linenomath}
\begin{equation}
\begin{split}
     &[\mathbf{Q},\mathbf{K},\mathbf{V}] = \mathbf{x}_e\mathbf{U}_{q,k,v} \ \ \ \ \mathbf{U}_{q,k,v}\in\mathbb{R}^{D\times D_{q,k,v}},\\
     &\mathbf{A} = \text{softmax}(\frac{\mathbf{Q}\mathbf{K}^\top}{\sqrt{D_k}}) \ \ \ \ \mathbf{A}\in\mathbb{R}^{N\times N},\\
     &SA(\mathbf{x}_e)=\mathbf{AV},
\end{split}
\end{equation}
\end{linenomath}
where $\mathbf{A}$ is the attention weight matrix, each element of $\mathbf{A}$ represents the pairwise similarity between two elements of the input sequence $\mathbf{x}_e$ and their respective query and key representations. In general, SA computes a normalized score for each input token based on the dot product of the Query and Key representations. The score is subsequently applied to the Value representation of the token, signifying to the network whether or not to focus on this token.

\paragraph{Multi-head self-attention} A Transformer employs multi-head self-attention (MSA) rather than a single attention function. MSA is an extension of self-attention in which $h$ self-attention operations (i.e., ``heads") are processed in parallel, thereby effectively increasing the number of trainable parameters. Then, the outputs of the SA operations are concatenated then projected onto a $D$-dimensional representation:
\begin{linenomath}
\begin{equation}
\begin{split}
    MSA(\mathbf{x}_e) = &[SA_1(\mathbf{x}_e);SA_2(\mathbf{x}_e);...;SA_h(\mathbf{x}_e)]\mathbf{U}_{MSA},
\end{split}
\end{equation}
\end{linenomath}
where $\mathbf{U}_{MSA}\in\mathbb{R}^{h\cdot D_h\times D}$, and $D_h$ is typically set to $D/h$ in order to keep the number of parameters constant before and after the MSA operation.

\subsection{Bayesian Deep Learning}
Uncertainty estimates help comprehend what a machine learning model does not know. They indicate the likelihood that a neural network may make an incorrect prediction. Because most deep neural networks are incapable of providing an estimate of the uncertainty in their output values, their predictions are frequently taken at face value and thought to be correct. \textit{Bayesian deep learning} estimates predictive uncertainty, providing a realistic paradigm for understanding uncertainty within deep neural networks (\cite{gal2016dropout}). The uncertainty caused by the parameters in a neural network is known as epistemic uncertainty, which is modeled by placing a prior distribution (e.g., a Gaussian prior distribution: $\mathbf{W}\sim \mathcal{N}(0,I)$) on the parameters of a network and then attempting to capture how much these weights vary given specific data. Recent efforts in this area include the \textit{Bayes by Backprop} (\cite{blundell2015weight}), its closely related mean-field variational inference by assuming a Gaussian prior distribution (\cite{Tolle2021}), stochastic batch normalization (\cite{atanov2018uncertainty}), and Monte-Carlo (MC) dropout (\cite{gal2016dropout, Kendall2017}). The applications of Bayesian deep learning in medical imaging expands on image denoising (\cite{Tolle2021, Laves2020}) and image segmentation (\cite{devries2018leveraging, baumgartner2019phiseg, mehrtash2020confidence}). In deep-learning-based image registration, the majority of methods provide a single, deterministic solution of the unknown geometric transformation. Knowing about epistemic uncertainty helps determine if and to what degree the registration results can be trusted and whether the input data is appropriate for the neural network. 

In general, two categories of registration uncertainty may be modeled using the epistemic uncertainty of a deep learning model: transformation uncertainty and appearance uncertainty (\cite{luo2019applicability, xu2021double}). Transformation uncertainty measures the local ambiguity of the spatial transformation (i.e., the deformation), whereas appearance uncertainty quantifies the uncertainty in the intensity values of registered voxels or the volumes of the registered organs. Transformation uncertainty estimates may be used for uncertainty-weighted registration (\cite{simpson2011longitudinal, kybic2009bootstrap}), surgical treatment planning, or directly visualized for qualitative evaluations (\cite{yang2017quicksilver}). Appearance uncertainty may be translated into dose uncertainties in cumulative dose for radiation or radiopharmaceutical therapy (\cite{risholm2011estimation, vickress2017representing, chetty2019deformable, gear2018eanm}). These registration uncertainty estimates also enable the assessment of operative risks and leads to better-informed clinical decisions~(\cite{luo2019applicability}). Cui et al. (\cite{cui2021bayesian}) and Yang et al. (\cite{yang2017quicksilver}) incorporated MC dropout layers in their registration network designs, which allows for the estimation of transformation uncertainty by sampling multiple deformation field predictions from the network.

The proposed image registration framework expands on these ideas. In particular, a new registration framework is presented that leverages a Transformer in the network design. We demonstrate that this framework can be readily adapted to several existing techniques to allow diffeomorphism for image registration, and incorporate Bayesian deep learning to estimate registration uncertainty.

%% file: medima-template.bbl
\begin{thebibliography}{129}
\expandafter\ifx\csname natexlab\endcsname\relax\def\natexlab#1{#1}\fi
\providecommand{\url}[1]{\texttt{#1}}
\providecommand{\href}[2]{#2}
\providecommand{\path}[1]{#1}
\providecommand{\DOIprefix}{doi:}
\providecommand{\ArXivprefix}{arXiv:}
\providecommand{\URLprefix}{URL: }
\providecommand{\Pubmedprefix}{pmid:}
\providecommand{\doi}[1]{\href{http://dx.doi.org/#1}{\path{#1}}}
\providecommand{\Pubmed}[1]{\href{pmid:#1}{\path{#1}}}
\providecommand{\bibinfo}[2]{#2}
\ifx\xfnm\relax \def\xfnm[#1]{\unskip,\space#1}\fi
\bibitem[{Alom et~al.(2018)Alom, Hasan, Yakopcic, Taha and
  Asari}]{alom2018recurrent}
\bibinfo{author}{Alom, M.Z.}, \bibinfo{author}{Hasan, M.},
  \bibinfo{author}{Yakopcic, C.}, \bibinfo{author}{Taha, T.M.},
  \bibinfo{author}{Asari, V.K.}, \bibinfo{year}{2018}.
\newblock \bibinfo{title}{Recurrent residual convolutional neural network based
  on u-net (r2u-net) for medical image segmentation}.
\newblock \bibinfo{journal}{arXiv preprint arXiv:1802.06955} .
\bibitem[{Armstrong(2014)}]{armstrong2014use}
\bibinfo{author}{Armstrong, R.A.}, \bibinfo{year}{2014}.
\newblock \bibinfo{title}{When to use the b onferroni correction}.
\newblock \bibinfo{journal}{Ophthalmic and Physiological Optics}
  \bibinfo{volume}{34}, \bibinfo{pages}{502--508}.
\bibitem[{Arsigny et~al.(2006)Arsigny, Commowick, Pennec and
  Ayache}]{arsigny2006log}
\bibinfo{author}{Arsigny, V.}, \bibinfo{author}{Commowick, O.},
  \bibinfo{author}{Pennec, X.}, \bibinfo{author}{Ayache, N.},
  \bibinfo{year}{2006}.
\newblock \bibinfo{title}{A log-euclidean framework for statistics on
  diffeomorphisms}, in: \bibinfo{booktitle}{International Conference on Medical
  Image Computing and Computer-Assisted Intervention},
  \bibinfo{organization}{Springer}. pp. \bibinfo{pages}{924--931}.
\bibitem[{Ashburner(2007)}]{ashburner2007fast}
\bibinfo{author}{Ashburner, J.}, \bibinfo{year}{2007}.
\newblock \bibinfo{title}{A fast diffeomorphic image registration algorithm}.
\newblock \bibinfo{journal}{Neuroimage} \bibinfo{volume}{38},
  \bibinfo{pages}{95--113}.
\bibitem[{Atanov et~al.(2018)Atanov, Ashukha, Molchanov, Neklyudov and
  Vetrov}]{atanov2018uncertainty}
\bibinfo{author}{Atanov, A.}, \bibinfo{author}{Ashukha, A.},
  \bibinfo{author}{Molchanov, D.}, \bibinfo{author}{Neklyudov, K.},
  \bibinfo{author}{Vetrov, D.}, \bibinfo{year}{2018}.
\newblock \bibinfo{title}{Uncertainty estimation via stochastic batch
  normalization}.
\newblock \bibinfo{journal}{arXiv preprint arXiv:1802.04893} .
\bibitem[{Avants et~al.(2008)Avants, Epstein, Grossman and
  Gee}]{avants2008symmetric}
\bibinfo{author}{Avants, B.B.}, \bibinfo{author}{Epstein, C.L.},
  \bibinfo{author}{Grossman, M.}, \bibinfo{author}{Gee, J.C.},
  \bibinfo{year}{2008}.
\newblock \bibinfo{title}{Symmetric diffeomorphic image registration with
  cross-correlation: evaluating automated labeling of elderly and
  neurodegenerative brain}.
\newblock \bibinfo{journal}{Medical image analysis} \bibinfo{volume}{12},
  \bibinfo{pages}{26--41}.
\bibitem[{Balakrishnan et~al.(2018)Balakrishnan, Zhao, Sabuncu, Guttag and
  Dalca}]{balakrishnan2018unsupervised}
\bibinfo{author}{Balakrishnan, G.}, \bibinfo{author}{Zhao, A.},
  \bibinfo{author}{Sabuncu, M.R.}, \bibinfo{author}{Guttag, J.},
  \bibinfo{author}{Dalca, A.V.}, \bibinfo{year}{2018}.
\newblock \bibinfo{title}{An unsupervised learning model for deformable medical
  image registration}, in: \bibinfo{booktitle}{Proceedings of the IEEE
  conference on computer vision and pattern recognition}, pp.
  \bibinfo{pages}{9252--9260}.
\bibitem[{Balakrishnan et~al.(2019)Balakrishnan, Zhao, Sabuncu, Guttag and
  Dalca}]{balakrishnan2019voxelmorph}
\bibinfo{author}{Balakrishnan, G.}, \bibinfo{author}{Zhao, A.},
  \bibinfo{author}{Sabuncu, M.R.}, \bibinfo{author}{Guttag, J.},
  \bibinfo{author}{Dalca, A.V.}, \bibinfo{year}{2019}.
\newblock \bibinfo{title}{Voxelmorph: a learning framework for deformable
  medical image registration}.
\newblock \bibinfo{journal}{IEEE transactions on medical imaging}
  \bibinfo{volume}{38}, \bibinfo{pages}{1788--1800}.
\bibitem[{Baumgartner et~al.(2019)Baumgartner, Tezcan, Chaitanya, H{\"o}tker,
  Muehlematter, Schawkat, Becker, Donati and Konukoglu}]{baumgartner2019phiseg}
\bibinfo{author}{Baumgartner, C.F.}, \bibinfo{author}{Tezcan, K.C.},
  \bibinfo{author}{Chaitanya, K.}, \bibinfo{author}{H{\"o}tker, A.M.},
  \bibinfo{author}{Muehlematter, U.J.}, \bibinfo{author}{Schawkat, K.},
  \bibinfo{author}{Becker, A.S.}, \bibinfo{author}{Donati, O.},
  \bibinfo{author}{Konukoglu, E.}, \bibinfo{year}{2019}.
\newblock \bibinfo{title}{Phiseg: Capturing uncertainty in medical image
  segmentation}, in: \bibinfo{booktitle}{International Conference on Medical
  Image Computing and Computer-Assisted Intervention},
  \bibinfo{organization}{Springer}. pp. \bibinfo{pages}{119--127}.
\bibitem[{Beg et~al.(2005)Beg, Miller, Trouv{\'e} and
  Younes}]{beg2005computing}
\bibinfo{author}{Beg, M.F.}, \bibinfo{author}{Miller, M.I.},
  \bibinfo{author}{Trouv{\'e}, A.}, \bibinfo{author}{Younes, L.},
  \bibinfo{year}{2005}.
\newblock \bibinfo{title}{Computing large deformation metric mappings via
  geodesic flows of diffeomorphisms}.
\newblock \bibinfo{journal}{International journal of computer vision}
  \bibinfo{volume}{61}, \bibinfo{pages}{139--157}.
\bibitem[{Bishop(2006)}]{bishop2006pattern}
\bibinfo{author}{Bishop, C.M.}, \bibinfo{year}{2006}.
\newblock \bibinfo{title}{Pattern recognition}.
\newblock \bibinfo{journal}{Machine learning} \bibinfo{volume}{128}.
\bibitem[{Blundell et~al.(2015)Blundell, Cornebise, Kavukcuoglu and
  Wierstra}]{blundell2015weight}
\bibinfo{author}{Blundell, C.}, \bibinfo{author}{Cornebise, J.},
  \bibinfo{author}{Kavukcuoglu, K.}, \bibinfo{author}{Wierstra, D.},
  \bibinfo{year}{2015}.
\newblock \bibinfo{title}{Weight uncertainty in neural network}, in:
  \bibinfo{booktitle}{International Conference on Machine Learning},
  \bibinfo{organization}{PMLR}. pp. \bibinfo{pages}{1613--1622}.
\bibitem[{Cao et~al.(2018)Cao, Yang, Wang, Xue, Wang and Shen}]{cao2018deep}
\bibinfo{author}{Cao, X.}, \bibinfo{author}{Yang, J.}, \bibinfo{author}{Wang,
  L.}, \bibinfo{author}{Xue, Z.}, \bibinfo{author}{Wang, Q.},
  \bibinfo{author}{Shen, D.}, \bibinfo{year}{2018}.
\newblock \bibinfo{title}{Deep learning based inter-modality image registration
  supervised by intra-modality similarity}, in:
  \bibinfo{booktitle}{International workshop on machine learning in medical
  imaging}, \bibinfo{organization}{Springer}. pp. \bibinfo{pages}{55--63}.
\bibitem[{Chen et~al.(2022)Chen, Frey and Du}]{chen2022unsupervised}
\bibinfo{author}{Chen, J.}, \bibinfo{author}{Frey, E.}, \bibinfo{author}{Du,
  Y.}, \bibinfo{year}{2022}.
\newblock \bibinfo{title}{Unsupervised learning of diffeomorphic image
  registration via transmorph}, in: \bibinfo{booktitle}{10th Internatioal
  Workshop on Biomedical Image Registration}.
\newblock \URLprefix \url{https://openreview.net/forum?id=uwIo__2xnTO}.
\bibitem[{Chen et~al.(2021a)Chen, He, Frey, Li and Du}]{chen2021vit}
\bibinfo{author}{Chen, J.}, \bibinfo{author}{He, Y.}, \bibinfo{author}{Frey,
  E.C.}, \bibinfo{author}{Li, Y.}, \bibinfo{author}{Du, Y.},
  \bibinfo{year}{2021}a.
\newblock \bibinfo{title}{Vit-v-net: Vision transformer for unsupervised
  volumetric medical image registration}.
\newblock \bibinfo{journal}{arXiv preprint arXiv:2104.06468} .
\bibitem[{Chen et~al.(2019)Chen, Jha and Frey}]{chen2019incorporating}
\bibinfo{author}{Chen, J.}, \bibinfo{author}{Jha, A.K.}, \bibinfo{author}{Frey,
  E.C.}, \bibinfo{year}{2019}.
\newblock \bibinfo{title}{Incorporating ct prior information in the robust
  fuzzy c-means algorithm for qspect image segmentation}, in:
  \bibinfo{booktitle}{Medical Imaging 2019: Image Processing},
  \bibinfo{organization}{International Society for Optics and Photonics}. p.
  \bibinfo{pages}{109491W}.
\bibitem[{Chen et~al.(2020)Chen, Li, Du and Frey}]{chen2020generating}
\bibinfo{author}{Chen, J.}, \bibinfo{author}{Li, Y.}, \bibinfo{author}{Du, Y.},
  \bibinfo{author}{Frey, E.C.}, \bibinfo{year}{2020}.
\newblock \bibinfo{title}{Generating anthropomorphic phantoms using fully
  unsupervised deformable image registration with convolutional neural
  networks}.
\newblock \bibinfo{journal}{Medical physics} .
\bibitem[{Chen et~al.(2021b)Chen, Lu, Yu, Luo, Adeli, Wang, Lu, Yuille and
  Zhou}]{chen2021transunet}
\bibinfo{author}{Chen, J.}, \bibinfo{author}{Lu, Y.}, \bibinfo{author}{Yu, Q.},
  \bibinfo{author}{Luo, X.}, \bibinfo{author}{Adeli, E.},
  \bibinfo{author}{Wang, Y.}, \bibinfo{author}{Lu, L.},
  \bibinfo{author}{Yuille, A.L.}, \bibinfo{author}{Zhou, Y.},
  \bibinfo{year}{2021}b.
\newblock \bibinfo{title}{Transunet: Transformers make strong encoders for
  medical image segmentation}.
\newblock \bibinfo{journal}{arXiv preprint arXiv:2102.04306} .
\bibitem[{Chetty and Rosu-Bubulac(2019)}]{chetty2019deformable}
\bibinfo{author}{Chetty, I.J.}, \bibinfo{author}{Rosu-Bubulac, M.},
  \bibinfo{year}{2019}.
\newblock \bibinfo{title}{Deformable registration for dose accumulation}, in:
  \bibinfo{booktitle}{Seminars in radiation oncology},
  \bibinfo{organization}{Elsevier}. pp. \bibinfo{pages}{198--208}.
\bibitem[{Christoffersen et~al.(2013)Christoffersen, Hansen, Poulsen and
  Sorensen}]{Christoffersen2013}
\bibinfo{author}{Christoffersen, C.P.}, \bibinfo{author}{Hansen, D.},
  \bibinfo{author}{Poulsen, P.}, \bibinfo{author}{Sorensen, T.S.},
  \bibinfo{year}{2013}.
\newblock \bibinfo{title}{{Registration-based reconstruction of
  four-dimensional cone beam computed tomography}}.
\newblock \bibinfo{journal}{IEEE Transactions on Medical Imaging}
  \bibinfo{volume}{32}, \bibinfo{pages}{2064--2077}.
\newblock \DOIprefix\doi{10.1109/TMI.2013.2272882}.
\bibitem[{Cui et~al.(2021)Cui, Fu, Li and Lin}]{cui2021bayesian}
\bibinfo{author}{Cui, K.}, \bibinfo{author}{Fu, P.}, \bibinfo{author}{Li, Y.},
  \bibinfo{author}{Lin, Y.}, \bibinfo{year}{2021}.
\newblock \bibinfo{title}{Bayesian fully convolutional networks for brain image
  registration}.
\newblock \bibinfo{journal}{Journal of Healthcare Engineering}
  \bibinfo{volume}{2021}.
\bibitem[{Dai et~al.(2021)Dai, Chen, Xiao, Chen, Liu, Yuan and
  Zhang}]{dai2021dynamic}
\bibinfo{author}{Dai, X.}, \bibinfo{author}{Chen, Y.}, \bibinfo{author}{Xiao,
  B.}, \bibinfo{author}{Chen, D.}, \bibinfo{author}{Liu, M.},
  \bibinfo{author}{Yuan, L.}, \bibinfo{author}{Zhang, L.},
  \bibinfo{year}{2021}.
\newblock \bibinfo{title}{Dynamic head: Unifying object detection heads with
  attentions}, in: \bibinfo{booktitle}{Proceedings of the IEEE/CVF Conference
  on Computer Vision and Pattern Recognition}, pp. \bibinfo{pages}{7373--7382}.
\bibitem[{Dalca et~al.(2019)Dalca, Balakrishnan, Guttag and
  Sabuncu}]{dalca2019unsupervised}
\bibinfo{author}{Dalca, A.V.}, \bibinfo{author}{Balakrishnan, G.},
  \bibinfo{author}{Guttag, J.}, \bibinfo{author}{Sabuncu, M.R.},
  \bibinfo{year}{2019}.
\newblock \bibinfo{title}{Unsupervised learning of probabilistic diffeomorphic
  registration for images and surfaces}.
\newblock \bibinfo{journal}{Medical image analysis} \bibinfo{volume}{57},
  \bibinfo{pages}{226--236}.
\bibitem[{Darken and Moody(1991)}]{darken1991towards}
\bibinfo{author}{Darken, C.}, \bibinfo{author}{Moody, J.},
  \bibinfo{year}{1991}.
\newblock \bibinfo{title}{Towards faster stochastic gradient search}.
\newblock \bibinfo{journal}{Advances in neural information processing systems}
  \bibinfo{volume}{4}.
\bibitem[{Devalla et~al.(2018)Devalla, Renukanand, Sreedhar, Subramanian,
  Zhang, Perera, Mari, Chin, Tun, Strouthidis et~al.}]{devalla2018drunet}
\bibinfo{author}{Devalla, S.K.}, \bibinfo{author}{Renukanand, P.K.},
  \bibinfo{author}{Sreedhar, B.K.}, \bibinfo{author}{Subramanian, G.},
  \bibinfo{author}{Zhang, L.}, \bibinfo{author}{Perera, S.},
  \bibinfo{author}{Mari, J.M.}, \bibinfo{author}{Chin, K.S.},
  \bibinfo{author}{Tun, T.A.}, \bibinfo{author}{Strouthidis, N.G.}, et~al.,
  \bibinfo{year}{2018}.
\newblock \bibinfo{title}{Drunet: a dilated-residual u-net deep learning
  network to segment optic nerve head tissues in optical coherence tomography
  images}.
\newblock \bibinfo{journal}{Biomedical optics express} \bibinfo{volume}{9},
  \bibinfo{pages}{3244--3265}.
\bibitem[{DeVries and Taylor(2018)}]{devries2018leveraging}
\bibinfo{author}{DeVries, T.}, \bibinfo{author}{Taylor, G.W.},
  \bibinfo{year}{2018}.
\newblock \bibinfo{title}{Leveraging uncertainty estimates for predicting
  segmentation quality}.
\newblock \bibinfo{journal}{arXiv preprint arXiv:1807.00502} .
\bibitem[{Dice(1945)}]{dice1945measures}
\bibinfo{author}{Dice, L.R.}, \bibinfo{year}{1945}.
\newblock \bibinfo{title}{Measures of the amount of ecologic association
  between species}.
\newblock \bibinfo{journal}{Ecology} \bibinfo{volume}{26},
  \bibinfo{pages}{297--302}.
\bibitem[{Dong et~al.(2021)Dong, Bao, Chen, Zhang, Yu, Yuan, Chen and
  Guo}]{dong2021cswin}
\bibinfo{author}{Dong, X.}, \bibinfo{author}{Bao, J.}, \bibinfo{author}{Chen,
  D.}, \bibinfo{author}{Zhang, W.}, \bibinfo{author}{Yu, N.},
  \bibinfo{author}{Yuan, L.}, \bibinfo{author}{Chen, D.}, \bibinfo{author}{Guo,
  B.}, \bibinfo{year}{2021}.
\newblock \bibinfo{title}{Cswin transformer: A general vision transformer
  backbone with cross-shaped windows}.
\newblock \bibinfo{journal}{arXiv preprint arXiv:2107.00652} .
\bibitem[{Dosovitskiy et~al.(2020)Dosovitskiy, Beyer, Kolesnikov, Weissenborn,
  Zhai, Unterthiner, Dehghani, Minderer, Heigold, Gelly
  et~al.}]{dosovitskiy2020image}
\bibinfo{author}{Dosovitskiy, A.}, \bibinfo{author}{Beyer, L.},
  \bibinfo{author}{Kolesnikov, A.}, \bibinfo{author}{Weissenborn, D.},
  \bibinfo{author}{Zhai, X.}, \bibinfo{author}{Unterthiner, T.},
  \bibinfo{author}{Dehghani, M.}, \bibinfo{author}{Minderer, M.},
  \bibinfo{author}{Heigold, G.}, \bibinfo{author}{Gelly, S.}, et~al.,
  \bibinfo{year}{2020}.
\newblock \bibinfo{title}{An image is worth 16x16 words: Transformers for image
  recognition at scale}.
\newblock \bibinfo{journal}{arXiv preprint arXiv:2010.11929} .
\bibitem[{Fischl(2012)}]{fischl2012freesurfer}
\bibinfo{author}{Fischl, B.}, \bibinfo{year}{2012}.
\newblock \bibinfo{title}{Freesurfer}.
\newblock \bibinfo{journal}{Neuroimage} \bibinfo{volume}{62},
  \bibinfo{pages}{774--781}.
\bibitem[{Foret et~al.(2020)Foret, Kleiner, Mobahi and
  Neyshabur}]{foret2020sharpness}
\bibinfo{author}{Foret, P.}, \bibinfo{author}{Kleiner, A.},
  \bibinfo{author}{Mobahi, H.}, \bibinfo{author}{Neyshabur, B.},
  \bibinfo{year}{2020}.
\newblock \bibinfo{title}{Sharpness-aware minimization for efficiently
  improving generalization}.
\newblock \bibinfo{journal}{arXiv preprint arXiv:2010.01412} .
\bibitem[{Friedman(2017)}]{friedman2017elements}
\bibinfo{author}{Friedman, J.H.}, \bibinfo{year}{2017}.
\newblock \bibinfo{title}{The elements of statistical learning: Data mining,
  inference, and prediction}.
\newblock \bibinfo{publisher}{springer open}.
\bibitem[{Fu et~al.(2021)Fu, Sharma, Abadi, Iliopoulos, Wang, Sun, Lo, Segars
  and Samei}]{fu2021iphantom}
\bibinfo{author}{Fu, W.}, \bibinfo{author}{Sharma, S.}, \bibinfo{author}{Abadi,
  E.}, \bibinfo{author}{Iliopoulos, A.S.}, \bibinfo{author}{Wang, Q.},
  \bibinfo{author}{Sun, X.}, \bibinfo{author}{Lo, J.Y.C.},
  \bibinfo{author}{Segars, W.P.}, \bibinfo{author}{Samei, E.},
  \bibinfo{year}{2021}.
\newblock \bibinfo{title}{iphantom: a framework for automated creation of
  individualized computational phantoms and its application to ct organ
  dosimetry}.
\newblock \bibinfo{journal}{IEEE Journal of Biomedical and Health Informatics}
  .
\bibitem[{Gal and Ghahramani(2016)}]{gal2016dropout}
\bibinfo{author}{Gal, Y.}, \bibinfo{author}{Ghahramani, Z.},
  \bibinfo{year}{2016}.
\newblock \bibinfo{title}{Dropout as a bayesian approximation: Representing
  model uncertainty in deep learning}, in: \bibinfo{booktitle}{international
  conference on machine learning}, \bibinfo{organization}{PMLR}. pp.
  \bibinfo{pages}{1050--1059}.
\bibitem[{Gear et~al.(2018)Gear, Cox, Gustafsson, Gleisner, Murray, Glatting,
  Konijnenberg and Flux}]{gear2018eanm}
\bibinfo{author}{Gear, J.I.}, \bibinfo{author}{Cox, M.G.},
  \bibinfo{author}{Gustafsson, J.}, \bibinfo{author}{Gleisner, K.S.},
  \bibinfo{author}{Murray, I.}, \bibinfo{author}{Glatting, G.},
  \bibinfo{author}{Konijnenberg, M.}, \bibinfo{author}{Flux, G.D.},
  \bibinfo{year}{2018}.
\newblock \bibinfo{title}{Eanm practical guidance on uncertainty analysis for
  molecular radiotherapy absorbed dose calculations}.
\newblock \bibinfo{journal}{European journal of nuclear medicine and molecular
  imaging} \bibinfo{volume}{45}, \bibinfo{pages}{2456--2474}.
\bibitem[{Goodfellow et~al.(2014)Goodfellow, Vinyals and
  Saxe}]{goodfellow2014qualitatively}
\bibinfo{author}{Goodfellow, I.J.}, \bibinfo{author}{Vinyals, O.},
  \bibinfo{author}{Saxe, A.M.}, \bibinfo{year}{2014}.
\newblock \bibinfo{title}{Qualitatively characterizing neural network
  optimization problems}.
\newblock \bibinfo{journal}{arXiv preprint arXiv:1412.6544} .
\bibitem[{Guo et~al.(2017)Guo, Pleiss, Sun and Weinberger}]{guo2017calibration}
\bibinfo{author}{Guo, C.}, \bibinfo{author}{Pleiss, G.}, \bibinfo{author}{Sun,
  Y.}, \bibinfo{author}{Weinberger, K.Q.}, \bibinfo{year}{2017}.
\newblock \bibinfo{title}{On calibration of modern neural networks}, in:
  \bibinfo{booktitle}{International Conference on Machine Learning},
  \bibinfo{organization}{PMLR}. pp. \bibinfo{pages}{1321--1330}.
\bibitem[{Ha et~al.(2020)Ha, Wilms and Heinrich}]{ha2020semantically}
\bibinfo{author}{Ha, I.Y.}, \bibinfo{author}{Wilms, M.},
  \bibinfo{author}{Heinrich, M.}, \bibinfo{year}{2020}.
\newblock \bibinfo{title}{Semantically guided large deformation estimation with
  deep networks}.
\newblock \bibinfo{journal}{Sensors} \bibinfo{volume}{20},
  \bibinfo{pages}{1392}.
\bibitem[{He et~al.(2016)He, Zhang, Ren and Sun}]{he2016deep}
\bibinfo{author}{He, K.}, \bibinfo{author}{Zhang, X.}, \bibinfo{author}{Ren,
  S.}, \bibinfo{author}{Sun, J.}, \bibinfo{year}{2016}.
\newblock \bibinfo{title}{Deep residual learning for image recognition}, in:
  \bibinfo{booktitle}{Proceedings of the IEEE conference on computer vision and
  pattern recognition}, pp. \bibinfo{pages}{770--778}.
\bibitem[{Heinrich et~al.(2013a)Heinrich, Jenkinson, Brady and
  Schnabel}]{heinrich2013mrf}
\bibinfo{author}{Heinrich, M.P.}, \bibinfo{author}{Jenkinson, M.},
  \bibinfo{author}{Brady, M.}, \bibinfo{author}{Schnabel, J.A.},
  \bibinfo{year}{2013}a.
\newblock \bibinfo{title}{Mrf-based deformable registration and ventilation
  estimation of lung ct}.
\newblock \bibinfo{journal}{IEEE transactions on medical imaging}
  \bibinfo{volume}{32}, \bibinfo{pages}{1239--1248}.
\bibitem[{Heinrich et~al.(2013b)Heinrich, Jenkinson, Papie{\.z}, Brady and
  Schnabel}]{heinrich2013towards}
\bibinfo{author}{Heinrich, M.P.}, \bibinfo{author}{Jenkinson, M.},
  \bibinfo{author}{Papie{\.z}, B.W.}, \bibinfo{author}{Brady, M.},
  \bibinfo{author}{Schnabel, J.A.}, \bibinfo{year}{2013}b.
\newblock \bibinfo{title}{Towards realtime multimodal fusion for image-guided
  interventions using self-similarities}, in: \bibinfo{booktitle}{International
  conference on medical image computing and computer-assisted intervention},
  \bibinfo{organization}{Springer}. pp. \bibinfo{pages}{187--194}.
\bibitem[{Heinrich et~al.(2015)Heinrich, Maier and Handels}]{heinrich2015multi}
\bibinfo{author}{Heinrich, M.P.}, \bibinfo{author}{Maier, O.},
  \bibinfo{author}{Handels, H.}, \bibinfo{year}{2015}.
\newblock \bibinfo{title}{Multi-modal multi-atlas segmentation using discrete
  optimisation and self-similarities.}
\newblock \bibinfo{journal}{VISCERAL Challenge@ ISBI} \bibinfo{volume}{1390},
  \bibinfo{pages}{27}.
\bibitem[{Hering et~al.(2021)Hering, Hansen, Mok, Chung, Siebert, H{\"a}ger,
  Lange, Kuckertz, Heldmann, Shao et~al.}]{hering2021learn2reg}
\bibinfo{author}{Hering, A.}, \bibinfo{author}{Hansen, L.},
  \bibinfo{author}{Mok, T.C.}, \bibinfo{author}{Chung, A.},
  \bibinfo{author}{Siebert, H.}, \bibinfo{author}{H{\"a}ger, S.},
  \bibinfo{author}{Lange, A.}, \bibinfo{author}{Kuckertz, S.},
  \bibinfo{author}{Heldmann, S.}, \bibinfo{author}{Shao, W.}, et~al.,
  \bibinfo{year}{2021}.
\newblock \bibinfo{title}{Learn2reg: comprehensive multi-task medical image
  registration challenge, dataset and evaluation in the era of deep learning}.
\newblock \bibinfo{journal}{arXiv preprint arXiv:2112.04489} .
\bibitem[{Hernandez et~al.(2009)Hernandez, Bossa and
  Olmos}]{hernandez2009registration}
\bibinfo{author}{Hernandez, M.}, \bibinfo{author}{Bossa, M.N.},
  \bibinfo{author}{Olmos, S.}, \bibinfo{year}{2009}.
\newblock \bibinfo{title}{Registration of anatomical images using paths of
  diffeomorphisms parameterized with stationary vector field flows}.
\newblock \bibinfo{journal}{International Journal of Computer Vision}
  \bibinfo{volume}{85}, \bibinfo{pages}{291--306}.
\bibitem[{Hoffmann et~al.(2020)Hoffmann, Billot, Iglesias, Fischl and
  Dalca}]{hoffmann2020learning}
\bibinfo{author}{Hoffmann, M.}, \bibinfo{author}{Billot, B.},
  \bibinfo{author}{Iglesias, J.E.}, \bibinfo{author}{Fischl, B.},
  \bibinfo{author}{Dalca, A.V.}, \bibinfo{year}{2020}.
\newblock \bibinfo{title}{Learning image registration without images}.
\newblock \bibinfo{journal}{arXiv preprint arXiv:2004.10282} .
\bibitem[{Hoopes et~al.(2021)Hoopes, Hoffmann, Fischl, Guttag and
  Dalca}]{hoopes2021hypermorph}
\bibinfo{author}{Hoopes, A.}, \bibinfo{author}{Hoffmann, M.},
  \bibinfo{author}{Fischl, B.}, \bibinfo{author}{Guttag, J.},
  \bibinfo{author}{Dalca, A.V.}, \bibinfo{year}{2021}.
\newblock \bibinfo{title}{Hypermorph: amortized hyperparameter learning for
  image registration}, in: \bibinfo{booktitle}{International Conference on
  Information Processing in Medical Imaging}, \bibinfo{organization}{Springer}.
  pp. \bibinfo{pages}{3--17}.
\bibitem[{Im et~al.(2016)Im, Tao and Branson}]{im2016empirical}
\bibinfo{author}{Im, D.J.}, \bibinfo{author}{Tao, M.},
  \bibinfo{author}{Branson, K.}, \bibinfo{year}{2016}.
\newblock \bibinfo{title}{An empirical analysis of deep network loss surfaces}
  .
\bibitem[{Isensee et~al.(2021)Isensee, Jaeger, Kohl, Petersen and
  Maier-Hein}]{isensee2021nnu}
\bibinfo{author}{Isensee, F.}, \bibinfo{author}{Jaeger, P.F.},
  \bibinfo{author}{Kohl, S.A.}, \bibinfo{author}{Petersen, J.},
  \bibinfo{author}{Maier-Hein, K.H.}, \bibinfo{year}{2021}.
\newblock \bibinfo{title}{nnu-net: a self-configuring method for deep
  learning-based biomedical image segmentation}.
\newblock \bibinfo{journal}{Nature methods} \bibinfo{volume}{18},
  \bibinfo{pages}{203--211}.
\bibitem[{Jaderberg et~al.(2015)Jaderberg, Simonyan, Zisserman
  et~al.}]{jaderberg2015spatial}
\bibinfo{author}{Jaderberg, M.}, \bibinfo{author}{Simonyan, K.},
  \bibinfo{author}{Zisserman, A.}, et~al., \bibinfo{year}{2015}.
\newblock \bibinfo{title}{Spatial transformer networks}.
\newblock \bibinfo{journal}{Advances in neural information processing systems}
  \bibinfo{volume}{28}, \bibinfo{pages}{2017--2025}.
\bibitem[{Jha et~al.(2019)Jha, Smedsrud, Riegler, Johansen, De~Lange, Halvorsen
  and Johansen}]{jha2019resunet++}
\bibinfo{author}{Jha, D.}, \bibinfo{author}{Smedsrud, P.H.},
  \bibinfo{author}{Riegler, M.A.}, \bibinfo{author}{Johansen, D.},
  \bibinfo{author}{De~Lange, T.}, \bibinfo{author}{Halvorsen, P.},
  \bibinfo{author}{Johansen, H.D.}, \bibinfo{year}{2019}.
\newblock \bibinfo{title}{Resunet++: An advanced architecture for medical image
  segmentation}, in: \bibinfo{booktitle}{2019 IEEE International Symposium on
  Multimedia (ISM)}, \bibinfo{organization}{IEEE}. pp.
  \bibinfo{pages}{225--2255}.
\bibitem[{Johnson and Christensen(2002)}]{johnson2002consistent}
\bibinfo{author}{Johnson, H.J.}, \bibinfo{author}{Christensen, G.E.},
  \bibinfo{year}{2002}.
\newblock \bibinfo{title}{Consistent landmark and intensity-based image
  registration}.
\newblock \bibinfo{journal}{IEEE transactions on medical imaging}
  \bibinfo{volume}{21}, \bibinfo{pages}{450--461}.
\bibitem[{Kendall and Gal(2017)}]{Kendall2017}
\bibinfo{author}{Kendall, A.}, \bibinfo{author}{Gal, Y.}, \bibinfo{year}{2017}.
\newblock \bibinfo{title}{{What uncertainties do we need in Bayesian deep
  learning for computer vision?}}, in: \bibinfo{booktitle}{Advances in Neural
  Information Processing Systems}, \bibinfo{publisher}{Neural information
  processing systems foundation}. pp. \bibinfo{pages}{5575--5585}.
\newblock \URLprefix \url{https://arxiv.org/abs/1703.04977v2},
  \href{http://arxiv.org/abs/1703.04977}{\tt arXiv:1703.04977}.
\bibitem[{Keskar et~al.(2016)Keskar, Mudigere, Nocedal, Smelyanskiy and
  Tang}]{keskar2016large}
\bibinfo{author}{Keskar, N.S.}, \bibinfo{author}{Mudigere, D.},
  \bibinfo{author}{Nocedal, J.}, \bibinfo{author}{Smelyanskiy, M.},
  \bibinfo{author}{Tang, P.T.P.}, \bibinfo{year}{2016}.
\newblock \bibinfo{title}{On large-batch training for deep learning:
  Generalization gap and sharp minima}.
\newblock \bibinfo{journal}{arXiv preprint arXiv:1609.04836} .
\bibitem[{Kim et~al.(2021)Kim, Kim, Park, Kim, Lee and Ye}]{kim2021cyclemorph}
\bibinfo{author}{Kim, B.}, \bibinfo{author}{Kim, D.H.}, \bibinfo{author}{Park,
  S.H.}, \bibinfo{author}{Kim, J.}, \bibinfo{author}{Lee, J.G.},
  \bibinfo{author}{Ye, J.C.}, \bibinfo{year}{2021}.
\newblock \bibinfo{title}{Cyclemorph: Cycle consistent unsupervised deformable
  image registration}.
\newblock \bibinfo{journal}{Medical Image Analysis} \bibinfo{volume}{71},
  \bibinfo{pages}{102036}.
\bibitem[{Kingma and Welling(2013)}]{kingma2013auto}
\bibinfo{author}{Kingma, D.P.}, \bibinfo{author}{Welling, M.},
  \bibinfo{year}{2013}.
\newblock \bibinfo{title}{Auto-encoding variational bayes}.
\newblock \bibinfo{journal}{arXiv preprint arXiv:1312.6114} .
\bibitem[{Krebs et~al.(2019)Krebs, Delingette, Mailh{\'e}, Ayache and
  Mansi}]{krebs2019learning}
\bibinfo{author}{Krebs, J.}, \bibinfo{author}{Delingette, H.},
  \bibinfo{author}{Mailh{\'e}, B.}, \bibinfo{author}{Ayache, N.},
  \bibinfo{author}{Mansi, T.}, \bibinfo{year}{2019}.
\newblock \bibinfo{title}{Learning a probabilistic model for diffeomorphic
  registration}.
\newblock \bibinfo{journal}{IEEE transactions on medical imaging}
  \bibinfo{volume}{38}, \bibinfo{pages}{2165--2176}.
\bibitem[{Kuleshov et~al.(2018)Kuleshov, Fenner and
  Ermon}]{kuleshov2018accurate}
\bibinfo{author}{Kuleshov, V.}, \bibinfo{author}{Fenner, N.},
  \bibinfo{author}{Ermon, S.}, \bibinfo{year}{2018}.
\newblock \bibinfo{title}{Accurate uncertainties for deep learning using
  calibrated regression}, in: \bibinfo{booktitle}{International Conference on
  Machine Learning}, \bibinfo{organization}{PMLR}. pp.
  \bibinfo{pages}{2796--2804}.
\bibitem[{Kybic(2009)}]{kybic2009bootstrap}
\bibinfo{author}{Kybic, J.}, \bibinfo{year}{2009}.
\newblock \bibinfo{title}{Bootstrap resampling for image registration
  uncertainty estimation without ground truth}.
\newblock \bibinfo{journal}{IEEE Transactions on Image Processing}
  \bibinfo{volume}{19}, \bibinfo{pages}{64--73}.
\bibitem[{Laves et~al.(2020a)Laves, Ihler, Fast, Kahrs and
  Ortmaier}]{pmlr-v121-laves20a}
\bibinfo{author}{Laves, M.H.}, \bibinfo{author}{Ihler, S.},
  \bibinfo{author}{Fast, J.F.}, \bibinfo{author}{Kahrs, L.A.},
  \bibinfo{author}{Ortmaier, T.}, \bibinfo{year}{2020}a.
\newblock \bibinfo{title}{Well-calibrated regression uncertainty in medical
  imaging with deep learning}, in: \bibinfo{editor}{Arbel, T.},
  \bibinfo{editor}{Ben~Ayed, I.}, \bibinfo{editor}{de~Bruijne, M.},
  \bibinfo{editor}{Descoteaux, M.}, \bibinfo{editor}{Lombaert, H.},
  \bibinfo{editor}{Pal, C.} (Eds.), \bibinfo{booktitle}{Proceedings of the
  Third Conference on Medical Imaging with Deep Learning},
  \bibinfo{publisher}{PMLR}. pp. \bibinfo{pages}{393--412}.
\newblock \URLprefix \url{https://proceedings.mlr.press/v121/laves20a.html}.
\bibitem[{Laves et~al.(2019)Laves, Ihler, Kortmann and
  Ortmaier}]{laves2019well}
\bibinfo{author}{Laves, M.H.}, \bibinfo{author}{Ihler, S.},
  \bibinfo{author}{Kortmann, K.P.}, \bibinfo{author}{Ortmaier, T.},
  \bibinfo{year}{2019}.
\newblock \bibinfo{title}{Well-calibrated model uncertainty with temperature
  scaling for dropout variational inference}.
\newblock \bibinfo{journal}{arXiv preprint arXiv:1909.13550} .
\bibitem[{Laves et~al.(2020b)Laves, T{\"{o}}lle and Ortmaier}]{Laves2020}
\bibinfo{author}{Laves, M.H.}, \bibinfo{author}{T{\"{o}}lle, M.},
  \bibinfo{author}{Ortmaier, T.}, \bibinfo{year}{2020}b.
\newblock \bibinfo{title}{{Uncertainty Estimation in Medical Image Denoising
  with Bayesian Deep Image Prior}}.
\newblock \bibinfo{journal}{Lecture Notes in Computer Science (including
  subseries Lecture Notes in Artificial Intelligence and Lecture Notes in
  Bioinformatics)} \bibinfo{volume}{12443 LNCS}, \bibinfo{pages}{81--96}.
\newblock \URLprefix \url{https://arxiv.org/abs/2008.08837v1},
  \href{http://arxiv.org/abs/2008.08837}{\tt arXiv:2008.08837}.
\bibitem[{Laves et~al.(2020c)Laves, T{\"o}lle and
  Ortmaier}]{laves2020uncertainty}
\bibinfo{author}{Laves, M.H.}, \bibinfo{author}{T{\"o}lle, M.},
  \bibinfo{author}{Ortmaier, T.}, \bibinfo{year}{2020}c.
\newblock \bibinfo{title}{Uncertainty estimation in medical image denoising
  with bayesian deep image prior}, in: \bibinfo{booktitle}{Uncertainty for Safe
  Utilization of Machine Learning in Medical Imaging, and Graphs in Biomedical
  Image Analysis}. \bibinfo{publisher}{Springer}, pp. \bibinfo{pages}{81--96}.
\bibitem[{Lei et~al.(2020)Lei, Fu, Wang, Liu, Patel, Curran, Liu and
  Yang}]{lei20204d}
\bibinfo{author}{Lei, Y.}, \bibinfo{author}{Fu, Y.}, \bibinfo{author}{Wang,
  T.}, \bibinfo{author}{Liu, Y.}, \bibinfo{author}{Patel, P.},
  \bibinfo{author}{Curran, W.J.}, \bibinfo{author}{Liu, T.},
  \bibinfo{author}{Yang, X.}, \bibinfo{year}{2020}.
\newblock \bibinfo{title}{4d-ct deformable image registration using multiscale
  unsupervised deep learning}.
\newblock \bibinfo{journal}{Physics in Medicine \& Biology}
  \bibinfo{volume}{65}, \bibinfo{pages}{085003}.
\bibitem[{Levi et~al.(2019)Levi, Gispan, Giladi and
  Fetaya}]{levi2019evaluating}
\bibinfo{author}{Levi, D.}, \bibinfo{author}{Gispan, L.},
  \bibinfo{author}{Giladi, N.}, \bibinfo{author}{Fetaya, E.},
  \bibinfo{year}{2019}.
\newblock \bibinfo{title}{Evaluating and calibrating uncertainty prediction in
  regression tasks}.
\newblock \bibinfo{journal}{arXiv preprint arXiv:1905.11659} .
\bibitem[{Li et~al.(2018)Li, Xu, Taylor, Studer and
  Goldstein}]{li2018visualizing}
\bibinfo{author}{Li, H.}, \bibinfo{author}{Xu, Z.}, \bibinfo{author}{Taylor,
  G.}, \bibinfo{author}{Studer, C.}, \bibinfo{author}{Goldstein, T.},
  \bibinfo{year}{2018}.
\newblock \bibinfo{title}{Visualizing the loss landscape of neural nets}.
\newblock \bibinfo{journal}{Advances in neural information processing systems}
  \bibinfo{volume}{31}.
\bibitem[{Li et~al.(2021)Li, Sui, Luo, Xu, Liu and Goh}]{li2021medical}
\bibinfo{author}{Li, S.}, \bibinfo{author}{Sui, X.}, \bibinfo{author}{Luo, X.},
  \bibinfo{author}{Xu, X.}, \bibinfo{author}{Liu, Y.}, \bibinfo{author}{Goh,
  R.S.M.}, \bibinfo{year}{2021}.
\newblock \bibinfo{title}{Medical image segmentation using
  squeeze-and-expansion transformers}.
\newblock \bibinfo{journal}{arXiv preprint arXiv:2105.09511} .
\bibitem[{Lian et~al.(2018)Lian, Liu, Zhang and Shen}]{lian2018hierarchical}
\bibinfo{author}{Lian, C.}, \bibinfo{author}{Liu, M.}, \bibinfo{author}{Zhang,
  J.}, \bibinfo{author}{Shen, D.}, \bibinfo{year}{2018}.
\newblock \bibinfo{title}{Hierarchical fully convolutional network for joint
  atrophy localization and alzheimer's disease diagnosis using structural mri}.
\newblock \bibinfo{journal}{IEEE transactions on pattern analysis and machine
  intelligence} \bibinfo{volume}{42}, \bibinfo{pages}{880--893}.
\bibitem[{Liu et~al.(2021a)Liu, Lin, Cao, Hu, Wei, Zhang, Lin and
  Guo}]{liu2021swin}
\bibinfo{author}{Liu, Z.}, \bibinfo{author}{Lin, Y.}, \bibinfo{author}{Cao,
  Y.}, \bibinfo{author}{Hu, H.}, \bibinfo{author}{Wei, Y.},
  \bibinfo{author}{Zhang, Z.}, \bibinfo{author}{Lin, S.}, \bibinfo{author}{Guo,
  B.}, \bibinfo{year}{2021}a.
\newblock \bibinfo{title}{Swin transformer: Hierarchical vision transformer
  using shifted windows}.
\newblock \bibinfo{journal}{arXiv preprint arXiv:2103.14030} .
\bibitem[{Liu et~al.(2022)Liu, Mao, Wu, Feichtenhofer, Darrell and
  Xie}]{liu2022convnet}
\bibinfo{author}{Liu, Z.}, \bibinfo{author}{Mao, H.}, \bibinfo{author}{Wu,
  C.Y.}, \bibinfo{author}{Feichtenhofer, C.}, \bibinfo{author}{Darrell, T.},
  \bibinfo{author}{Xie, S.}, \bibinfo{year}{2022}.
\newblock \bibinfo{title}{A convnet for the 2020s}.
\newblock \bibinfo{journal}{arXiv preprint arXiv:2201.03545} .
\bibitem[{Liu et~al.(2021b)Liu, Ning, Cao, Wei, Zhang, Lin and
  Hu}]{liu2021video}
\bibinfo{author}{Liu, Z.}, \bibinfo{author}{Ning, J.}, \bibinfo{author}{Cao,
  Y.}, \bibinfo{author}{Wei, Y.}, \bibinfo{author}{Zhang, Z.},
  \bibinfo{author}{Lin, S.}, \bibinfo{author}{Hu, H.}, \bibinfo{year}{2021}b.
\newblock \bibinfo{title}{Video swin transformer}.
\newblock \bibinfo{journal}{arXiv preprint arXiv:2106.13230} .
\bibitem[{Long et~al.(2015)Long, Shelhamer and Darrell}]{long2015fully}
\bibinfo{author}{Long, J.}, \bibinfo{author}{Shelhamer, E.},
  \bibinfo{author}{Darrell, T.}, \bibinfo{year}{2015}.
\newblock \bibinfo{title}{Fully convolutional networks for semantic
  segmentation}, in: \bibinfo{booktitle}{Proceedings of the IEEE conference on
  computer vision and pattern recognition}, pp. \bibinfo{pages}{3431--3440}.
\bibitem[{Looney(1996)}]{looney1996stabilization}
\bibinfo{author}{Looney, C.G.}, \bibinfo{year}{1996}.
\newblock \bibinfo{title}{Stabilization and speedup of convergence in training
  feedforward neural networks}.
\newblock \bibinfo{journal}{Neurocomputing} \bibinfo{volume}{10},
  \bibinfo{pages}{7--31}.
\bibitem[{Luo et~al.(2019)Luo, Sedghi, Popuri, Cobzas, Zhang, Preiswerk, Toews,
  Golby, Sugiyama, Wells et~al.}]{luo2019applicability}
\bibinfo{author}{Luo, J.}, \bibinfo{author}{Sedghi, A.},
  \bibinfo{author}{Popuri, K.}, \bibinfo{author}{Cobzas, D.},
  \bibinfo{author}{Zhang, M.}, \bibinfo{author}{Preiswerk, F.},
  \bibinfo{author}{Toews, M.}, \bibinfo{author}{Golby, A.},
  \bibinfo{author}{Sugiyama, M.}, \bibinfo{author}{Wells, W.M.}, et~al.,
  \bibinfo{year}{2019}.
\newblock \bibinfo{title}{On the applicability of registration uncertainty},
  in: \bibinfo{booktitle}{International Conference on Medical Image Computing
  and Computer-Assisted Intervention}, \bibinfo{organization}{Springer}. pp.
  \bibinfo{pages}{410--419}.
\bibitem[{Luo et~al.(2016)Luo, Li, Urtasun and Zemel}]{luo2016understanding}
\bibinfo{author}{Luo, W.}, \bibinfo{author}{Li, Y.}, \bibinfo{author}{Urtasun,
  R.}, \bibinfo{author}{Zemel, R.}, \bibinfo{year}{2016}.
\newblock \bibinfo{title}{Understanding the effective receptive field in deep
  convolutional neural networks}, in: \bibinfo{booktitle}{Proceedings of the
  30th International Conference on Neural Information Processing Systems}, pp.
  \bibinfo{pages}{4905--4913}.
\bibitem[{Lv et~al.(2022)Lv, Wang, Shi, Zhang, Wang, Wang and Li}]{lv2022joint}
\bibinfo{author}{Lv, J.}, \bibinfo{author}{Wang, Z.}, \bibinfo{author}{Shi,
  H.}, \bibinfo{author}{Zhang, H.}, \bibinfo{author}{Wang, S.},
  \bibinfo{author}{Wang, Y.}, \bibinfo{author}{Li, Q.}, \bibinfo{year}{2022}.
\newblock \bibinfo{title}{Joint progressive and coarse-to-fine registration of
  brain mri via deformation field integration and non-rigid feature fusion}.
\newblock \bibinfo{journal}{IEEE Transactions on Medical Imaging} .
\bibitem[{Maas et~al.(2013)Maas, Hannun, Ng et~al.}]{maas2013rectifier}
\bibinfo{author}{Maas, A.L.}, \bibinfo{author}{Hannun, A.Y.},
  \bibinfo{author}{Ng, A.Y.}, et~al., \bibinfo{year}{2013}.
\newblock \bibinfo{title}{Rectifier nonlinearities improve neural network
  acoustic models}, in: \bibinfo{booktitle}{Proc. icml},
  \bibinfo{organization}{Citeseer}. p.~\bibinfo{pages}{3}.
\bibitem[{Marcus et~al.(2007)Marcus, Wang, Parker, Csernansky, Morris and
  Buckner}]{marcus2007open}
\bibinfo{author}{Marcus, D.S.}, \bibinfo{author}{Wang, T.H.},
  \bibinfo{author}{Parker, J.}, \bibinfo{author}{Csernansky, J.G.},
  \bibinfo{author}{Morris, J.C.}, \bibinfo{author}{Buckner, R.L.},
  \bibinfo{year}{2007}.
\newblock \bibinfo{title}{Open access series of imaging studies (oasis):
  cross-sectional mri data in young, middle aged, nondemented, and demented
  older adults}.
\newblock \bibinfo{journal}{Journal of cognitive neuroscience}
  \bibinfo{volume}{19}, \bibinfo{pages}{1498--1507}.
\bibitem[{Mehrtash et~al.(2020)Mehrtash, Wells, Tempany, Abolmaesumi and
  Kapur}]{mehrtash2020confidence}
\bibinfo{author}{Mehrtash, A.}, \bibinfo{author}{Wells, W.M.},
  \bibinfo{author}{Tempany, C.M.}, \bibinfo{author}{Abolmaesumi, P.},
  \bibinfo{author}{Kapur, T.}, \bibinfo{year}{2020}.
\newblock \bibinfo{title}{Confidence calibration and predictive uncertainty
  estimation for deep medical image segmentation}.
\newblock \bibinfo{journal}{IEEE transactions on medical imaging}
  \bibinfo{volume}{39}, \bibinfo{pages}{3868--3878}.
\bibitem[{Milletari et~al.(2016)Milletari, Navab and Ahmadi}]{milletari2016v}
\bibinfo{author}{Milletari, F.}, \bibinfo{author}{Navab, N.},
  \bibinfo{author}{Ahmadi, S.A.}, \bibinfo{year}{2016}.
\newblock \bibinfo{title}{V-net: Fully convolutional neural networks for
  volumetric medical image segmentation}, in: \bibinfo{booktitle}{2016 fourth
  international conference on 3D vision (3DV)}, \bibinfo{organization}{IEEE}.
  pp. \bibinfo{pages}{565--571}.
\bibitem[{Modat et~al.(2010)Modat, Ridgway, Taylor, Lehmann, Barnes, Hawkes,
  Fox and Ourselin}]{modat2010fast}
\bibinfo{author}{Modat, M.}, \bibinfo{author}{Ridgway, G.R.},
  \bibinfo{author}{Taylor, Z.A.}, \bibinfo{author}{Lehmann, M.},
  \bibinfo{author}{Barnes, J.}, \bibinfo{author}{Hawkes, D.J.},
  \bibinfo{author}{Fox, N.C.}, \bibinfo{author}{Ourselin, S.},
  \bibinfo{year}{2010}.
\newblock \bibinfo{title}{Fast free-form deformation using graphics processing
  units}.
\newblock \bibinfo{journal}{Computer methods and programs in biomedicine}
  \bibinfo{volume}{98}, \bibinfo{pages}{278--284}.
\bibitem[{Mok and Chung(2020)}]{mok2020fast}
\bibinfo{author}{Mok, T.C.}, \bibinfo{author}{Chung, A.}, \bibinfo{year}{2020}.
\newblock \bibinfo{title}{Fast symmetric diffeomorphic image registration with
  convolutional neural networks}, in: \bibinfo{booktitle}{Proceedings of the
  IEEE/CVF conference on computer vision and pattern recognition}, pp.
  \bibinfo{pages}{4644--4653}.
\bibitem[{Mok and Chung(2021)}]{mok2021conditional}
\bibinfo{author}{Mok, T.C.}, \bibinfo{author}{Chung, A.}, \bibinfo{year}{2021}.
\newblock \bibinfo{title}{Conditional deformable image registration with
  convolutional neural network}, in: \bibinfo{booktitle}{International
  Conference on Medical Image Computing and Computer-Assisted Intervention},
  \bibinfo{organization}{Springer}. pp. \bibinfo{pages}{35--45}.
\bibitem[{Onofrey et~al.(2013)Onofrey, Staib and
  Papademetris}]{onofrey2013semi}
\bibinfo{author}{Onofrey, J.A.}, \bibinfo{author}{Staib, L.H.},
  \bibinfo{author}{Papademetris, X.}, \bibinfo{year}{2013}.
\newblock \bibinfo{title}{Semi-supervised learning of nonrigid deformations for
  image registration}, in: \bibinfo{booktitle}{International MICCAI Workshop on
  Medical Computer Vision}, \bibinfo{organization}{Springer}. pp.
  \bibinfo{pages}{13--23}.
\bibitem[{Pace et~al.(2013)Pace, Aylward and Niethammer}]{pace2013locally}
\bibinfo{author}{Pace, D.F.}, \bibinfo{author}{Aylward, S.R.},
  \bibinfo{author}{Niethammer, M.}, \bibinfo{year}{2013}.
\newblock \bibinfo{title}{A locally adaptive regularization based on
  anisotropic diffusion for deformable image registration of sliding organs}.
\newblock \bibinfo{journal}{IEEE transactions on medical imaging}
  \bibinfo{volume}{32}, \bibinfo{pages}{2114--2126}.
\bibitem[{Park and Kim(2022)}]{park2022vision}
\bibinfo{author}{Park, N.}, \bibinfo{author}{Kim, S.}, \bibinfo{year}{2022}.
\newblock \bibinfo{title}{How do vision transformers work?}
\newblock \bibinfo{journal}{arXiv preprint arXiv:2202.06709} .
\bibitem[{Paszke et~al.(2019)Paszke, Gross, Massa, Lerer, Bradbury, Chanan,
  Killeen, Lin, Gimelshein, Antiga et~al.}]{paszke2019pytorch}
\bibinfo{author}{Paszke, A.}, \bibinfo{author}{Gross, S.},
  \bibinfo{author}{Massa, F.}, \bibinfo{author}{Lerer, A.},
  \bibinfo{author}{Bradbury, J.}, \bibinfo{author}{Chanan, G.},
  \bibinfo{author}{Killeen, T.}, \bibinfo{author}{Lin, Z.},
  \bibinfo{author}{Gimelshein, N.}, \bibinfo{author}{Antiga, L.}, et~al.,
  \bibinfo{year}{2019}.
\newblock \bibinfo{title}{Pytorch: An imperative style, high-performance deep
  learning library}.
\newblock \bibinfo{journal}{Advances in neural information processing systems}
  \bibinfo{volume}{32}, \bibinfo{pages}{8026--8037}.
\bibitem[{Phan et~al.(2018)Phan, Salay, Czarnecki, Abdelzad, Denouden and
  Vernekar}]{phan2018calibrating}
\bibinfo{author}{Phan, B.}, \bibinfo{author}{Salay, R.},
  \bibinfo{author}{Czarnecki, K.}, \bibinfo{author}{Abdelzad, V.},
  \bibinfo{author}{Denouden, T.}, \bibinfo{author}{Vernekar, S.},
  \bibinfo{year}{2018}.
\newblock \bibinfo{title}{Calibrating uncertainties in object localization
  task}.
\newblock \bibinfo{journal}{arXiv preprint arXiv:1811.11210} .
\bibitem[{Qiu et~al.(2021)Qiu, Qin, Schuh, Hammernik and
  Rueckert}]{qiu2021learning}
\bibinfo{author}{Qiu, H.}, \bibinfo{author}{Qin, C.}, \bibinfo{author}{Schuh,
  A.}, \bibinfo{author}{Hammernik, K.}, \bibinfo{author}{Rueckert, D.},
  \bibinfo{year}{2021}.
\newblock \bibinfo{title}{Learning diffeomorphic and modality-invariant
  registration using b-splines}, in: \bibinfo{booktitle}{Medical Imaging with
  Deep Learning}.
\bibitem[{Raghu et~al.(2021)Raghu, Unterthiner, Kornblith, Zhang and
  Dosovitskiy}]{raghu2021vision}
\bibinfo{author}{Raghu, M.}, \bibinfo{author}{Unterthiner, T.},
  \bibinfo{author}{Kornblith, S.}, \bibinfo{author}{Zhang, C.},
  \bibinfo{author}{Dosovitskiy, A.}, \bibinfo{year}{2021}.
\newblock \bibinfo{title}{Do vision transformers see like convolutional neural
  networks?}
\newblock \bibinfo{journal}{arXiv preprint arXiv:2108.08810} .
\bibitem[{Redmon et~al.(2016)Redmon, Divvala, Girshick and
  Farhadi}]{redmon2016you}
\bibinfo{author}{Redmon, J.}, \bibinfo{author}{Divvala, S.},
  \bibinfo{author}{Girshick, R.}, \bibinfo{author}{Farhadi, A.},
  \bibinfo{year}{2016}.
\newblock \bibinfo{title}{You only look once: Unified, real-time object
  detection}, in: \bibinfo{booktitle}{Proceedings of the IEEE conference on
  computer vision and pattern recognition}, pp. \bibinfo{pages}{779--788}.
\bibitem[{Risholm et~al.(2011)Risholm, Balter and
  Wells}]{risholm2011estimation}
\bibinfo{author}{Risholm, P.}, \bibinfo{author}{Balter, J.},
  \bibinfo{author}{Wells, W.M.}, \bibinfo{year}{2011}.
\newblock \bibinfo{title}{Estimation of delivered dose in radiotherapy: the
  influence of registration uncertainty}, in: \bibinfo{booktitle}{International
  Conference on Medical Image Computing and Computer-Assisted Intervention},
  \bibinfo{organization}{Springer}. pp. \bibinfo{pages}{548--555}.
\bibitem[{Risholm et~al.(2013)Risholm, Janoos, Norton, Golby and
  Wells~III}]{risholm2013bayesian}
\bibinfo{author}{Risholm, P.}, \bibinfo{author}{Janoos, F.},
  \bibinfo{author}{Norton, I.}, \bibinfo{author}{Golby, A.J.},
  \bibinfo{author}{Wells~III, W.M.}, \bibinfo{year}{2013}.
\newblock \bibinfo{title}{Bayesian characterization of uncertainty in
  intra-subject non-rigid registration}.
\newblock \bibinfo{journal}{Medical image analysis} \bibinfo{volume}{17},
  \bibinfo{pages}{538--555}.
\bibitem[{Roh{\'e} et~al.(2017)Roh{\'e}, Datar, Heimann, Sermesant and
  Pennec}]{rohe2017svf}
\bibinfo{author}{Roh{\'e}, M.M.}, \bibinfo{author}{Datar, M.},
  \bibinfo{author}{Heimann, T.}, \bibinfo{author}{Sermesant, M.},
  \bibinfo{author}{Pennec, X.}, \bibinfo{year}{2017}.
\newblock \bibinfo{title}{Svf-net: Learning deformable image registration using
  shape matching}, in: \bibinfo{booktitle}{International conference on medical
  image computing and computer-assisted intervention},
  \bibinfo{organization}{Springer}. pp. \bibinfo{pages}{266--274}.
\bibitem[{Ronneberger et~al.(2015)Ronneberger, Fischer and
  Brox}]{ronneberger2015u}
\bibinfo{author}{Ronneberger, O.}, \bibinfo{author}{Fischer, P.},
  \bibinfo{author}{Brox, T.}, \bibinfo{year}{2015}.
\newblock \bibinfo{title}{U-net: Convolutional networks for biomedical image
  segmentation}, in: \bibinfo{booktitle}{International Conference on Medical
  image computing and computer-assisted intervention},
  \bibinfo{organization}{Springer}. pp. \bibinfo{pages}{234--241}.
\bibitem[{Rueckert et~al.(1999)Rueckert, Sonoda, Hayes, Hill, Leach and
  Hawkes}]{rueckert1999nonrigid}
\bibinfo{author}{Rueckert, D.}, \bibinfo{author}{Sonoda, L.I.},
  \bibinfo{author}{Hayes, C.}, \bibinfo{author}{Hill, D.L.},
  \bibinfo{author}{Leach, M.O.}, \bibinfo{author}{Hawkes, D.J.},
  \bibinfo{year}{1999}.
\newblock \bibinfo{title}{Nonrigid registration using free-form deformations:
  application to breast mr images}.
\newblock \bibinfo{journal}{IEEE transactions on medical imaging}
  \bibinfo{volume}{18}, \bibinfo{pages}{712--721}.
\bibitem[{Santurkar et~al.(2018)Santurkar, Tsipras, Ilyas and
  Madry}]{santurkar2018does}
\bibinfo{author}{Santurkar, S.}, \bibinfo{author}{Tsipras, D.},
  \bibinfo{author}{Ilyas, A.}, \bibinfo{author}{Madry, A.},
  \bibinfo{year}{2018}.
\newblock \bibinfo{title}{How does batch normalization help optimization?}
\newblock \bibinfo{journal}{Advances in neural information processing systems}
  \bibinfo{volume}{31}.
\bibitem[{Segars et~al.(2013)Segars, Bond, Frush, Hon, Eckersley, Williams,
  Feng, Tward, Ratnanather, Miller et~al.}]{segars2013population}
\bibinfo{author}{Segars, W.}, \bibinfo{author}{Bond, J.},
  \bibinfo{author}{Frush, J.}, \bibinfo{author}{Hon, S.},
  \bibinfo{author}{Eckersley, C.}, \bibinfo{author}{Williams, C.H.},
  \bibinfo{author}{Feng, J.}, \bibinfo{author}{Tward, D.J.},
  \bibinfo{author}{Ratnanather, J.}, \bibinfo{author}{Miller, M.}, et~al.,
  \bibinfo{year}{2013}.
\newblock \bibinfo{title}{Population of anatomically variable 4d xcat adult
  phantoms for imaging research and optimization}.
\newblock \bibinfo{journal}{Medical physics} \bibinfo{volume}{40},
  \bibinfo{pages}{043701}.
\bibitem[{Segars et~al.(2010)Segars, Sturgeon, Mendonca, Grimes and
  Tsui}]{segars20104d}
\bibinfo{author}{Segars, W.P.}, \bibinfo{author}{Sturgeon, G.},
  \bibinfo{author}{Mendonca, S.}, \bibinfo{author}{Grimes, J.},
  \bibinfo{author}{Tsui, B.M.}, \bibinfo{year}{2010}.
\newblock \bibinfo{title}{4d xcat phantom for multimodality imaging research}.
\newblock \bibinfo{journal}{Medical physics} \bibinfo{volume}{37},
  \bibinfo{pages}{4902--4915}.
\bibitem[{Siebert et~al.(2021)Siebert, Hansen and Heinrich}]{siebert2021fast}
\bibinfo{author}{Siebert, H.}, \bibinfo{author}{Hansen, L.},
  \bibinfo{author}{Heinrich, M.P.}, \bibinfo{year}{2021}.
\newblock \bibinfo{title}{Fast 3d registration with accurate optimisation and
  little learning for learn2reg 2021}.
\newblock \bibinfo{journal}{arXiv preprint arXiv:2112.03053} .
\bibitem[{Simpson et~al.(2011)Simpson, Woolrich, Groves and
  Schnabel}]{simpson2011longitudinal}
\bibinfo{author}{Simpson, I.J.}, \bibinfo{author}{Woolrich, M.},
  \bibinfo{author}{Groves, A.R.}, \bibinfo{author}{Schnabel, J.A.},
  \bibinfo{year}{2011}.
\newblock \bibinfo{title}{Longitudinal brain mri analysis with uncertain
  registration}, in: \bibinfo{booktitle}{International Conference on Medical
  Image Computing and Computer-Assisted Intervention},
  \bibinfo{organization}{Springer}. pp. \bibinfo{pages}{647--654}.
\bibitem[{Smith and Topin(2019)}]{smith2019super}
\bibinfo{author}{Smith, L.N.}, \bibinfo{author}{Topin, N.},
  \bibinfo{year}{2019}.
\newblock \bibinfo{title}{Super-convergence: Very fast training of neural
  networks using large learning rates}, in: \bibinfo{booktitle}{Artificial
  intelligence and machine learning for multi-domain operations applications},
  \bibinfo{organization}{International Society for Optics and Photonics}. p.
  \bibinfo{pages}{1100612}.
\bibitem[{Sokooti et~al.(2017)Sokooti, De~Vos, Berendsen, Lelieveldt,
  I{\v{s}}gum and Staring}]{sokooti2017nonrigid}
\bibinfo{author}{Sokooti, H.}, \bibinfo{author}{De~Vos, B.},
  \bibinfo{author}{Berendsen, F.}, \bibinfo{author}{Lelieveldt, B.P.},
  \bibinfo{author}{I{\v{s}}gum, I.}, \bibinfo{author}{Staring, M.},
  \bibinfo{year}{2017}.
\newblock \bibinfo{title}{Nonrigid image registration using multi-scale 3d
  convolutional neural networks}, in: \bibinfo{booktitle}{International
  conference on medical image computing and computer-assisted intervention},
  \bibinfo{organization}{Springer}. pp. \bibinfo{pages}{232--239}.
\bibitem[{Sutskever et~al.(2013)Sutskever, Martens, Dahl and
  Hinton}]{sutskever2013importance}
\bibinfo{author}{Sutskever, I.}, \bibinfo{author}{Martens, J.},
  \bibinfo{author}{Dahl, G.}, \bibinfo{author}{Hinton, G.},
  \bibinfo{year}{2013}.
\newblock \bibinfo{title}{On the importance of initialization and momentum in
  deep learning}, in: \bibinfo{booktitle}{International conference on machine
  learning}, \bibinfo{organization}{PMLR}. pp. \bibinfo{pages}{1139--1147}.
\bibitem[{T{\"{o}}lle et~al.(2021)T{\"{o}}lle, Laves and Schlaefer}]{Tolle2021}
\bibinfo{author}{T{\"{o}}lle, M.}, \bibinfo{author}{Laves, M.H.},
  \bibinfo{author}{Schlaefer, A.}, \bibinfo{year}{2021}.
\newblock \bibinfo{title}{{A Mean-Field Variational Inference Approach to Deep
  Image Prior for Inverse Problems in Medical Imaging}}.
\newblock \bibinfo{journal}{Medical Imaging with Deep Learning} ,
  \bibinfo{pages}{698--713}\URLprefix
  \url{https://openreview.net/forum?id=DvV%5C_blKLiB4}.
\bibitem[{Vaswani et~al.(2017)Vaswani, Shazeer, Parmar, Uszkoreit, Jones,
  Gomez, Kaiser and Polosukhin}]{vaswani2017attention}
\bibinfo{author}{Vaswani, A.}, \bibinfo{author}{Shazeer, N.},
  \bibinfo{author}{Parmar, N.}, \bibinfo{author}{Uszkoreit, J.},
  \bibinfo{author}{Jones, L.}, \bibinfo{author}{Gomez, A.N.},
  \bibinfo{author}{Kaiser, L.}, \bibinfo{author}{Polosukhin, I.},
  \bibinfo{year}{2017}.
\newblock \bibinfo{title}{Attention is all you need}.
\newblock \bibinfo{journal}{arXiv preprint arXiv:1706.03762} .
\bibitem[{Vercauteren et~al.(2009)Vercauteren, Pennec, Perchant and
  Ayache}]{vercauteren2009diffeomorphic}
\bibinfo{author}{Vercauteren, T.}, \bibinfo{author}{Pennec, X.},
  \bibinfo{author}{Perchant, A.}, \bibinfo{author}{Ayache, N.},
  \bibinfo{year}{2009}.
\newblock \bibinfo{title}{Diffeomorphic demons: Efficient non-parametric image
  registration}.
\newblock \bibinfo{journal}{NeuroImage} \bibinfo{volume}{45},
  \bibinfo{pages}{S61--S72}.
\bibitem[{Vickress et~al.(2017)Vickress, Battista, Barnett and
  Yartsev}]{vickress2017representing}
\bibinfo{author}{Vickress, J.}, \bibinfo{author}{Battista, J.},
  \bibinfo{author}{Barnett, R.}, \bibinfo{author}{Yartsev, S.},
  \bibinfo{year}{2017}.
\newblock \bibinfo{title}{Representing the dosimetric impact of deformable
  image registration errors}.
\newblock \bibinfo{journal}{Physics in Medicine \& Biology}
  \bibinfo{volume}{62}, \bibinfo{pages}{N391}.
\bibitem[{Viola and Wells~III(1997)}]{viola1997alignment}
\bibinfo{author}{Viola, P.}, \bibinfo{author}{Wells~III, W.M.},
  \bibinfo{year}{1997}.
\newblock \bibinfo{title}{Alignment by maximization of mutual information}.
\newblock \bibinfo{journal}{International journal of computer vision}
  \bibinfo{volume}{24}, \bibinfo{pages}{137--154}.
\bibitem[{Vishnevskiy et~al.(2016)Vishnevskiy, Gass, Szekely, Tanner and
  Goksel}]{vishnevskiy2016isotropic}
\bibinfo{author}{Vishnevskiy, V.}, \bibinfo{author}{Gass, T.},
  \bibinfo{author}{Szekely, G.}, \bibinfo{author}{Tanner, C.},
  \bibinfo{author}{Goksel, O.}, \bibinfo{year}{2016}.
\newblock \bibinfo{title}{Isotropic total variation regularization of
  displacements in parametric image registration}.
\newblock \bibinfo{journal}{IEEE transactions on medical imaging}
  \bibinfo{volume}{36}, \bibinfo{pages}{385--395}.
\bibitem[{de~Vos et~al.(2019)de~Vos, Berendsen, Viergever, Sokooti, Staring and
  I{\v{s}}gum}]{de2019deep}
\bibinfo{author}{de~Vos, B.D.}, \bibinfo{author}{Berendsen, F.F.},
  \bibinfo{author}{Viergever, M.A.}, \bibinfo{author}{Sokooti, H.},
  \bibinfo{author}{Staring, M.}, \bibinfo{author}{I{\v{s}}gum, I.},
  \bibinfo{year}{2019}.
\newblock \bibinfo{title}{A deep learning framework for unsupervised affine and
  deformable image registration}.
\newblock \bibinfo{journal}{Medical image analysis} \bibinfo{volume}{52},
  \bibinfo{pages}{128--143}.
\bibitem[{de~Vos et~al.(2017)de~Vos, Berendsen, Viergever, Staring and
  I{\v{s}}gum}]{de2017end}
\bibinfo{author}{de~Vos, B.D.}, \bibinfo{author}{Berendsen, F.F.},
  \bibinfo{author}{Viergever, M.A.}, \bibinfo{author}{Staring, M.},
  \bibinfo{author}{I{\v{s}}gum, I.}, \bibinfo{year}{2017}.
\newblock \bibinfo{title}{End-to-end unsupervised deformable image registration
  with a convolutional neural network}, in: \bibinfo{booktitle}{Deep learning
  in medical image analysis and multimodal learning for clinical decision
  support}. \bibinfo{publisher}{Springer}, pp. \bibinfo{pages}{204--212}.
\bibitem[{Wang et~al.(2021a)Wang, Wu and Yu}]{wang2021ted}
\bibinfo{author}{Wang, D.}, \bibinfo{author}{Wu, Z.}, \bibinfo{author}{Yu, H.},
  \bibinfo{year}{2021}a.
\newblock \bibinfo{title}{Ted-net: Convolution-free t2t vision
  transformer-based encoder-decoder dilation network for low-dose ct
  denoising}.
\newblock \bibinfo{journal}{arXiv preprint arXiv:2106.04650} .
\bibitem[{Wang et~al.(2021b)Wang, Chen, Ding, Li, Yu and
  Zha}]{wang2021transbts}
\bibinfo{author}{Wang, W.}, \bibinfo{author}{Chen, C.}, \bibinfo{author}{Ding,
  M.}, \bibinfo{author}{Li, J.}, \bibinfo{author}{Yu, H.},
  \bibinfo{author}{Zha, S.}, \bibinfo{year}{2021}b.
\newblock \bibinfo{title}{Transbts: Multimodal brain tumor segmentation using
  transformer}.
\newblock \bibinfo{journal}{arXiv preprint arXiv:2103.04430} .
\bibitem[{Wang et~al.(2021c)Wang, Xie, Li, Fan, Song, Liang, Lu, Luo and
  Shao}]{wang2021pyramid}
\bibinfo{author}{Wang, W.}, \bibinfo{author}{Xie, E.}, \bibinfo{author}{Li,
  X.}, \bibinfo{author}{Fan, D.P.}, \bibinfo{author}{Song, K.},
  \bibinfo{author}{Liang, D.}, \bibinfo{author}{Lu, T.}, \bibinfo{author}{Luo,
  P.}, \bibinfo{author}{Shao, L.}, \bibinfo{year}{2021}c.
\newblock \bibinfo{title}{Pyramid vision transformer: A versatile backbone for
  dense prediction without convolutions}.
\newblock \bibinfo{journal}{arXiv preprint arXiv:2102.12122} .
\bibitem[{Wang et~al.(2004)Wang, Bovik, Sheikh and Simoncelli}]{wang2004image}
\bibinfo{author}{Wang, Z.}, \bibinfo{author}{Bovik, A.C.},
  \bibinfo{author}{Sheikh, H.R.}, \bibinfo{author}{Simoncelli, E.P.},
  \bibinfo{year}{2004}.
\newblock \bibinfo{title}{Image quality assessment: from error visibility to
  structural similarity}.
\newblock \bibinfo{journal}{IEEE transactions on image processing}
  \bibinfo{volume}{13}, \bibinfo{pages}{600--612}.
\bibitem[{Wolberg and Zokai(2000)}]{wolberg2000robust}
\bibinfo{author}{Wolberg, G.}, \bibinfo{author}{Zokai, S.},
  \bibinfo{year}{2000}.
\newblock \bibinfo{title}{Robust image registration using log-polar transform},
  in: \bibinfo{booktitle}{Proceedings 2000 International Conference on Image
  Processing (Cat. No. 00CH37101)}, \bibinfo{organization}{IEEE}. pp.
  \bibinfo{pages}{493--496}.
\bibitem[{Xie et~al.(2021)Xie, Zhang, Shen and Xia}]{xie2021cotr}
\bibinfo{author}{Xie, Y.}, \bibinfo{author}{Zhang, J.}, \bibinfo{author}{Shen,
  C.}, \bibinfo{author}{Xia, Y.}, \bibinfo{year}{2021}.
\newblock \bibinfo{title}{Cotr: Efficiently bridging cnn and transformer for 3d
  medical image segmentation}.
\newblock \bibinfo{journal}{arXiv preprint arXiv:2103.03024} .
\bibitem[{Xu et~al.(2022)Xu, Luo, Lu, Yan, Frisken, Jagadeesan, Wells~III, Li,
  Zheng and Tong}]{xu2021double}
\bibinfo{author}{Xu, Z.}, \bibinfo{author}{Luo, J.}, \bibinfo{author}{Lu, D.},
  \bibinfo{author}{Yan, J.}, \bibinfo{author}{Frisken, S.},
  \bibinfo{author}{Jagadeesan, J.}, \bibinfo{author}{Wells~III, W.},
  \bibinfo{author}{Li, X.}, \bibinfo{author}{Zheng, Y.}, \bibinfo{author}{Tong,
  R.}, \bibinfo{year}{2022}.
\newblock \bibinfo{title}{Double-uncertainty guided spatial and temporal
  consistency regularization weighting for learning-based abdominal
  registration}, in: \bibinfo{booktitle}{International Conference on Medical
  Image Computing and Computer-Assisted Intervention},
  \bibinfo{organization}{Springer}.
\bibitem[{Yang et~al.(2016)Yang, Kwitt and Niethammer}]{yang2016fast}
\bibinfo{author}{Yang, X.}, \bibinfo{author}{Kwitt, R.},
  \bibinfo{author}{Niethammer, M.}, \bibinfo{year}{2016}.
\newblock \bibinfo{title}{Fast predictive image registration}, in:
  \bibinfo{booktitle}{Deep Learning and Data Labeling for Medical
  Applications}. \bibinfo{publisher}{Springer}, pp. \bibinfo{pages}{48--57}.
\bibitem[{Yang et~al.(2017a)Yang, Kwitt, Styner and Niethammer}]{yang2017fast}
\bibinfo{author}{Yang, X.}, \bibinfo{author}{Kwitt, R.},
  \bibinfo{author}{Styner, M.}, \bibinfo{author}{Niethammer, M.},
  \bibinfo{year}{2017}a.
\newblock \bibinfo{title}{Fast predictive multimodal image registration}, in:
  \bibinfo{booktitle}{2017 IEEE 14th International Symposium on Biomedical
  Imaging (ISBI 2017)}, \bibinfo{organization}{IEEE}. pp.
  \bibinfo{pages}{858--862}.
\bibitem[{Yang et~al.(2017b)Yang, Kwitt, Styner and
  Niethammer}]{yang2017quicksilver}
\bibinfo{author}{Yang, X.}, \bibinfo{author}{Kwitt, R.},
  \bibinfo{author}{Styner, M.}, \bibinfo{author}{Niethammer, M.},
  \bibinfo{year}{2017}b.
\newblock \bibinfo{title}{Quicksilver: Fast predictive image registration--a
  deep learning approach}.
\newblock \bibinfo{journal}{NeuroImage} \bibinfo{volume}{158},
  \bibinfo{pages}{378--396}.
\bibitem[{Zeiler et~al.(2013)Zeiler, Ranzato, Monga, Mao, Yang, Le, Nguyen,
  Senior, Vanhoucke, Dean et~al.}]{zeiler2013rectified}
\bibinfo{author}{Zeiler, M.D.}, \bibinfo{author}{Ranzato, M.},
  \bibinfo{author}{Monga, R.}, \bibinfo{author}{Mao, M.},
  \bibinfo{author}{Yang, K.}, \bibinfo{author}{Le, Q.V.},
  \bibinfo{author}{Nguyen, P.}, \bibinfo{author}{Senior, A.},
  \bibinfo{author}{Vanhoucke, V.}, \bibinfo{author}{Dean, J.}, et~al.,
  \bibinfo{year}{2013}.
\newblock \bibinfo{title}{On rectified linear units for speech processing}, in:
  \bibinfo{booktitle}{2013 IEEE International Conference on Acoustics, Speech
  and Signal Processing}, \bibinfo{organization}{IEEE}. pp.
  \bibinfo{pages}{3517--3521}.
\bibitem[{Zhai et~al.(2022)Zhai, Kolesnikov, Houlsby and
  Beyer}]{zhai2022scaling}
\bibinfo{author}{Zhai, X.}, \bibinfo{author}{Kolesnikov, A.},
  \bibinfo{author}{Houlsby, N.}, \bibinfo{author}{Beyer, L.},
  \bibinfo{year}{2022}.
\newblock \bibinfo{title}{Scaling vision transformers}, in:
  \bibinfo{booktitle}{Proceedings of the IEEE/CVF Conference on Computer Vision
  and Pattern Recognition}, pp. \bibinfo{pages}{12104--12113}.
\bibitem[{Zhang(2018)}]{zhang2018inverse}
\bibinfo{author}{Zhang, J.}, \bibinfo{year}{2018}.
\newblock \bibinfo{title}{Inverse-consistent deep networks for unsupervised
  deformable image registration}.
\newblock \bibinfo{journal}{arXiv preprint arXiv:1809.03443} .
\bibitem[{Zhang et~al.(2017)Zhang, Ma, Iyengar, Zhong and Wang}]{zhang2017new}
\bibinfo{author}{Zhang, Y.}, \bibinfo{author}{Ma, J.},
  \bibinfo{author}{Iyengar, P.}, \bibinfo{author}{Zhong, Y.},
  \bibinfo{author}{Wang, J.}, \bibinfo{year}{2017}.
\newblock \bibinfo{title}{A new ct reconstruction technique using adaptive
  deformation recovery and intensity correction (adric)}.
\newblock \bibinfo{journal}{Medical physics} \bibinfo{volume}{44},
  \bibinfo{pages}{2223--2241}.
\bibitem[{Zhang et~al.(2021)Zhang, Yu, Liang, Zhao and Xing}]{zhang2021transct}
\bibinfo{author}{Zhang, Z.}, \bibinfo{author}{Yu, L.}, \bibinfo{author}{Liang,
  X.}, \bibinfo{author}{Zhao, W.}, \bibinfo{author}{Xing, L.},
  \bibinfo{year}{2021}.
\newblock \bibinfo{title}{Transct: Dual-path transformer for low dose computed
  tomography}.
\newblock \bibinfo{journal}{arXiv preprint arXiv:2103.00634} .
\bibitem[{Zhou et~al.(2021)Zhou, Guo, Zhang, Yu, Wang and
  Yu}]{zhou2021nnformer}
\bibinfo{author}{Zhou, H.Y.}, \bibinfo{author}{Guo, J.},
  \bibinfo{author}{Zhang, Y.}, \bibinfo{author}{Yu, L.}, \bibinfo{author}{Wang,
  L.}, \bibinfo{author}{Yu, Y.}, \bibinfo{year}{2021}.
\newblock \bibinfo{title}{nnformer: Interleaved transformer for volumetric
  segmentation}.
\newblock \href{http://arxiv.org/abs/2109.03201}{\tt arXiv:2109.03201}.
\bibitem[{Zhou et~al.(2019)Zhou, Siddiquee, Tajbakhsh and
  Liang}]{zhou2019unet++}
\bibinfo{author}{Zhou, Z.}, \bibinfo{author}{Siddiquee, M.M.R.},
  \bibinfo{author}{Tajbakhsh, N.}, \bibinfo{author}{Liang, J.},
  \bibinfo{year}{2019}.
\newblock \bibinfo{title}{Unet++: Redesigning skip connections to exploit
  multiscale features in image segmentation}.
\newblock \bibinfo{journal}{IEEE transactions on medical imaging}
  \bibinfo{volume}{39}, \bibinfo{pages}{1856--1867}.
\bibitem[{Zhu et~al.(2018)Zhu, Liu, Cauley, Rosen and Rosen}]{zhu2018image}
\bibinfo{author}{Zhu, B.}, \bibinfo{author}{Liu, J.Z.},
  \bibinfo{author}{Cauley, S.F.}, \bibinfo{author}{Rosen, B.R.},
  \bibinfo{author}{Rosen, M.S.}, \bibinfo{year}{2018}.
\newblock \bibinfo{title}{Image reconstruction by domain-transform manifold
  learning}.
\newblock \bibinfo{journal}{Nature} \bibinfo{volume}{555},
  \bibinfo{pages}{487--492}.

\end{thebibliography}
